\documentclass[final,1p,times]{elsarticle}
\usepackage{lscape}

\usepackage{atlasphysics} 
                          
\usepackage{multirow} 
\usepackage{subfigure}
\RequirePackage{xspace}
 \usepackage{amsmath}
\usepackage{rotating}
\usepackage{hyperref}

\journal{Nuclear Physics B}

\begin{document}

\begin{frontmatter}
\begin{flushleft}
{\bf CERN-PH-EP-2011-041}
\end{flushleft}

\title{Measurement of the differential cross-sections of inclusive,
  prompt and non-prompt {$J/\psi$} production in proton-proton collisions at $\sqrt{s}=7$~TeV}

\author{The ATLAS Collaboration}

\begin{abstract}
{ 
\noindent The inclusive $J/\psi$ production cross-section and fraction of $J/\psi$ mesons produced in $B$-hadron decays are measured 
in proton-proton collisions at $\sqrt{s}=7$\;TeV with the ATLAS detector at the LHC, as a function of the transverse momentum and rapidity of the $J/\psi$, 
using 2.3~pb$^{-1}$ of integrated luminosity. The cross-section is measured from a minimum $p_T$ of 1~GeV to a maximum of 70~GeV and for rapidities within
$|y|<2.4$ giving the widest reach of any measurement of $J/\psi$ production to date. The differential production cross-sections
of prompt and non-prompt $J/\psi$ are separately determined and are compared to Colour Singlet NNLO$^\star$, Colour Evaporation Model, and FONLL predictions.
}
\end{abstract}

\end{frontmatter}

\section{Introduction}
\label{section:intro}

The production of heavy quarkonium at hadron colliders provides particular challenges and opportunity for insight into the theory of Quantum Chromodynamics (QCD) as its mechanisms 
of production operate at the boundary of the perturbative and non-perturbative regimes. Despite being among the most studied of the bound-quark systems, there is still no clear understanding of the mechanisms in the production of quarkonium states like the $J/\psi$ that can consistently explain both the production 
cross-section and spin-alignment measurements in $e^+e^-$, heavy-ion and hadron-hadron collisions 
(see review articles\,\cite{reviews} and references therein).

Data obtained by the Large Hadron Collider (LHC) collaborations can help to test existing theoretical models of both quarkonium production and $b$-production in a new energy regime, at higher transverse momenta and in wider rapidity ranges than have previously been studied. 
Furthermore, quarkonium production in proton-proton collisions plays a key role as a reference point to understand heavy ion collisions and to understand the interplay between production and suppression mechanisms in such collisions\,\cite{jpsiHI}.

This paper presents a measurement of the inclusive $J/\psi$ production cross-section and the production fraction $f_B$ of non-prompt $J/\psi$ (produced via the decay of a $B$-hadron) 
to inclusively-produced $J/\psi$ (hereafter referred to as the {\em non-prompt fraction}):
\begin{equation}
\label{eqn:fraction}
f_B \equiv \frac{\sigma(pp\to B+X\to J/\psi X^{\prime})}{\sigma(pp\xrightarrow[]{\textrm{Inclusive}} J/\psi X^{\prime\prime})}
\end{equation}
in the decay channel $J/\psi\to\mu^+\mu^-$ as a function of both $J/\psi$
transverse momentum and rapidity in $pp$ collisions at the LHC at a centre-of-mass energy of 7~TeV and with an integrated luminosity of up to 2.3~pb$^{-1}$.
The fraction has the advantage that acceptances and many efficiencies are the same for the numerator and denominator, and so systematic effects are reduced. 
The results of these analyses are compared to those made by the CMS Collaboration~\cite{CMS} with 314~nb${}^{-1}$ of integrated luminosity and those
from the CDF Collaboration\,\cite{CDF} where appropriate.

From these measurements, the prompt $J/\psi$ production cross-section ($\sigma(pp \to J/\psi X^{\prime})$, produced directly from the proton-proton collisions 
or from decays of heavier charmonium states like the $\chi_c$ or $\psi(2S)$), and the non-prompt ($\sigma(pp\to B+X\to J/\psi X^{\prime})$) $J/\psi$ production cross-section, are extracted.
These results are compared to corresponding predictions made by the Colour Evaporation Model~\cite{CEM_RHIC}, Fixed-Order Next-to-Leading Log (FONLL)~\cite{Cacciari} 
and Colour Singlet NNLO$^\star$ calculations~\cite{NNLO_upsilon,onia_prod}.
Further details of the results of measurements presented here may be found in reference~\cite{HEPDATA}.

\hyphenation{spectrometer spectro-meter}

\section{The ATLAS Detector and Data Processing}
\label{section:Atlasdata}
In this section, the collection and processing of the data used in the paper are outlined. This involves a description of the most 
relevant subsystems of the ATLAS detector\,\cite{ATLAS}: the trigger system, 
the muon system and the inner tracking detector.
Also specified are the triggers used and the offline data processing, in particular the selection of candidate muons.

\subsection{The ATLAS detector}
\label{section:Atlasdet}

The ATLAS detector covers almost the full solid angle around the collision point with layers of 
tracking detectors, calorimeters and muon chambers. For the measurements presented in this paper, the trigger system, the inner detector tracking devices (ID) and the muon spectrometer (MS) are of particular importance. 

The ID covers the pseudorapidity range $|\eta |<$ 2.5. It consists of a silicon pixel detector, a silicon strip detector (SCT) and a transition radiation tracker (TRT). These detectors are located at a radial distance from the beam axis between 50.5\,mm
and 1066\,mm and are immersed in a 2 T solenoidal magnetic field. The ID barrel consists of 3 pixel layers, 4 layers of double-sided silicon strip modules and 73 layers of TRT straws. The ID end-cap has $2\times 3$ pixel layers, $2\times 9$ layers of silicon strips and $2\times 160$ layers of TRT straws.

The MS is located inside a toroidal magnetic field which provides 2.5 Tm of bending power in the barrel and 5 Tm in the end-caps. It consists of four detectors using different technologies and 
is divided into a barrel region ($|\eta|<1.05$) and two end-cap regions ($1.05<|\eta|<2.7$). 
Precise muon measurements are made using monitored drift tube chambers (MDT) in both the barrel and end-cap sections and using Cathode Strip Chambers (CSC) in the end-caps; fast triggers are obtained from resistive
plate chambers (RPC) in the barrel and thin gap chambers (TGC) in the end-caps. The chambers are arranged in three layers, so high-\pt\ particles leave at least three measurement points with a lever arm of several metres.

\subsection{Trigger}
\label{section:AtlasTrig}
The ATLAS detector has a three-level trigger system: level 1 (L1), level 2 (L2) and the event filter (EF). For the measurements presented here, the trigger relies on the Minimum Bias Trigger Scintillators (MBTS) and the muon trigger chambers. 

The MBTS are mounted in front of each liquid argon endcap calorimeter cryostat at $z=\pm3.56$ m and are segmented into eight sectors in azimuth and two rings in pseudorapidity ($2.09<|\eta|< 2.82$ and $2.82<|\eta|<3.84$). The MBTS trigger is configured to require two hits above threshold from either side of the detector. A dedicated muon trigger at the EF level is required to confirm the candidate events chosen for these measurements. This is initiated by the MBTS L1 trigger and searches for the presence of at least one track in the entire MS. This trigger is referred to as the EF minimum bias trigger; it has an adjustable threshold on the reconstructed muon $p_T$ above which events are accepted and can be prescaled to accept a pre-determined fraction of events meeting
the trigger condition.  

The L1 muon trigger is based on RPCs for the barrel and TGCs for the end-caps \cite{ATLAS}.  It seeks hit coincidences within different RPC or TGC detector layers inside programmed geometrical windows which define the muon candidate $p_T$, then selects candidates above six programmable thresholds and provides a rough estimate of their positions \cite{confmuontrigger}. For the earlier data used in this analysis, the muon trigger corresponds to the lowest \pt\ threshold trigger which requires a simple two-layer time coincidence within a region of $~0.1\times 0$.1 in $\eta$-$\phi$.
 No further geometrical constraint is applied.  

As the instantaneous luminosity of the collider increases, the trigger requirement switches from the EF minimum bias trigger to the L1 muon trigger. Later data periods make use of triggers seeded by this L1 trigger but with additional $p_T$ cuts applied at the EF stage (these are referred to henceforth as the EF muon triggers). 

\subsection{Muon identification and reconstruction}
\label{section:MuonRecon}
Muon identification and reconstruction extends to $|\eta|<2.7$, covering a \pt\ range from 1 GeV up to more than 1 TeV.
``Standalone MS tracks'' are constructed entirely based on the signal hits collected in the MS. The track parameters are obtained from the MS track and are extrapolated to the interaction point,
  taking into account multiple scattering and energy loss in the traversed material. In this analysis, two categories of reconstructed muons are then defined:
\begin{itemize}
\item {\bf Muons from combined reconstruction:}
  the {\em combined} muon reconstruction relies on a statistical combination of both a standalone MS track and an ID track.
  Due to ID 
  coverage, the combined reconstruction covers $|\eta|<2.5$. 
\item {\bf Muons from ID track tagging:}
  a {\em tagged} muon is formed by MS track segments which are not formed into a complete MS track, but which are matched to ID tracks extrapolated to the MS.
  Such a reconstructed muon adopts the measured parameters of the associated ID track. 
In this paper, the muon tagging is limited to $|\eta|<2$, in order to ensure high quality tracking
  and a reduction of fake muon candidates.
\end{itemize}

\noindent  
The muon track helix parameters are taken from the ID measurement alone, since the MS does not add much to the precision in the lower momentum range relevant for the $J/\psi$ measurements presented here. 

\section{Data and Monte Carlo Samples}
\label{section:samples}
Proton-proton collision data, at a centre-of-mass energy of 7 TeV,
are included in this analysis if taken during stable beam periods and when the MS, ID and magnet systems were collecting data of a sufficiently high quality to be suitable for physics analysis. 

Monte Carlo samples are used for determining acceptance corrections,
as part of the trigger efficiency studies and in systematic cross-checks.
They are generated using \textsc{Pythia 6}\, \cite{Pythia6} and tuned
using the ATLAS MC09 tune\,\cite{MC09} which uses the MRST LO$^\star$ parton distribution functions \cite{MRSTLO}.
The passage of the generated particles through the detector is simulated with \textsc{Geant4}\ \cite{Geant} and the data are fully reconstructed
with the same software that is used to process the data from the detector.
For the signal $J/\psi$ Monte Carlo (used to derive the kinematic
acceptance corrections), the \textsc{Pythia} implementation of prompt
$J/\psi$ production sub-processes in the
NRQCD Colour Octet Mechanism framework \cite{JpsiTune} is used.

Prompt $J/\psi$ production includes {\em direct} production from the
hard interaction, as well as charmonium feed-down from excited states.
These {\em prompt} production modes are distinct from
{\em non-prompt} production that is characterised by the
production of $J/\psi$ via the decay of a $B$-hadron.

All samples are generated with polar and azimuthal isotropy
in the decay of the $J/\psi$ (the default in \textsc{Pythia})
and are reweighted at the particle level according to their respective angular dependencies in order to describe a number of
different spin-alignment scenarios
(see Section~\ref{sec:acceptance}). 
The $J/\psi$ spin-alignment is not measured in this analysis, so the reweighted MC samples are used 
to provide an uncertainty band on the measurement of the production cross-section, determined by the
maximum variation in acceptance across the full allowed range of $J/\psi$ spin alignment.

\subsection{Event and candidate selection}
\label{section:eventsel}

The analyses presented in this paper make use of the triggers described in Section~\ref{section:AtlasTrig}. 
For the inclusive cross-section, in a given data taking period an event is retained or discarded based on the decision of a single specific trigger, without reference to any other triggers. For data from the initial running with lower instantaneous luminosity, the 
L1 muon trigger is used. During later periods, with higher instantaneous luminosity, a more selective EF muon
trigger with a 4~GeV \pt\ threshold is required, and eventually, this is increased to a 6~GeV \pt\
threshold. The sample collected by these triggers and passing the data quality selections corresponds to an integrated luminosity of $2.2~\textrm{pb}^{-1}$.

For the measurement of the $B\to J/\psi$ non-prompt fraction (see Equation~\ref{eqn:fraction}), two additional triggers 
are employed, and rather than using a single trigger to veto or accept events, several triggers are used simultaneously such that any one of them having fired results in the event being included. From the initial period, events triggering either the L1 muon trigger or the EF minimum bias trigger are used (whereas only the L1 muon trigger is used for the cross section). For intermediate instantaneous luminosities the L1 muon trigger is used alone since the EF minimum bias trigger is highly prescaled at this stage. For the highest instantaneous luminosities, events are accepted which pass any of the EF muon triggers with \pt\ thresholds of 4, 6  or 10~GeV. During the runs with the highest instantaneous luminosities, the triggers with $4$ and $6$ GeV are prescaled; however, the $10~\textrm{GeV}$ threshold trigger is not. The inclusion of this unprescaled trigger along with the addition of the EF minimum bias trigger 
for the $B\to J/\psi$ non-prompt fraction measurement results in a slightly higher integrated luminosity of $2.3~\textrm{pb}^{-1}$.

To veto cosmic rays, events passing the trigger selection are required to have at least three tracks associated with the same reconstructed primary vertex. The three tracks must each have at least one hit in the pixel system and at least six hits in the SCT. 

Each remaining event is required to contain at least one pair of reconstructed muons. Only muons associated with ID tracks that have at least one hit in the pixels and six in the SCT are accepted. 
Di-muon pairs with opposite charges are considered to be $J/\psi$ candidates if at least one combined muon is present in the pair. 
At least one reconstructed muon candidate is required to match a muon trigger (that is, at least one muon from the $J/\psi$ candidate should have fired the trigger). For the early data, when the trigger is essentially based on the L1 muon trigger, at least one of the offline muons is required to match the trigger muon candidate to within 
$\Delta R=\sqrt{\Delta\eta^2+\Delta\phi^2}<0.4$ at the MS plane; for the later data taking, where the EF muon trigger is used, 
the offline and trigger muons are required to match within $\Delta R<0.005$. 

The two ID tracks from each pair of muons passing these selections are fitted to a common vertex ~\cite{VKalVrt}. No constraints are applied in the fit and a very loose vertex quality requirement, which retains over $99\%$ of the candidates, is used.  

For the $B\to J/\psi$ non-prompt fraction analysis, where 
lifetime information
is an important element of the fit,
additional requirements are made on the $J/\psi\to\mu^+\mu^-$ candidates. The probability of the 
fit to the $J/\psi$ vertex is required to be greater than $0.005$.
For this measurement $J/\psi$ candidates are rejected if the two muon candidate tracks were used to build different primary vertices in the offline reconstruction (so that there is an ambiguity as to which primary vertex to use in the lifetime calculation). This rejects fewer than $0.2\%$ of the $J/\psi$ candidates. This selection is not applied for the cross-section analysis. 

\def\ru{\ensuremath{\rightarrow}}
\newcommand{\curlyL}{ \ensuremath{L}}
\newcommand{\GammaNeg}{\ensuremath {  \Gamma_{-} } \space} 
\newcommand{\ppFrac}{\ensuremath { pp_{fr} } \space}
\newcommand{\bbFrac}{\ensuremath { bb_{fr} } \space}
\newcommand{\mphi}{ \ensuremath { \phi }}
\newcommand{\Eff}{ \ensuremath { \epsilon }}
\newcommand{\wtf}{ \ensuremath { wf }}
\newcommand{\bquark}{\ensuremath{b}}
\newcommand{\abquark}{\bbar  \space}

\newcommand{\bab}{\bbbar \space }
\newcommand{\pp}{ pp   \space }
\newcommand{\cac}{ \ccbar} 
\newcommand{\bbJpsi}{\ensuremath{ \bbbar \ra  \Jpsi X  } \space}
\newcommand{\ppJpsi}{\ensuremath{ pp \ra \Jpsi X  } \space}

\newcommand{\Jpsipa}{\Jpsi \ }
\newcommand{\Jpsipan}{\Jpsi } 
\newcommand{\Upspa}{\Ups}
\newcommand{\jpsimass}{ \ensuremath{M (\Jpsipan) }\space}

\newcommand{\jpsieta}{ \ensuremath{\eta (\Jpsipan) }\space}
\newcommand{\jpsiphi}{ \ensuremath{\phi (\Jpsipan) }\space}
\newcommand{\jpsiPDGMass}{ \ensuremath{M_{PGD} (\Jpsipan) }\space}
\newcommand{\jpsiPDGmass}{ \ensuremath{M_{PGD} (\Jpsipan) }\space}
\newcommand{\jpsilxy}{\ensuremath{L_{xy} (\Jpsipan) }\space}
\newcommand{\lxyjpsi}{\ensuremath{L_{xy} (\Jpsipan) }\space}
\newcommand{\lxy}{\ensuremath{L_{xy} }\space}

\newcommand{\lxyb}{\ensuremath{L_{xy} (B) }\space}
                                                  
\newcommand{\psubt}{ \pt \space} 
\newcommand{\jpsipt}{\pt (\Jpsipan) \space} 
\newcommand{ \ptb}{\pt (B) \space} 
\newcommand{ \bpt}{\pt (B) \space}
\newcommand{ \bdpt}{\pt (Bd) \space}
\newcommand{ \tauB}{ $\tau$ (B) \space}

\newcommand{\lumilow}{\begL\   }
\newcommand{\lumimed}{\begM\   }
\newcommand{\lumihigh}{\lowL\   }
\newcommand{\lumihihh}{\highL\space} 
\newcommand{\pico}{\ipb\   }
\newcommand{\femto}{\ifb\space} 
\newcommand{\psubts}{\ensuremath{\pt > \GeV }\space} 

\newcommand{\absetadp}{\ensuremath{\abseta < 2.4 }\space} 
\newcommand{\Hz}{\ensuremath{\rm Hz}\   }
\newcommand{\ps}{\ensuremath{\rm ps}\   }

\newcommand{\Jpsitomm}{\ensuremath{ \Jpsi \ra \mumu }\space} 
\newcommand{\Jpsitoee}{\ensuremath{ \Jpsi \ra \epem}\space} 
\newcommand{\Jpsitol}{\ensuremath{\Jpsi \ra \mumu}\space} 
\newcommand{\mumuppi}{\ensuremath{\mumu p\pi^{-}}\space} 
\newcommand{\mueeppi}{\ensuremath{\mu eep\pi^{-}\ }\space} 
\newcommand{\bbarb}{\ensuremath{ pp \ra \bbbar X\ }\space} 
\newcommand{\phiKK}{\ensuremath{ \phi \ra \kplus \kminus}\space} 
\newcommand{\KnstoKpi}{\ensuremath{ \Kns \ra K^{\pm}\pi^{\pm}\ }\space} 
\newcommand{\Kpi}{\ensuremath{ K^{\pm}\pi^{\pm}\space} \space}

\section{Inclusive {\Jpsitomm} Differential Production Cross-Section}
\label{section:mass}

The measurement of the inclusive differential cross-section is determined as 
\begin{equation}
\frac{d^2\sigma(J/\psi)}{dp_Tdy} Br(J/\psi \to \mu^+\mu^-) = \frac{N^{J/\psi}_{corr}}{{\cal{L}}\cdot\Delta p_T \Delta y}
\end{equation}
where $N^{J/\psi}_{corr}$ is the $J/\psi$ yield in a given $p_T-y$ bin after continuum background subtraction and correction for detector efficiency, bin migration and acceptance effects,
${\cal L}$ is the integrated luminosity of the data sample and $\Delta p_T$ and $\Delta y$ are the $p_T$ and rapidity bin widths.
The probability $P$ that a $J/\psi\to\mu\mu$ decay is reconstructed depends on the kinematics of the decay, as well as the muon reconstruction and trigger efficiencies.
In order to recover the true number $N^{J/\psi}_{corr}$ of such decays produced in the collisions, 
a weight $w$ is applied to each observed $J/\psi$ candidate, defined as the inverse of 
that probability and calculated as follows:
\begin{eqnarray}
\label{weights}
P &=& w^{-1} = {\cal {A}} \cdot {\cal M} \cdot {\cal {E}}^2_{\mathrm{trk}}
\cdot {\cal {E}^+}_{\mu}(p_T^{+},\eta^{+}) 
\cdot {\cal {E}^-}_{\mu}(p_T^{-},\eta^{-})
\cdot {\cal {E}}_{\mathrm {trig}}
\end{eqnarray}
where ${\cal{A}}$ is the kinematic acceptance, ${\cal M}$ is a correction factor for bin migrations due to finite detector resolution, ${\cal {E}}_{\mathrm{trk}}$
is the ID tracking efficiency and ${\cal{E}}_{\mu}$ is the
single-muon offline reconstruction efficiency. Here $p_T^{\pm}$ and $\eta^{\pm}$ are the transverse momenta and pseudorapidities
of the positive and negative muons from the $J/\psi$ decay. The trigger
efficiency ${\cal{E}}_{\mathrm{trig}}$ for a given $J/\psi$ candidate is
calculated from single-muon trigger efficiencies
${\cal{E}}^{\pm}_{\mathrm{trig}}(p_T^{\pm},\eta^{\pm})$ as follows:
\begin{equation}
{\cal{E}}_{\mathrm{trig}} = 1 -
\left(1-{\cal{E}}^+_{\mathrm{trig}}(p_T^+,\eta^+)\right)\cdot
\left(1-{\cal{E}}^-_{\mathrm{trig}}(p_T^-,\eta^-)\right).
\end{equation}
The resultant weighted invariant mass peak is then fitted (see Section~\ref{sec:MLfit}) to extract $N^{J/\psi}_{corr}$.

\subsection{Acceptance} 
\label{sec:acceptance}
The kinematic acceptance ${\cal {A}}(p_T,y)$ is the probability that the muons
from a $J/\psi$ with transverse momentum $p_T$ and rapidity $y$ 
fall into the fiducial volume of the detector. This is calculated using
generator-level Monte Carlo, applying cuts on the momenta and
pseudorapidities of the muons to emulate the detector geometry. Global
cuts of 
$|\vec{p}_+|, |\vec{p}_-| > 3$\;GeV for $|\eta_+|,|\eta_-|<2.5$ are
supplemented by finer $p_T$ thresholds in slices of $\eta$ to
ensure that regions of the detector where the values of offline and trigger
efficiencies are so low as to be compatible with zero within the uncertainties (approximately 10\%)
are excluded from the analysis.

\begin{figure}[htbp]
  \begin{center}
    \includegraphics[width=0.4\textwidth]{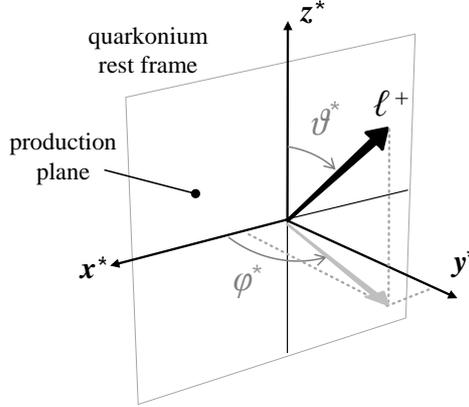}
  \end{center}
  \caption{Definitions of the $J/\psi$ spin-alignment angles, in the $J/\psi$
    decay frame. 
   $\theta^\star$ is the angle between the direction of
    the positive muon in that frame and the direction of
    $J/\psi$ in the laboratory frame, which is directed along the
    $z^\star$-axis. $\phi^\star$ is the angle between the $J/\psi$ production
    ($x^\star-z^\star$) plane and its decay plane formed by the direction of the 
    $J/\psi$ and the lepton $\ell^+$ (from \cite{faccioli}).
  }
 \label{fig:coordinates}
\end{figure}

The acceptance also depends on the spin-alignment of the $J/\psi$, which
is not known for LHC conditions. The general angular distribution
for the decay $J/\psi\to\mu\mu$ in the $J/\psi$ decay frame is given by:
\begin{equation}
\label{eqn:spinalign}
\frac{d^2N}{d\cos\theta^{\star} d\phi^{\star}}\propto 1+
\lambda_{\theta}\cos^2\theta^\star+
\lambda_{\phi}\sin^2\theta^\star\cos2\phi^\star+
\lambda_{\theta\phi}\sin2\theta^\star\cos\phi^\star
\end{equation}
where $\theta^\star$ is the angle between the 
direction of the positive muon
momentum in the $J/\psi$ decay frame and the $J/\psi$ line of flight, while
$\phi^\star$ is defined as the angle between the $J/\psi$ production and decay planes in the lab frame (see Figure \ref{fig:coordinates}, 
reference~\cite{faccioli} and references therein).

\begin{figure}[htbp]
  \begin{center}
    \subfigure[$\lambda_\theta=\lambda_\phi=\lambda_{\theta\phi}=0$]{
      \label{fig:acc_2d_FLAT}
      \includegraphics[width=0.3\textwidth]{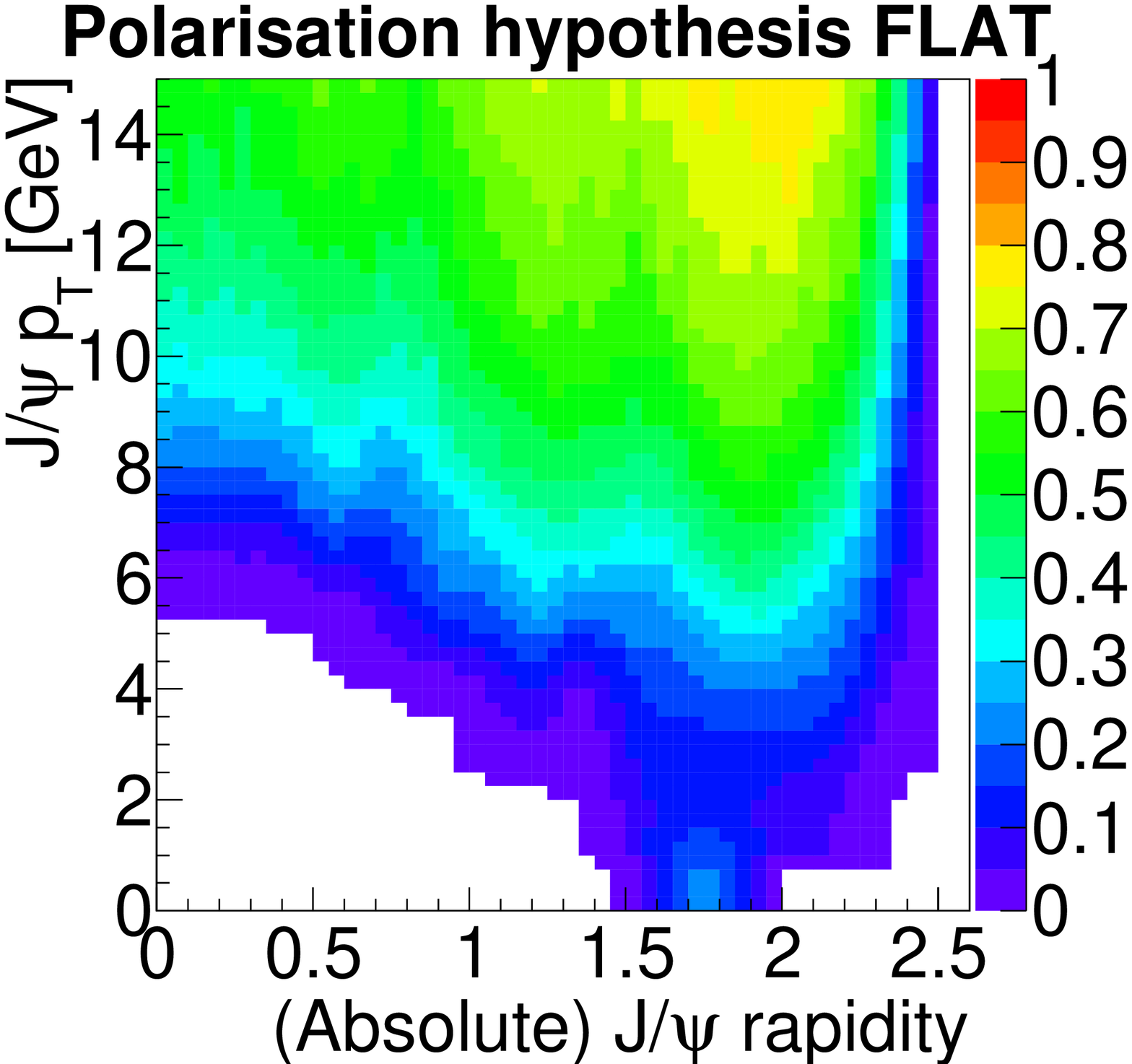}
    }
    \subfigure[$\lambda_\theta=+1, \lambda_\phi=\lambda_{\theta\phi}=0$]{
      \label{fig:acc_2d_TRP0}
      \includegraphics[width=0.3\textwidth]{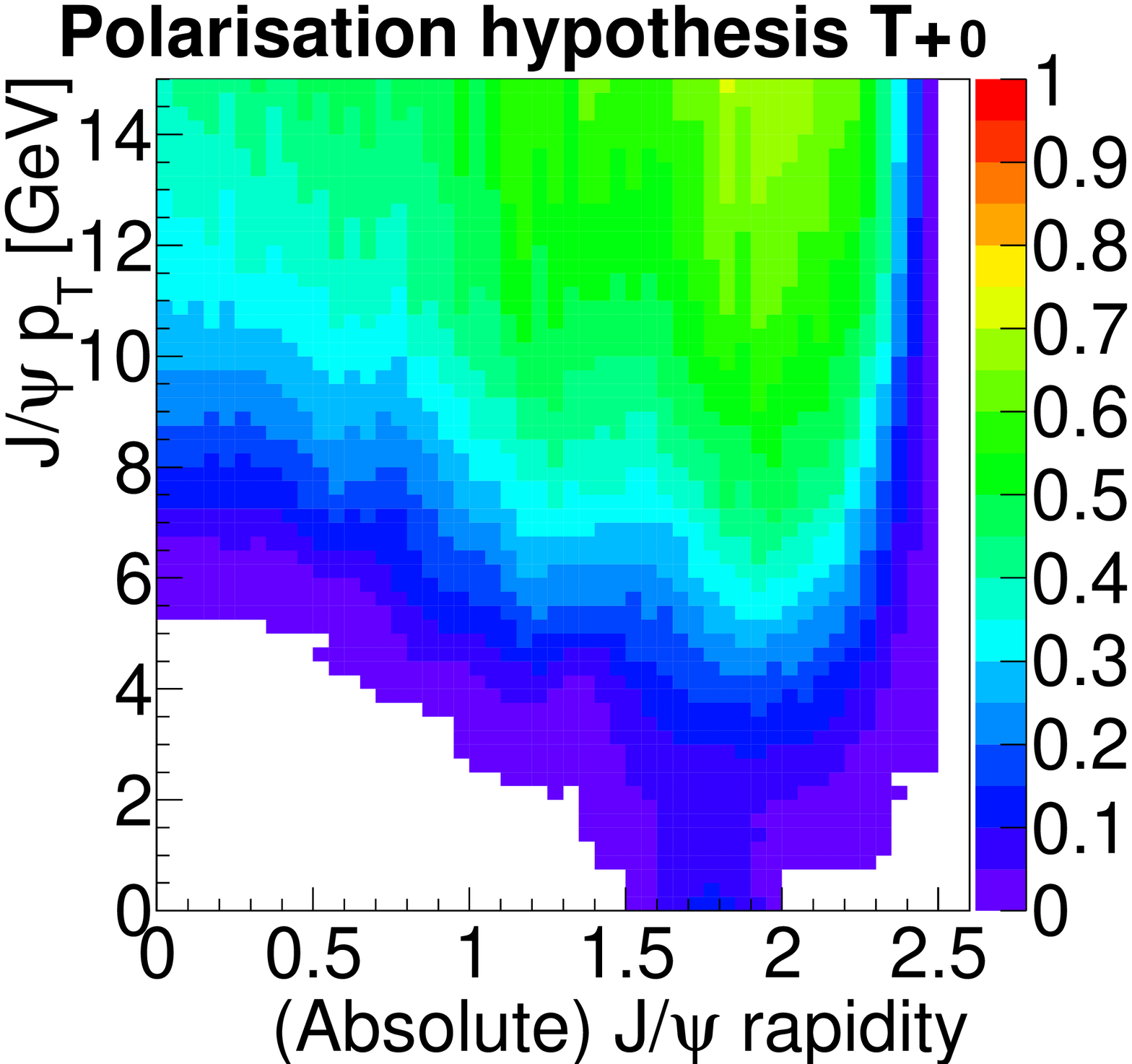}
    }
    \subfigure[$\lambda_\theta=-1, \lambda_\phi=\lambda_{\theta\phi}=0$]{
      \label{fig:acc_2d_LONG}
      \includegraphics[width=0.3\textwidth]{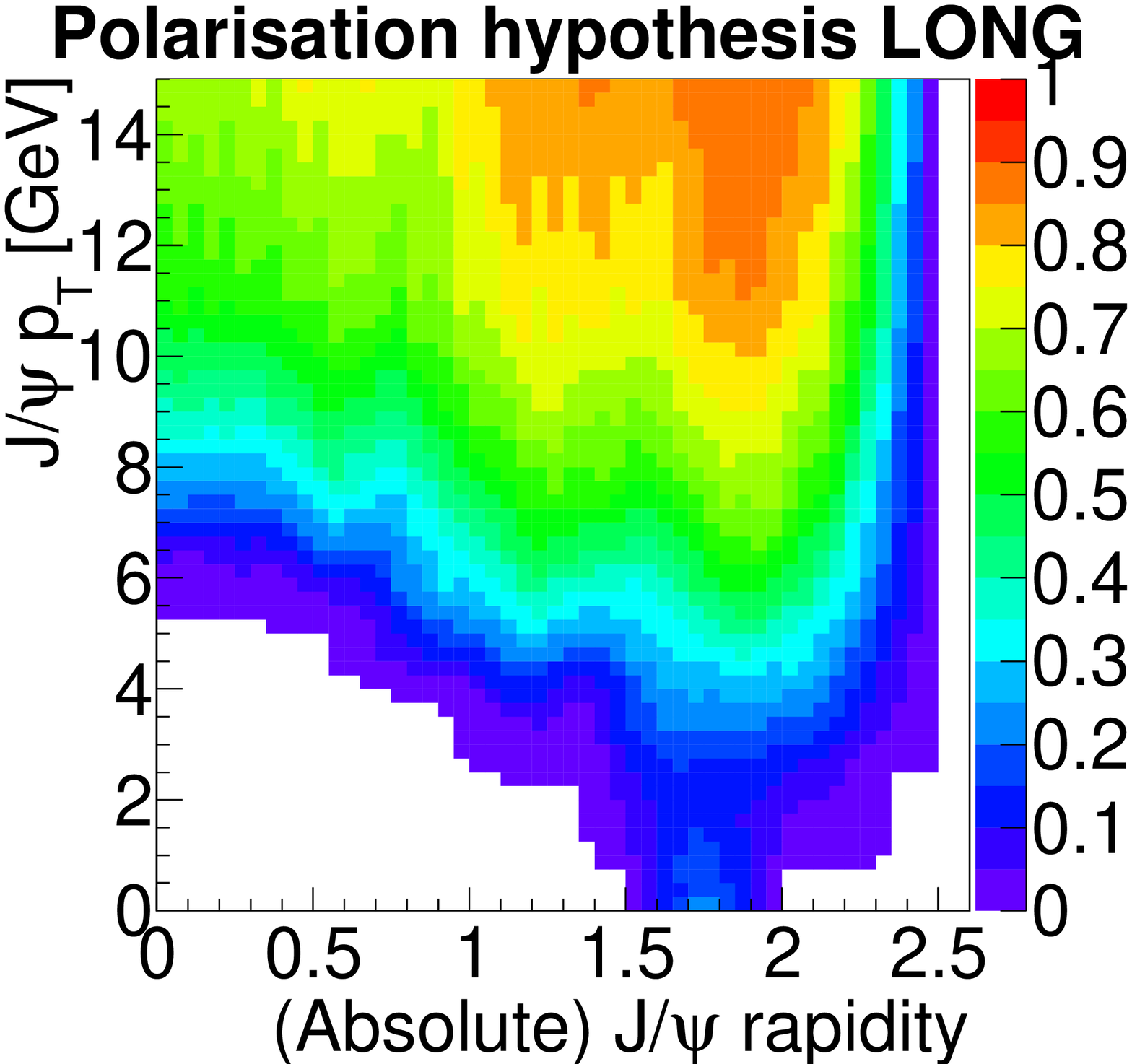}
    }
    \subfigure[$\lambda_\theta=+1, \lambda_\phi=+1, \lambda_{\theta\phi}=0$]{
      \label{fig:acc_2d_TRPP}
      \includegraphics[width=0.3\textwidth]{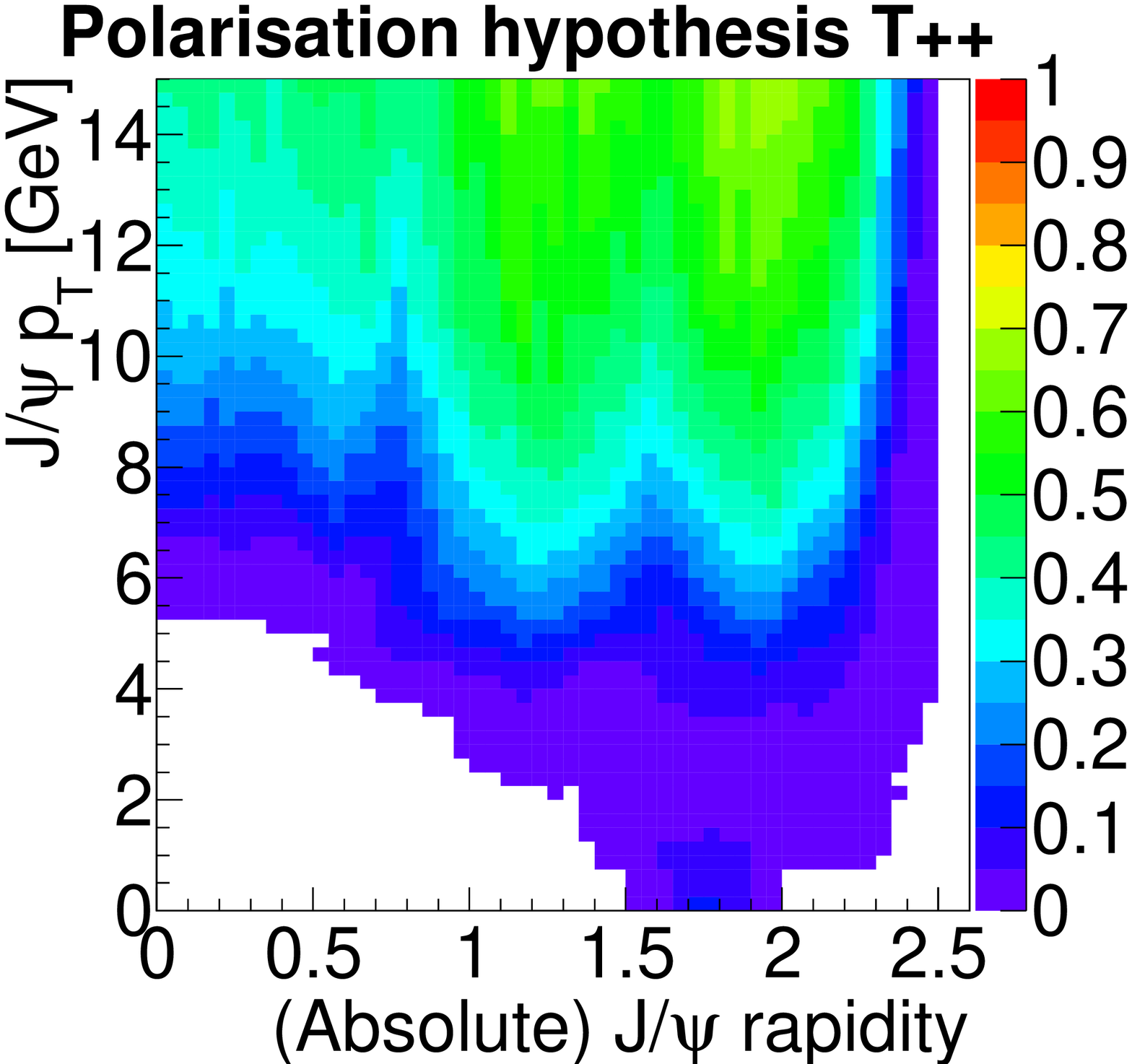}
    }
    \subfigure[$\lambda_\theta=+1, \lambda_\phi=-1, \lambda_{\theta\phi}=0$]{
      \label{fig:acc_2d_TRPM}
      \includegraphics[width=0.3\textwidth]{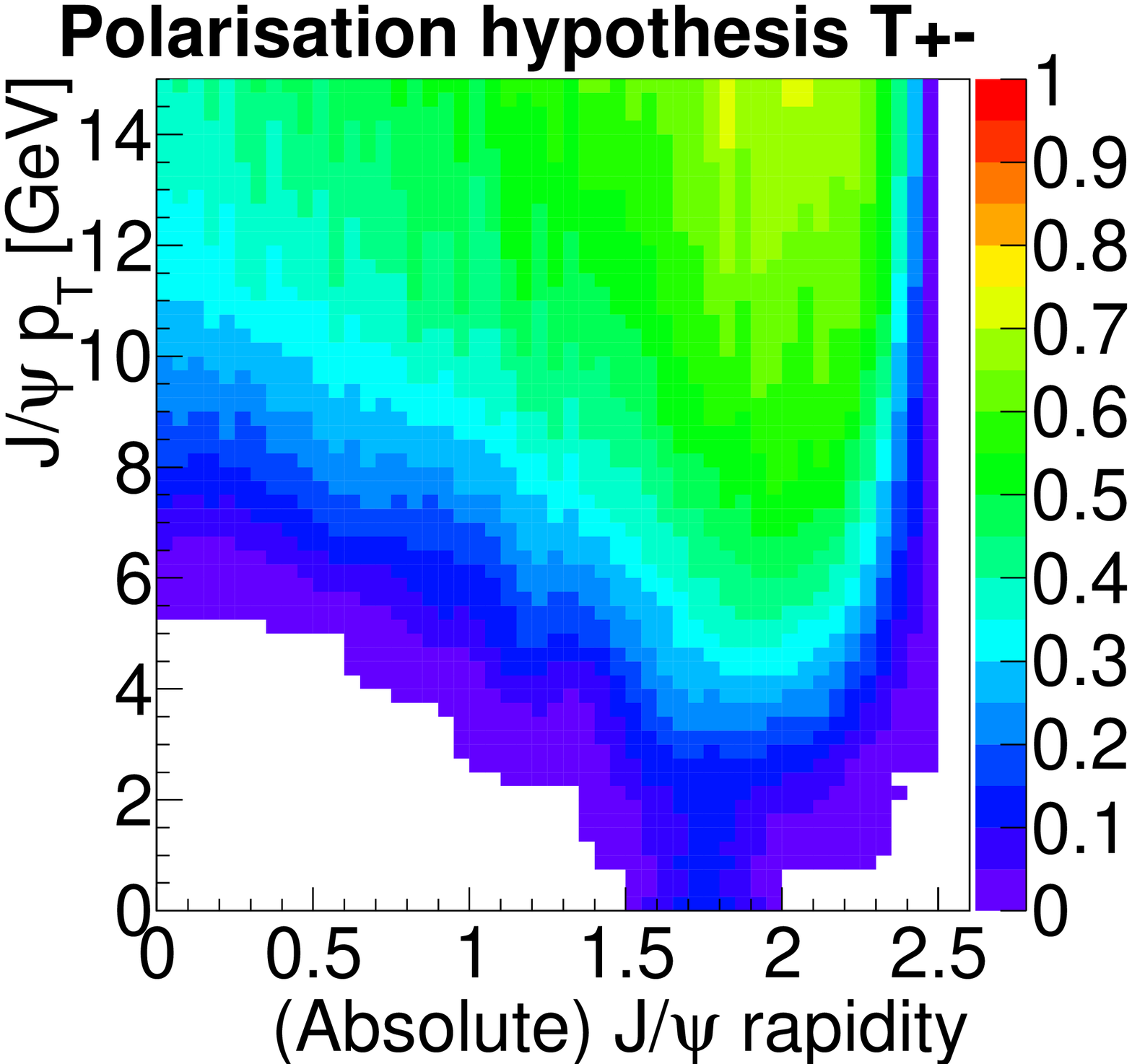}
    }
    \caption{Kinematic acceptance maps as a function of $J/\psi$ transverse momentum and rapidity
      for specific spin-alignment scenarios considered,
      which are representative of the extrema of the variation of the
      measured
      cross-section due to spin-alignment configurations.
      Differences in acceptance behaviour, particularly at low $p_T$,  occur between scenarios and
      can significantly influence the cross-section measurement in a
      given bin.
      \label{fig:acc_2d}
    }
  \end{center}
\end{figure}

A large number of possible combinations of the coefficients $\lambda_{\theta}, \lambda_{\phi}, \lambda_{\theta\phi}$
have been studied, including some with $\lambda_{\theta\phi}\neq 0$. Five extreme cases have been identified that lead to the biggest variation of acceptance
within the kinematics of the ATLAS detector and define an envelope in which the results may vary under all possible polarisation assumptions:
\begin{enumerate}
\item
Isotropic distribution, independent of $\theta^\star$ and $\phi^\star$, with
$\lambda_{\theta}=\lambda_{\phi}=\lambda_{\theta\phi}=0$, labelled as ``FLAT". 
This is used as the main (central) hypothesis.
\item
Full longitudinal alignment with 
$\lambda_{\theta}=-1, \lambda_{\phi}=\lambda_{\theta\phi}=0$, labelled as ``LONG".
\item
Transverse alignment with
$\lambda_{\theta}=+1, \lambda_{\phi}=\lambda_{\theta\phi}=0$, labelled as $\textrm{T}_{+0}$.
\item
Transverse alignment with 
$\lambda_{\theta}=+1, \lambda_{\phi}=+1, \lambda_{\theta\phi}=0$, labelled as $\textrm{T}_{++}$.
\item
Transverse alignment with 
$\lambda_{\theta}=+1, \lambda_{\phi}=-1, \lambda_{\theta\phi}=0$, labelled as $\textrm{T}_{+-}$.
\end{enumerate}
Two-dimensional acceptance maps are produced in bins of $p_T$ and $y$ of the $J/\psi$,
for each of these five scenarios, and are illustrated in Figure~\ref{fig:acc_2d}. The maps are obtained by reweighting the flat distribution
at the generator level using Equation~\ref{eqn:spinalign}. The central
value for the cross-section measurement is obtained using the flat
distribution, and the measurement is repeated using
the other scenarios to provide an envelope of maximum variation, which
is stated as a separate uncertainty.

\subsection{Bin migration corrections} 
\label{sec:binmigration}
\noindent
The measured efficiency and acceptance corrected $J/\psi$ $p_T$ distribution is parameterised in each rapidity slice by a smooth analytic function smeared 
with a Gaussian distribution, with resolution derived from the data.
This function is integrated numerically over each analysis bin, both with and without smearing applied, and the ratio of the two integrals is assigned
as the correction factor. The effects of this correction are minimal at low $p_T$ and at low rapidities (around 0.1\%) but increase at higher $p_T$ and at higher 
rapidities (reflecting the decreasing momentum resolution) to a maximum of approximately 3\%.

\subsection{Muon trigger and reconstruction efficiency} 
\label{sec:trigeff}
\noindent
The offline single muon reconstruction efficiencies are obtained from data using a tag and probe method~\cite{tagandprobe}, 
where muons are paired with ID tracks (``probes") of opposite charge. The pairs are divided into two categories: those in 
which the probe is reconstructed as a muon (``matched'') and those in which it is not (``unmatched''). Both sets of pairs 
are binned according to the $p_T$ and $\eta$ of the probe. In each of these bins, the muon reconstruction efficiency is obtained 
as the ratio of the number of $J/\psi$ candidates in the peak of the matched distribution to the total number of candidates in the 
two mass distributions. The efficiency is extracted as a parameter of a simultaneous fit to both distributions. The dependence of 
the offline reconstruction efficiency on the muon charge is well described by MC within the acceptance. This procedure is 
repeated separately for combined and tagged muons.
At higher $p_T$ (for muons with $p_T$ above 6~GeV), the efficiency determination is supported by additional tag and probe 
$Z\to\mu^+\mu^-$ data\,\cite{Zmumu_tagandprobe} for improved precision in the efficiency plateau region.

A hybrid Monte Carlo and data-derived (tag and probe) scheme is used
to provide trigger efficiencies for the analysis with finer binning
than would be possible with the available data statistics. This is
necessary to avoid significant biases that would otherwise appear in
the analysis with coarsely binned efficiencies across rapidly-changing
efficiency regions. Due to significant charge dependence at low $p_T$ and high pseudorapidity,
separate trigger efficiency maps are produced for positive and negative muons. Fully simulated
samples of prompt $pp\rightarrow J/\psi\left(\mu^+\mu^-\right)X$ decays
are used to populate the $J/\psi$ $p_T-y$ plane, using a fine binning. For
each bin, the probability of a muon activating the trigger is
determined. 
The derived efficiencies are then reweighted to match the data efficiencies in
the reconstructed bins in cases where discrepancies exist between the data and Monte Carlo,
 and uncertainties from data are assigned.

Muon reconstruction efficiencies have been determined relative to reconstructed ID tracks.
Inner Detector tracks associated to muons and having the selection cuts used in this analysis have a reconstruction efficiency ${\cal {E}}_{\mathrm{trk}}$ 
of $99.5\%\pm 0.5\%$  per track (with no significant pseudorapidity or $p_T$ dependence observed within the phase space probed with this analysis),
which is applied as an additional correction to the $J/\psi$ candidate yields.

\subsection{Fit of {\Jpsi} candidate mass distributions}
\label{sec:MLfit}
\begin{figure}[htb]
  \begin{center}        
      \hfill\includegraphics[width=0.55\textwidth]{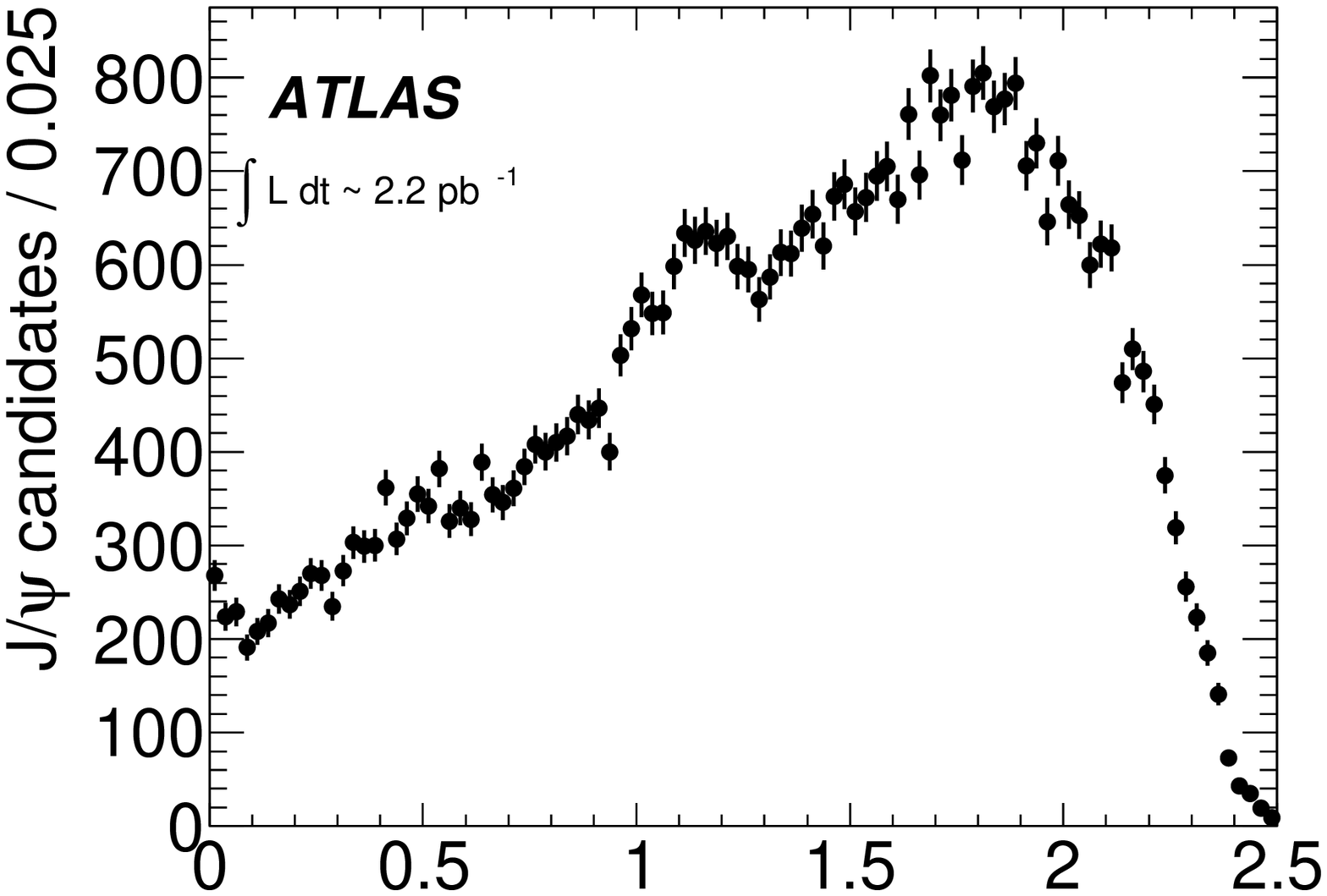}\\     
      \hfill\includegraphics[height=0.33\textwidth, width=0.42\textwidth, angle=90]{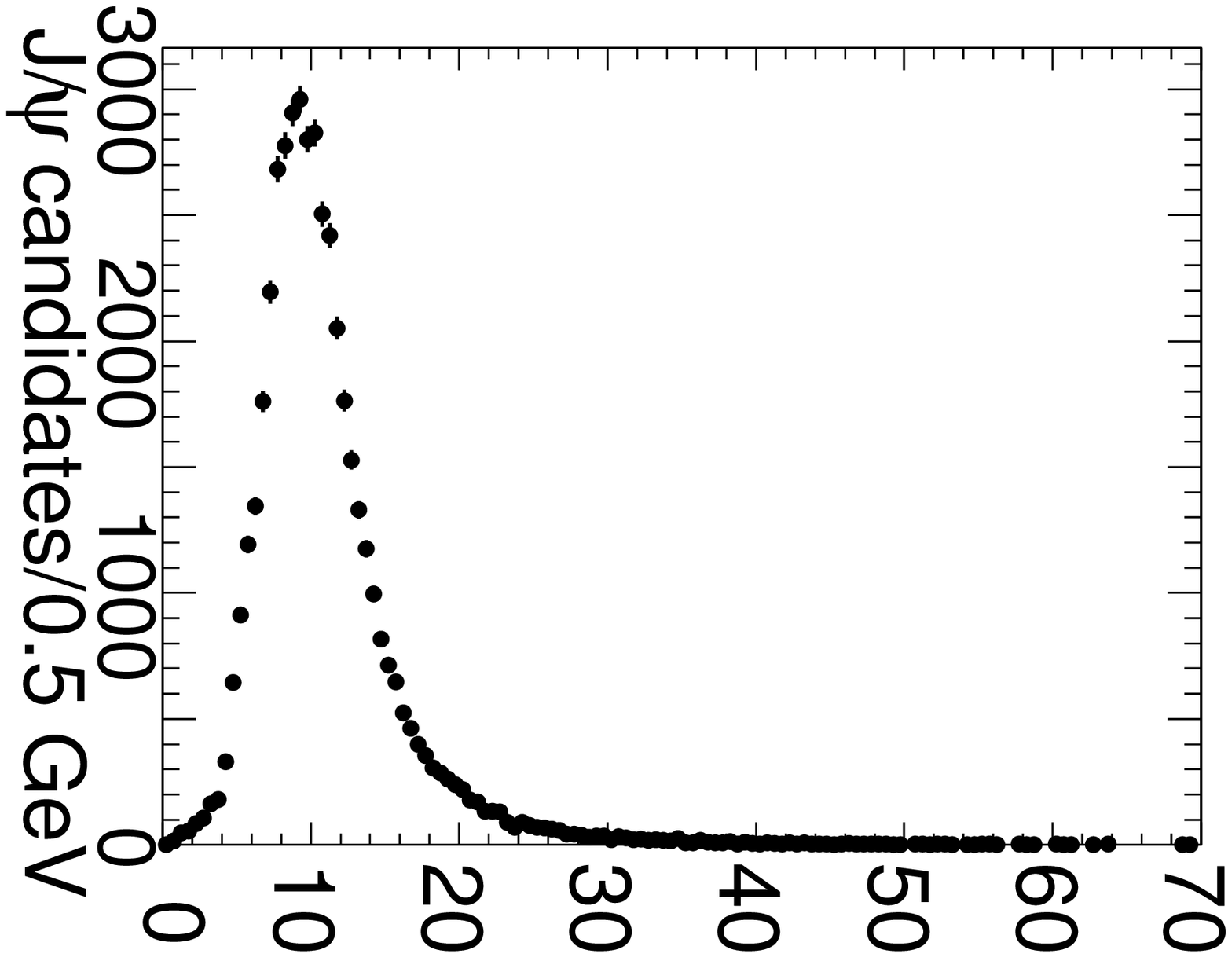}     
      \includegraphics[width=0.55\textwidth]{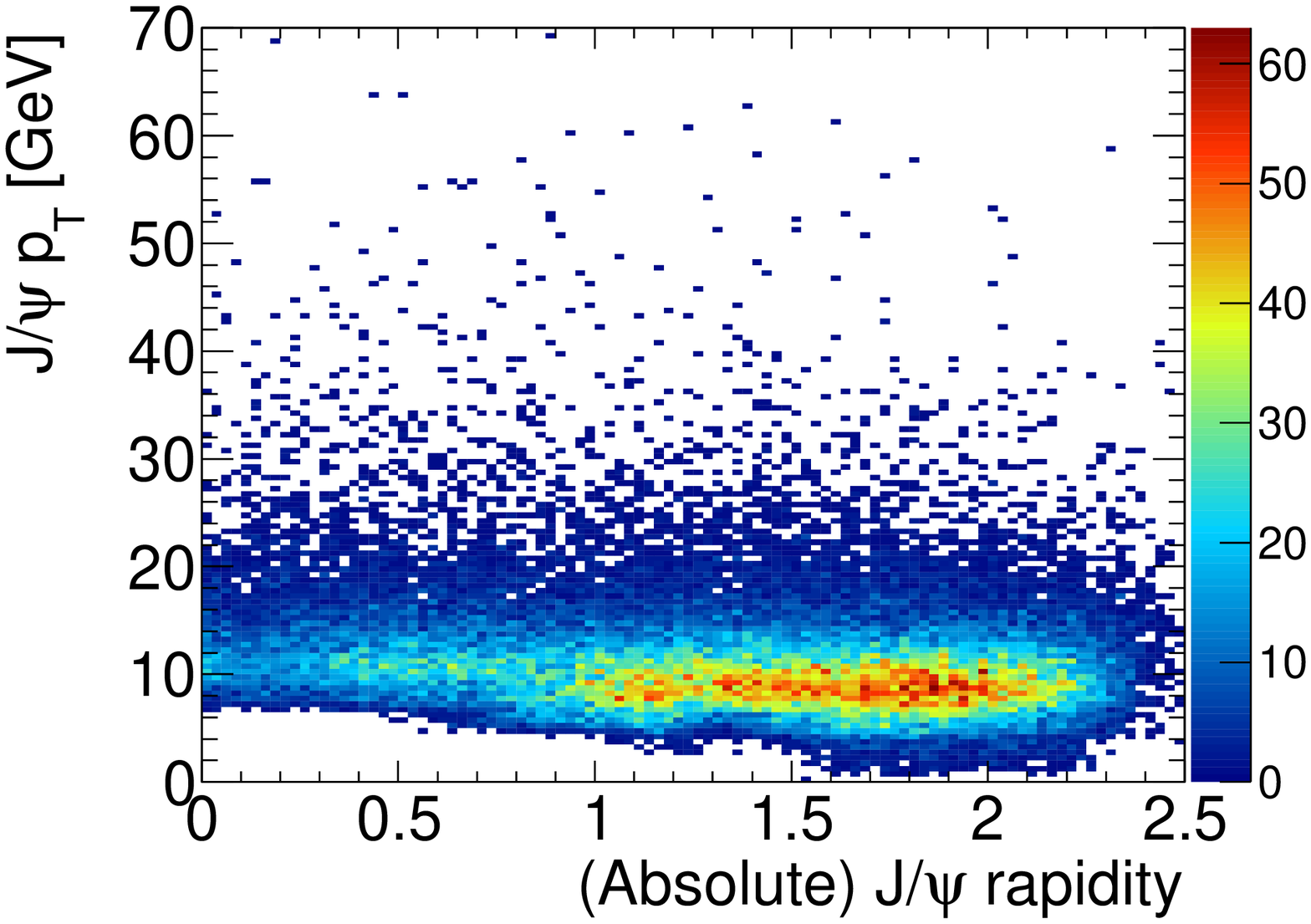}\\     
  \end{center}
  \caption{Distribution of reconstructed $J/\psi$ candidates (in the invariant mass interval $2.7<m_{J/\psi} < 3.5$ GeV) as a function of $J/\psi$ $p_T$ and rapidity.}
 \label{fig:yieldmap}
\end{figure}

The distribution of reconstructed $J/\psi$ candidates over the candidate
$p_T - y$ plane is shown in Figure~\ref{fig:yieldmap}.
The majority of $J/\psi$ candidates are reconstructed in intermediate-$p_T$, high-$y$ areas, 
as at lower $p_T$ values the acceptance of the detector is limited.

\begin{figure}[!htb]
  \begin{center}
      \label{fig:Jpsimass1}
      \includegraphics[width=0.45\textwidth]{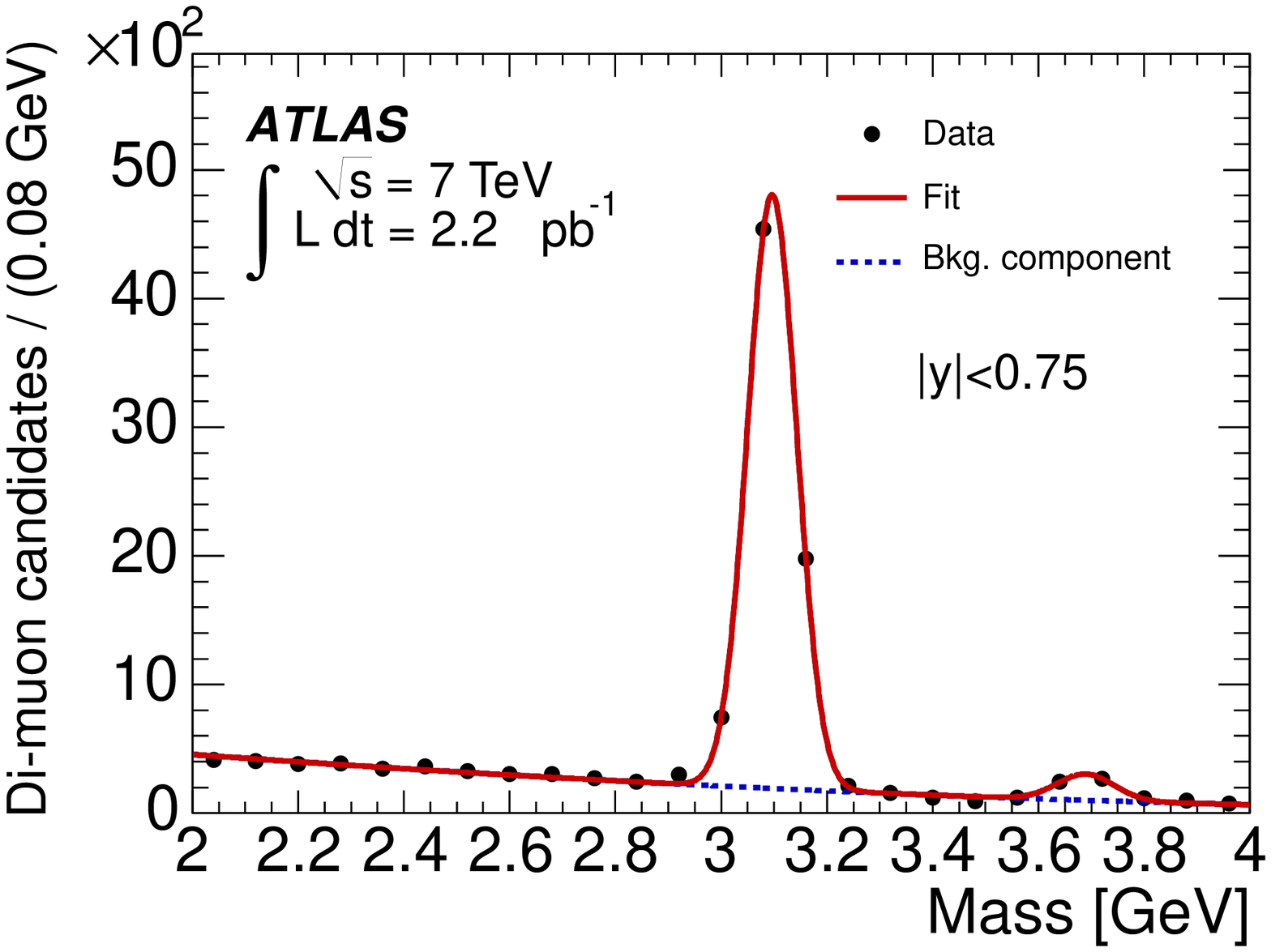}
      \label{fig:Jpsimass2}
      \includegraphics[width=0.45\textwidth]{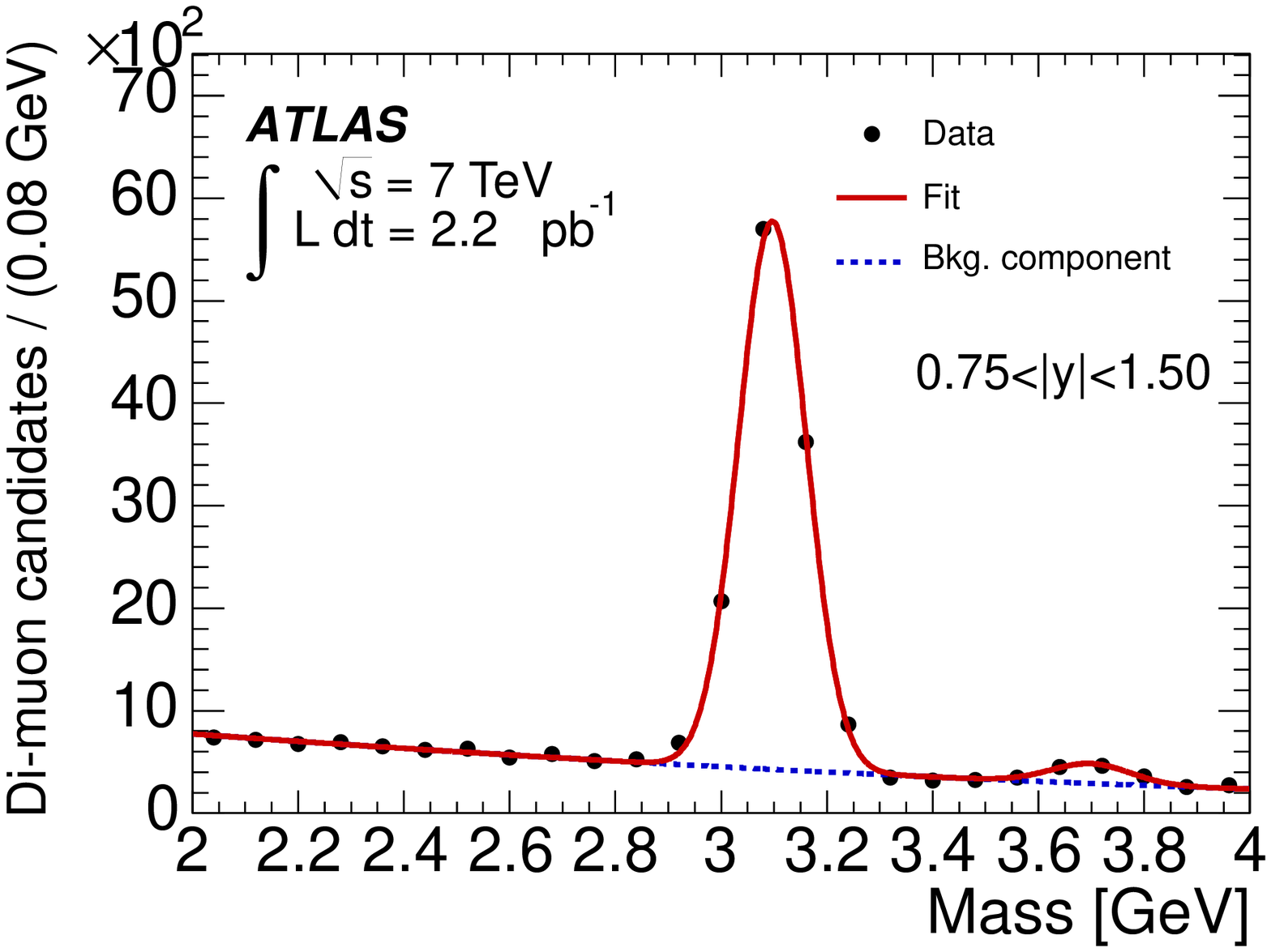}
      \label{fig:Jpsimass3}
      \includegraphics[width=0.45\textwidth]{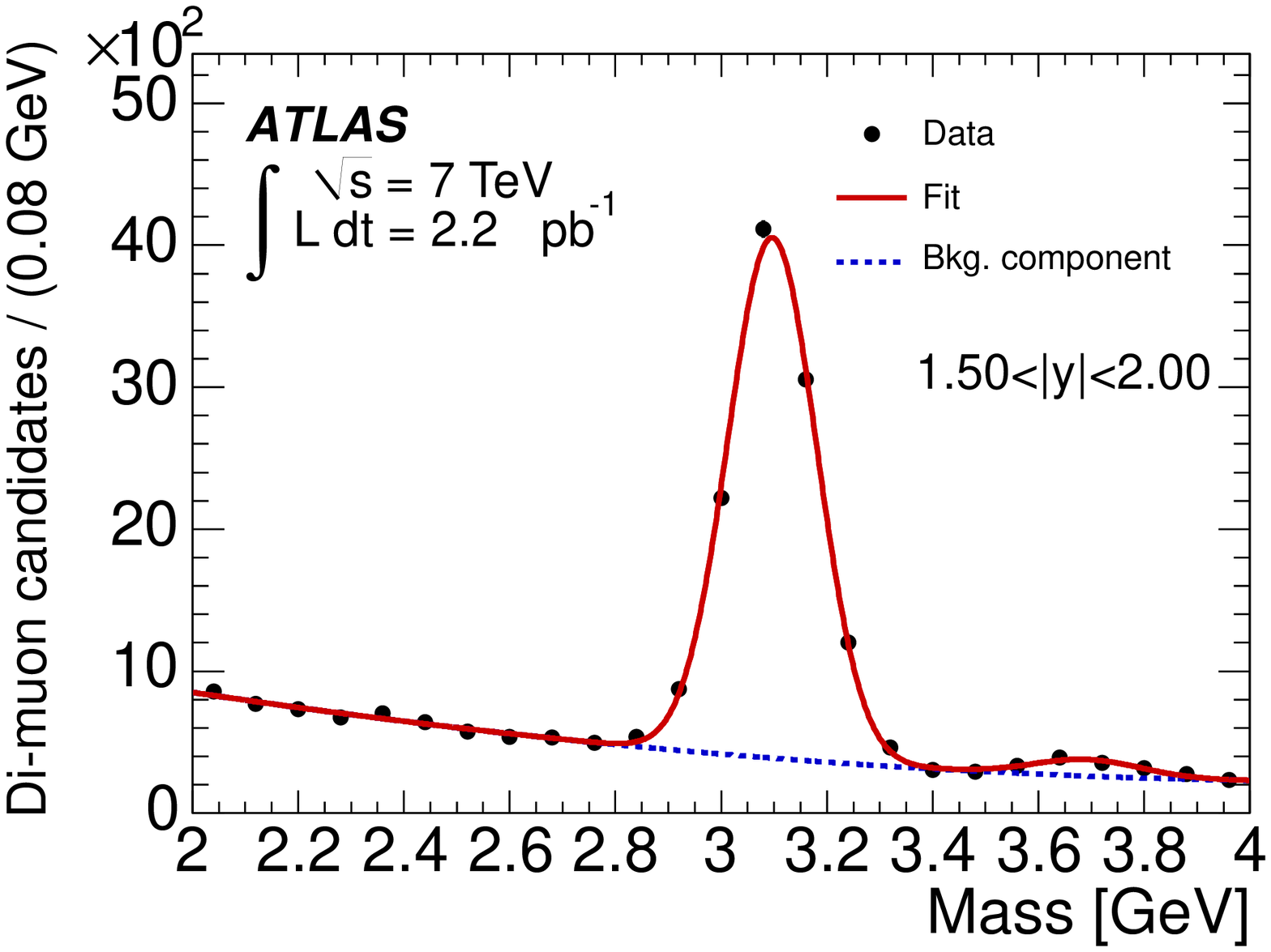}
      \label{fig:Jpsimass4}
      \includegraphics[width=0.45\textwidth]{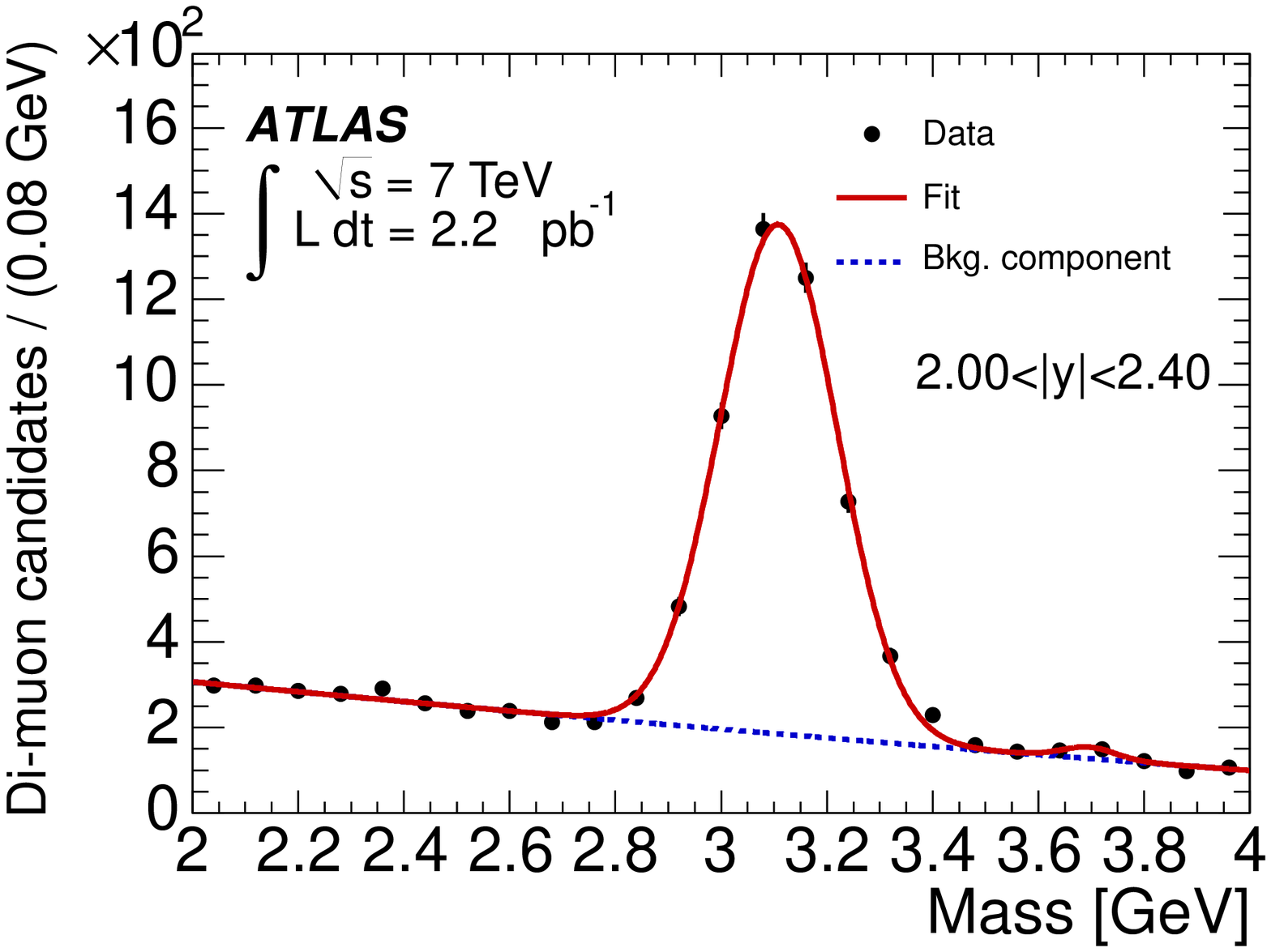}
    \caption{Invariant mass  distributions of reconstructed $\ensuremath{ J/\psi\to \mumu}$ 
candidates used in the cross-section analysis, corresponding to  an integrated luminosity of $2.2$ pb$^{-1}$. 
The points are data, and the uncertainties indicated are statistical only. The solid lines are the result of 
the fit described in the text.}
      \label{fig:Jpsimass}
  \end{center}
\end{figure}

The inclusive $J/\psi$ production cross-section is determined in four slices
of $J/\psi$ rapidity: $|y|<0.75, 0.75<|y|<1.5,1.5<|y|<2.0$ and $2.0<|y|<2.4$. 
In Figure~\ref{fig:Jpsimass}, the invariant mass distributions for all 
oppositely charged muon pairs passing the selection for 
the differential cross-section measurement are shown, 
before acceptance and efficiency corrections, for the four
rapidity slices. Table~\ref{tab:Jpsimass} presents the results of the 
combined signal and background fits. 
In these fits the $J/\psi$ and $\psi$(2S) peaks are represented
by Gaussians, while the background is described by a 
quadratic polynomial. 

\begin{table}[!htb]
\caption{Fitted mass, resolution and yields of $J/\psi$ candidates reconstructed in four $J/\psi$ rapidity bins. All uncertainties quoted are statistical only. The shift in mass away from the world average in the highest rapidity bin reflects the few-per-mille uncertainty in the tracking $p_T$ scale at the extreme ends of the detector.}
\label{tab:Jpsimass}
\begin{center}
\begin{tabular}{r|cccc}
\hline\hline
        & \multicolumn{4}{|c}{$J/\psi$ rapidity range} \\
        & $|y|<0.75$ & $0.75<|y|<1.5$ & $1.5<|y|<2.0$ & $2.0<|y|<2.4$ \\
\hline
Signal yield & $6710\pm 90$ & $10710\pm 120$ & $9630\pm 130$ & $4130\pm 90$ \\ 
Fitted mass (GeV)  & $3.096\pm 0.001$ & $3.097\pm 0.001$ & $3.097\pm 0.001$ & $3.109\pm 0.002$ \\
Fitted resolution (MeV) & $46\pm 1$  & $64\pm 1$ & $84\pm 1$ & $111\pm 2$ \\
\hline\hline
\end{tabular}
\end{center}
\end{table}

The invariant mass distribution of $J/\psi\to\mu^+\mu^-$ candidates
in each $p_T-y$ bin is fitted using a binned minimum-$\chi^2$ method. The
$J/\psi$ and $\psi$(2S) signals are described by single Gaussians, while the background is
treated as a straight line. 

\begin{figure}[!htb]
\begin{center}
\subfigure{
\label{fig:jpsiMassBins_Weight_r0_p00mt}
\includegraphics[width=0.45\textwidth]{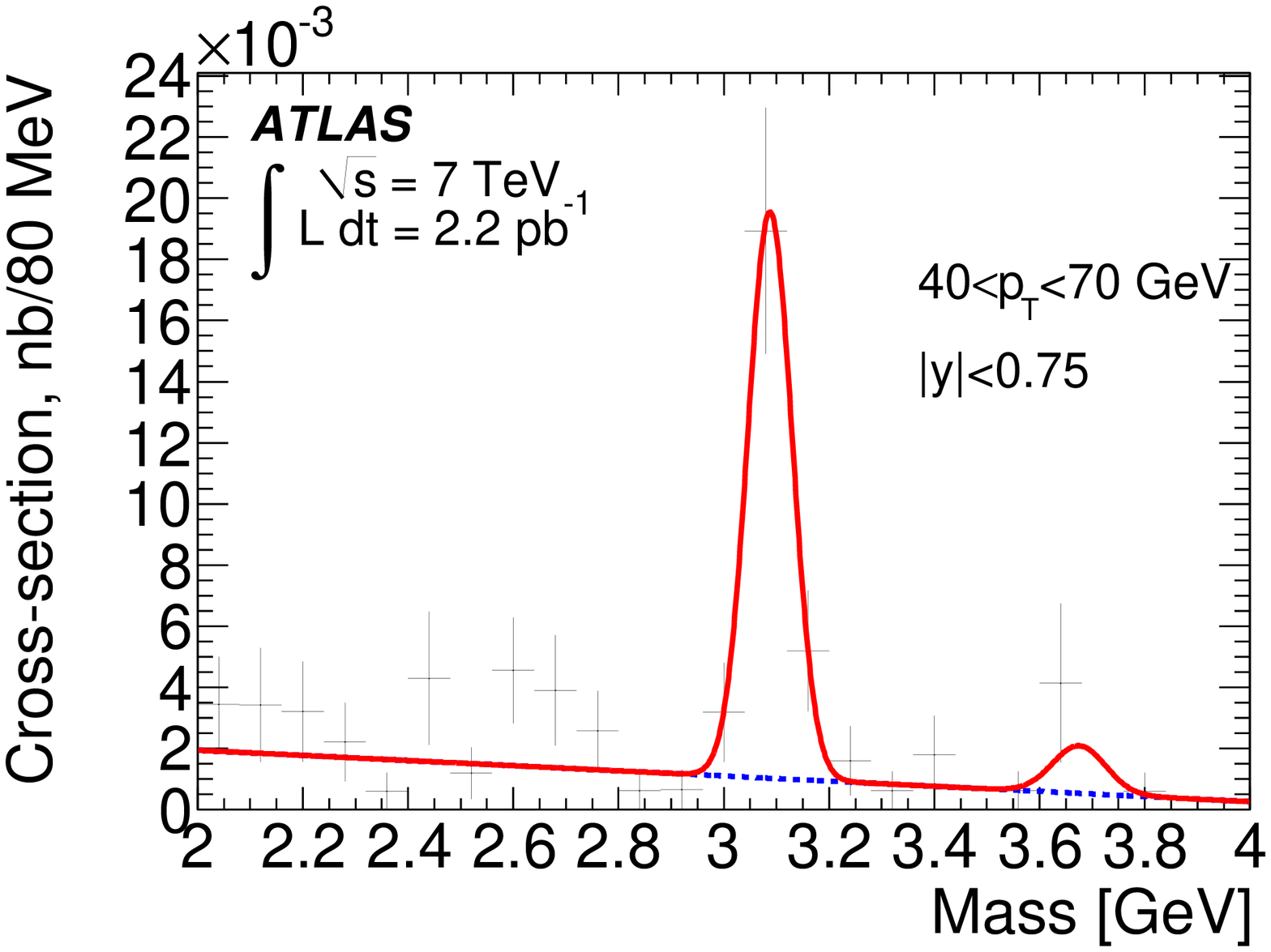}
}
\subfigure{
\label{fig:jpsiMassBins_Weight_r3_p02mt}
\includegraphics[width=0.45\textwidth]{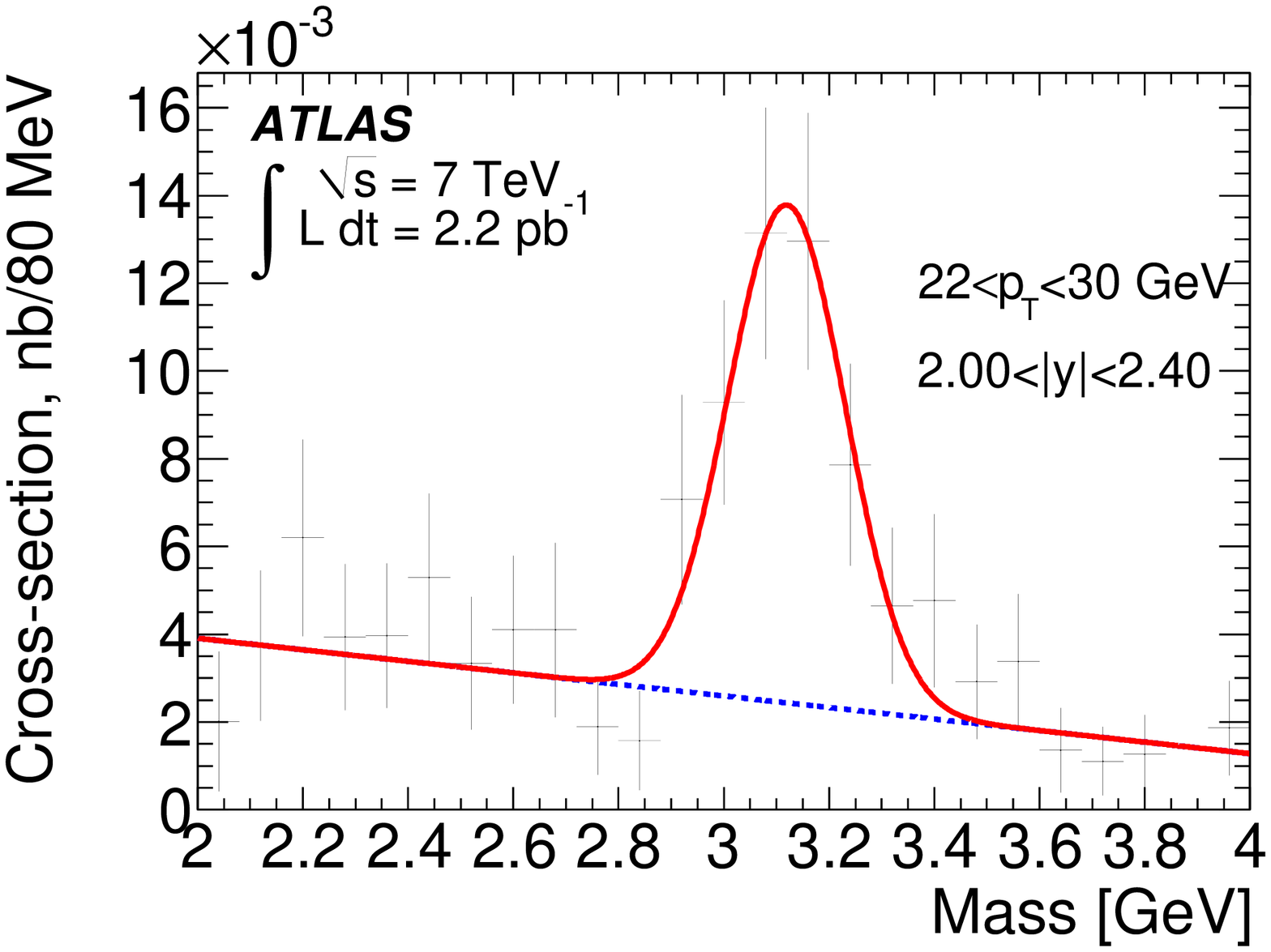}
}
\subfigure{
\label{fig:jpsiMassBins_Weight_r0_p13mt}
\includegraphics[width=0.45\textwidth]{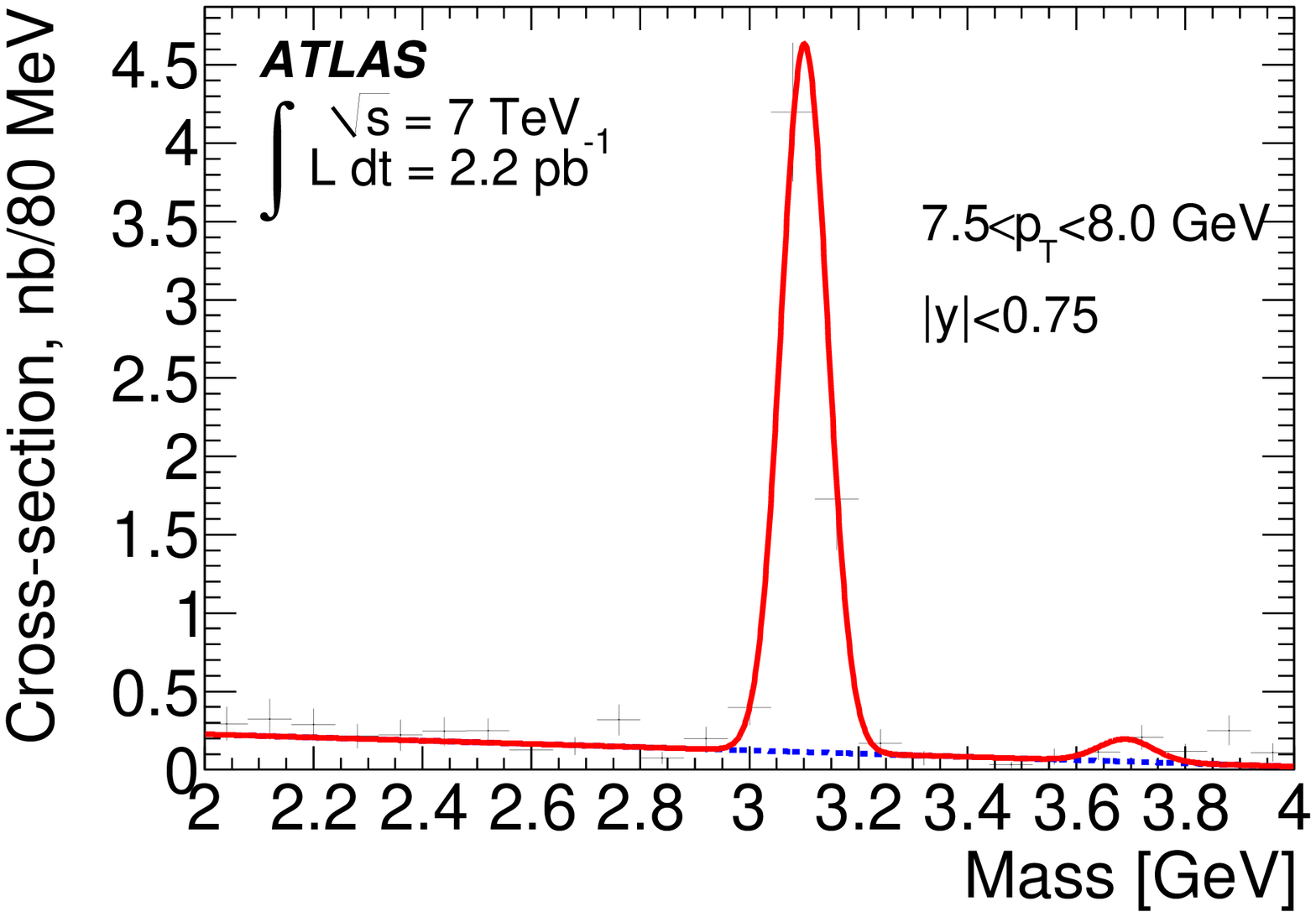}
}
\subfigure{
\label{fig:jpsiMassBins_Weight_r3_p16mt}
\includegraphics[width=0.45\textwidth]{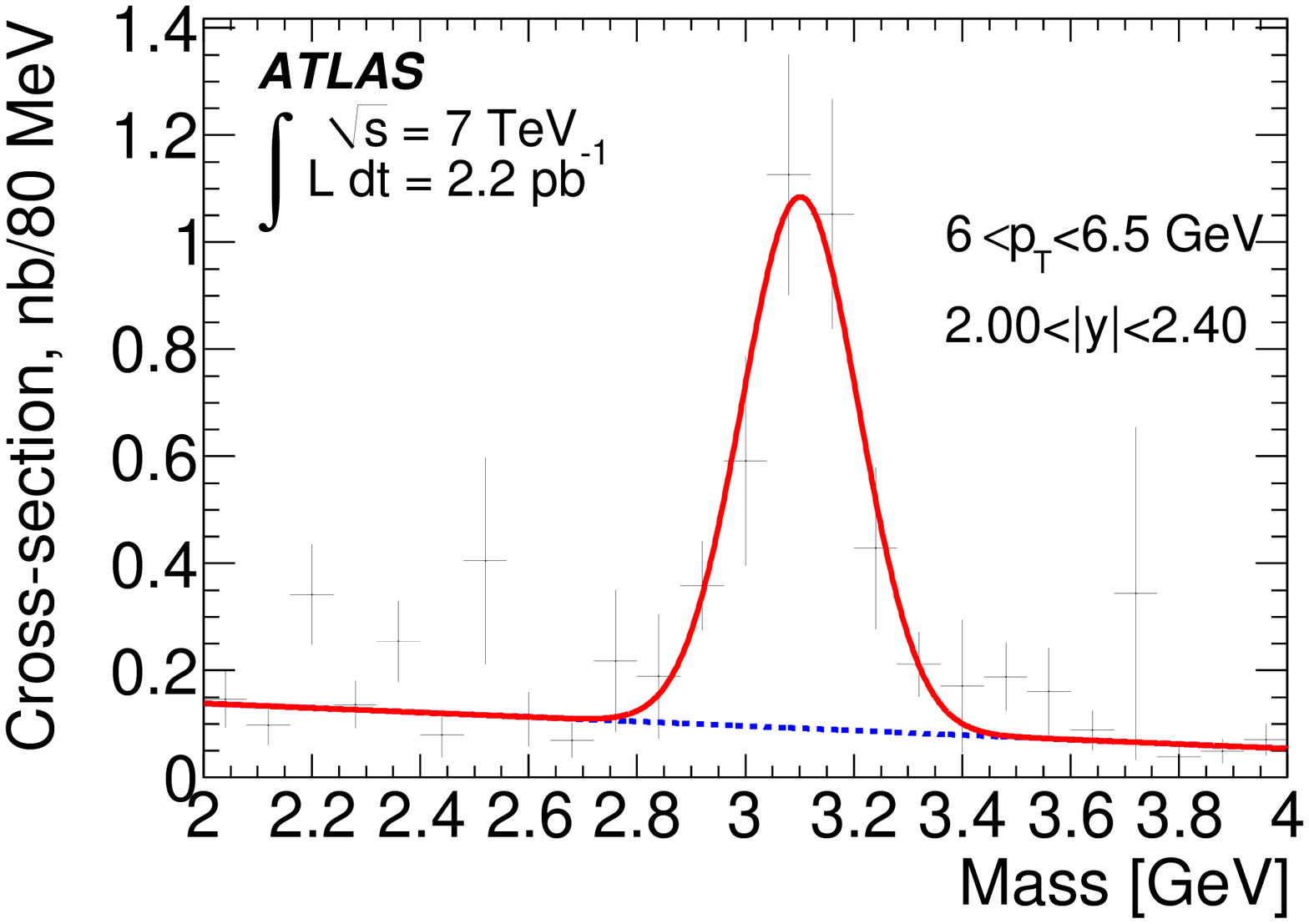}
}
\caption{Acceptance- and efficiency-corrected invariant di-muon mass distributions scaled by integrated luminosity for selected bins in $J/\psi$ rapidity and transverse momentum.
  Low- and high- $p_{T}$ bins are shown here for the central and forward rapidity ranges, to represent the complete sample. Statistical uncertainties and systematic uncertainties 
due to efficiency and acceptance corrections are shown, combined in quadrature.}
\label{fig:weightedMass3mt}
\end{center}
\end{figure}

For the differential cross-section measurement,
the correction weight $w$ defined in Equation~\ref{weights} is applied to each
candidate, and a new binned minimum-$\chi^2$ fit is performed 
in each bin. The yields of $J/\psi$ determined from these fits, 
divided by the integrated luminosity, give the inclusive production
cross-section for a given bin. Representative invariant mass
distributions are shown in Figure~\ref{fig:weightedMass3mt}.
The $\chi^2$ probability distribution of the weighted fits across all bins 
is found to be consistent with the statistical expectation.

The cross-sections obtained for each bin are listed in Table~\ref{tab:Ainclxsec1}, 
the systematic uncertainties considered are displayed in Figure~\ref{fig:totalsystPlot} and the cross-section results are presented in Figure~\ref{fig:xsec_result}. The measurement in each $p_T-y$ analysis bin is positioned at the average $p_T$ for $J/\psi$ candidates in that bin.   
Various tests of the method described above are performed using simulated samples
of known composition, and the number of $J/\psi$ in each analysis bin is successfully recovered within expectations in all cases.

\subsection{Systematic uncertainties}

Studies are performed to assess all relevant sources of systematic uncertainty on 
the measurement of the $J/\psi$ inclusive production cross-section. Sources of uncertainty are listed below, ordered according to the approximate size of their contribution (starting with the largest).

\begin{enumerate}

\item{\bf Spin-alignment:}
Kinematic acceptance depends on the spin-alignment state
of the $J/\psi$ and hence affects the corrected yield. Five spin-alignment 
scenarios are considered, which correspond to the extreme cases for the 
acceptance corrections within the kinematics accessible in ATLAS.
In each bin, the maximal deviations in either direction are assigned as
the systematic uncertainty due to the unknown spin-alignment of the $J/\psi$. These uncertainties are regarded as theoretical rather than experimental, and are quoted independently of the statistical and experimental systematic uncertainties.

\item
{\bf Muon reconstruction:}
The single muon efficiency maps are obtained  from
the data using the tag and probe method, in bins of muon transverse
momentum and pseudorapidity. Each efficiency has an uncertainty (predominantly statistical in nature, but with a systematic component from the tag and probe method)
associated with it. In order to obtain an estimate on the effects of uncertainties within these bins, the relative
uncertainties (due to systematic and statistical components) on all $J/\psi$ candidates in a bin are averaged. 
Inner Detector tracks originating from muons and having the selection cuts used in this analysis have a reconstruction efficiency of $99.5\%\pm 0.5\%$ per track.
The results are corrected for this efficiency, and a systematic uncertainty on the efficiency is assigned for each track, propagated linearly into the cross-section systematic.

\item
{\bf Trigger:}
The uncertainty on the trigger efficiency has components from the data-derived efficiency determination method
(again largely statistical in nature) and from the reweighting of MC maps to the data-driven (tag and probe)
efficiency values. These errors are treated similarly to the reconstruction efficiency uncertainties.

\item{\bf Luminosity:}
The uncertainty on the integrated luminosity used for this measurement
is determined to be 3.4\%\,\cite{conflumidet}, fully correlated between bins.

\item
{\bf Acceptance:}       
\begin{itemize}                 
\item Monte Carlo statistics:
The acceptance maps are obtained from dedicated Monte Carlo production, in bins of $J/\psi$ transverse momentum and
rapidity. The acceptance in each bin has an uncertainty due to Monte Carlo statistics. The relative error on the acceptance correction
for each $J/\psi$ candidate contributing to a particular analysis bin is averaged in quadrature to evaluate the systematic effect of these
errors on the cross-section measurement in that bin.       
\item Kinematic dependence:
The impact of any discrepancies in the underlying kinematic distribution modelling in the Monte Carlo used to build the maps, or differences in the $p_T$ dependence of
the prompt and non-prompt components to the overall inclusive cross-section are studied. A correction to the acceptance maps is made based on the measured non-prompt
to prompt fraction to ensure proper correction of the two populations, and an uncertainty is assigned based on the difference in yields from using the 
corrected and uncorrected maps. This uncertainty is significantly below 1\% in most analysis bins, reaching a maximum of 1.5\% in some high $p_T$, low rapidity bins.
\item Bin migration: 
The changes to the measured cross-section due to the migration of entries
between the $p_T$ bins is determined by analytically smearing the efficiency and acceptance corrected $p_T$ spectrum with a 
Gaussian resolution function with width based on muon $p_T$ resolutions, taken from data. The correction needed to the central value due to bin migrations
is as small as 0.1\% at low $p_T$ and low rapidity and rises to $\sim 3\%$ at high $p_T$ and high rapidity. The variation of the bin migration correction
within a given analysis bin (due to changing detector resolution and parameterisation of the $p_T$ spectrum) is taken as a systematic.
\item Final-State Radiation: 
The acceptance maps correct the measured cross-section back to the $J/\psi$ kinematics, rather than the final-state muon kinematics, in order to allow proper comparison
with theoretical predictions. Emission of QED final-state radiation is known to high accuracy, so the relative uncertainty on the modelling of this correction is
determined to be less than 0.1\%.
\end{itemize}

\item
{\bf Fit:}
Invariant mass distributions for a large number of pseudo-experiments are constructed for each $p_T-y$ bin of the analysis,
with the bin contents for each pseudo-experiment being an independently Poisson-fluctuated value with mean equal to the measured
data, and uncertainty in the bin determining the variance of the fluctuations.
Within these pseudo-experiments, the candidate yields from the central fit procedure and yields from varied fitting models
are determined, and the shift per pseudo-experiment calculated. The variation in fitting models include
signal and background fitting functions and inclusion/exclusion of the $\psi(2S)$ mass region.
The means of the resultant shifts across all pseudo-experiments for each fit model are used to evaluate 
the systematic uncertainty. The largest mean variation in that bin is assigned as a systematic uncertainty
due to the fit procedure.

\item{\bf {\boldmath $J/\psi$} vertex-finding efficiency:}
The loose vertex quality requirement retains over $99.9\%$ of di-muon candidates used in the analysis, 
so any correction and systematics associated to the vertexing are neglected.

\end{enumerate}

\begin{figure}[!htb]
  \begin{center}
    \includegraphics[width=0.94\textwidth]{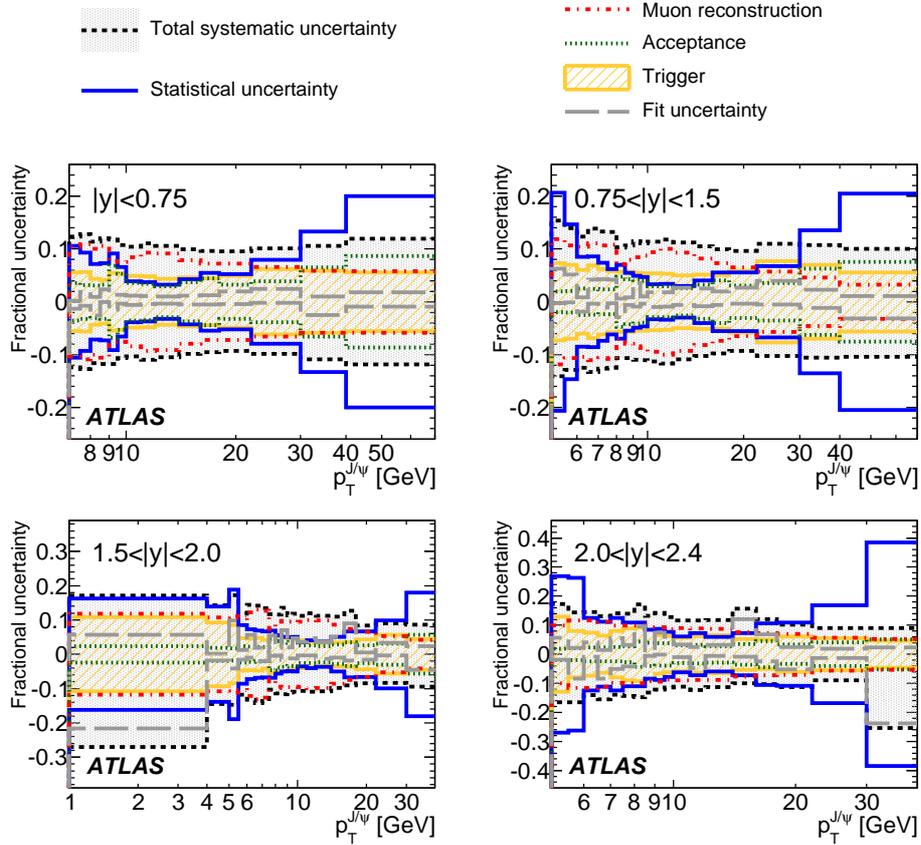}
  \end{center}
  \caption{Summary of the contributions from various sources to the systematic uncertainty
    on the inclusive differential cross-section, in the $J/\psi$ $p_T$ and rapidity
    bins of the analysis. The total systematic and statistical uncertainties are
    also overlaid. The theoretical uncertainty due to the unknown spin alignment is not included on these plots.
  }
 \label{fig:totalsystPlot}
\end{figure}
 
\noindent A summary of the various contributions to the systematic uncertainties
on the measurement in each rapidity slice as a function of $J/\psi$
$p_T$ is shown in Figure~\ref{fig:totalsystPlot}. The uncertainty due to the luminosity (3.4\%) is not shown, nor is
the spin-alignment envelope which represents a full range of variation due to the unknown spin-alignment state.

\subsection{Inclusive {$J/\psi$} cross-section results} 

\begin{sidewaysfigure}[!phtb]
  \begin{center}      
      \includegraphics[width=0.49\textwidth]{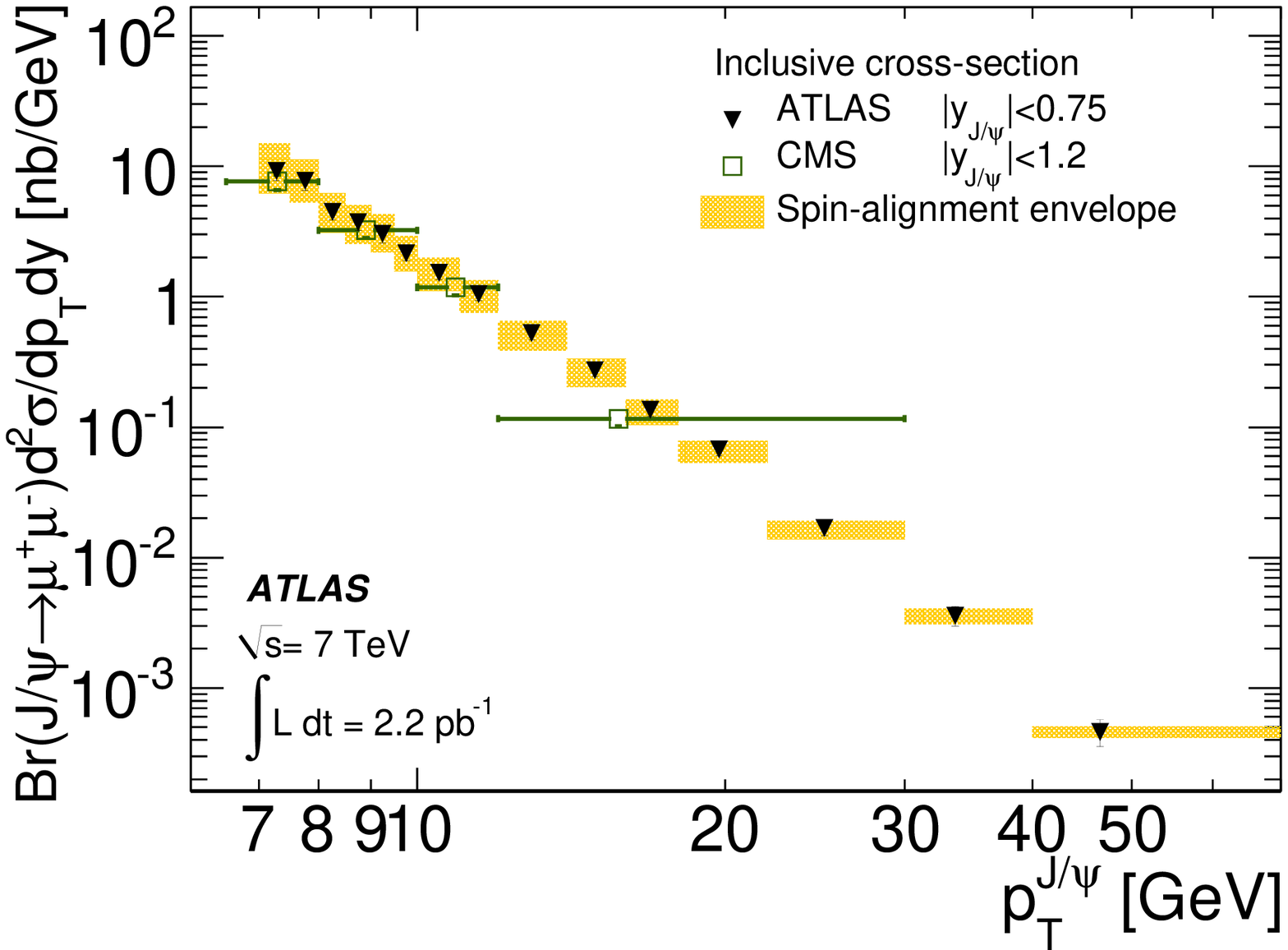}
      \includegraphics[width=0.49\textwidth]{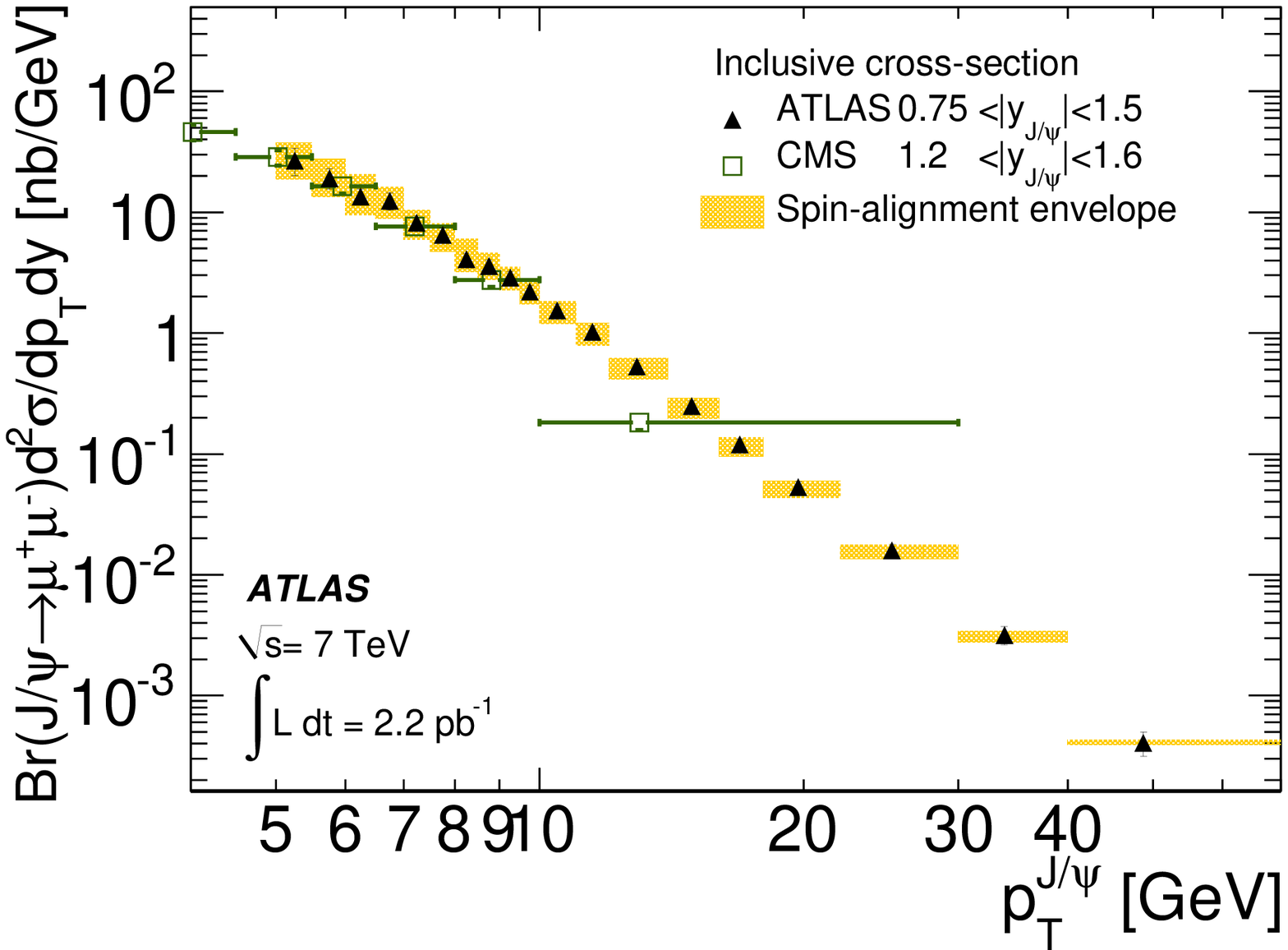}
      \includegraphics[width=0.49\textwidth]{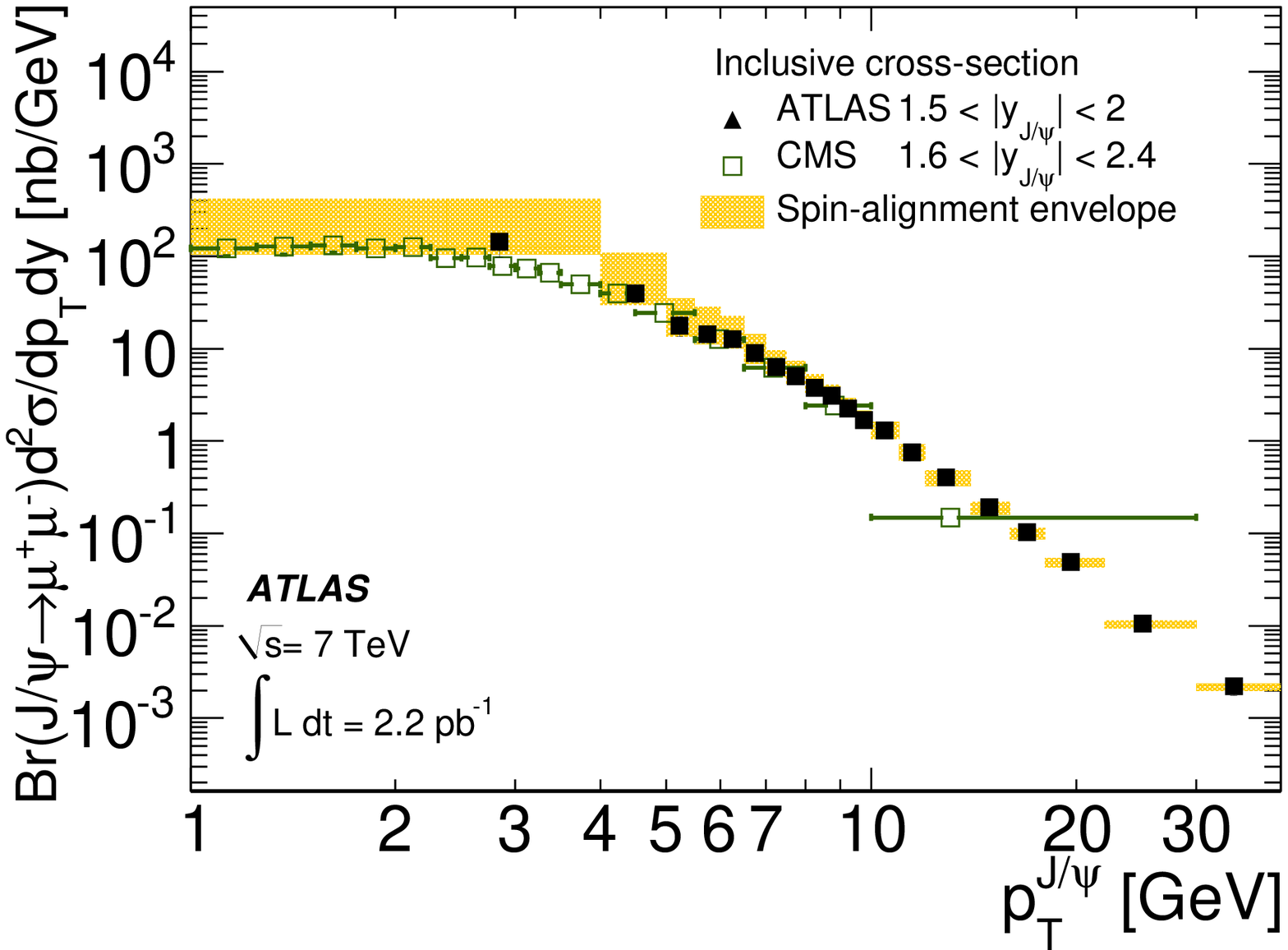}
      \includegraphics[width=0.49\textwidth]{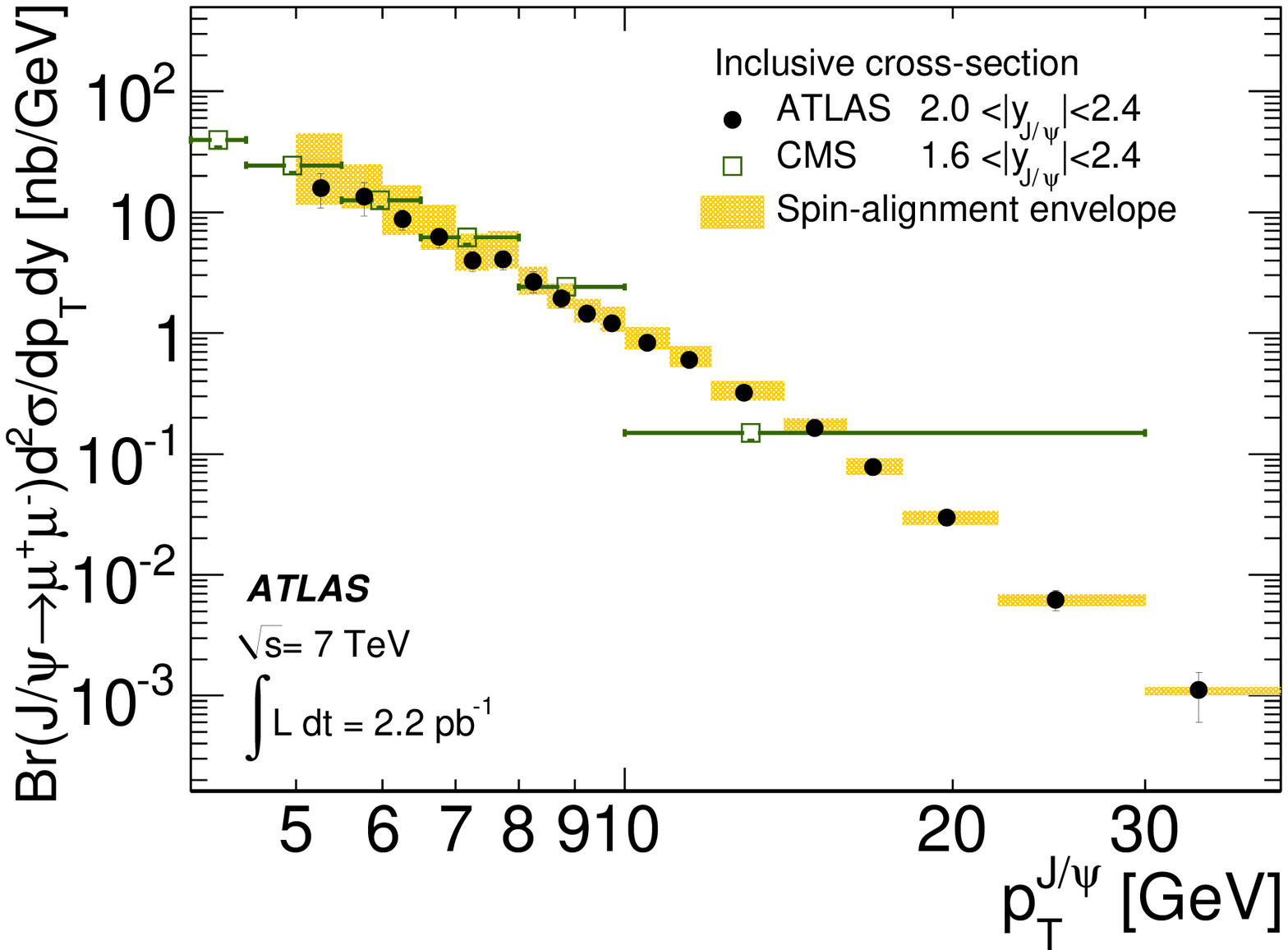}
    \caption{Inclusive $J/\psi$ production cross-section as
      a function of $J/\psi$ transverse momentum in the four rapidity bins.
      Overlaid is a band representing the variation of the result
      under various spin-alignment scenarios (see text) representing a
      theoretical uncertainty. The equivalent results from CMS
      \cite{CMS} are overlaid. The luminosity uncertainty (3.4\%) is not shown. 
      \label{fig:xsec_result}
    }
  \end{center}
\end{sidewaysfigure}

The results of the inclusive double-differential $J/\psi$ production
cross-section measurement are given in Table~\ref{tab:Ainclxsec1}. They are compared to CMS results\,\cite{CMS}
in Figure~\ref{fig:xsec_result} for cases where the rapidity ranges are close enough to permit
comparison. The two sets of results show good agreement within experimental uncertainties 
and provide complementary measurements at low (CMS) and high (ATLAS) $p_T$, together providing a measurement over
a large kinematic range. 

The systematics are dominated by the data-driven muon reconstruction
efficiency uncertainties, which are in turn dominated by their statistical uncertainties.
There is an additional overall uncertainty of $\pm 3.4\%$ (fully correlated
between bins) due to the luminosity measurement
uncertainty.
The measurement of the differential cross-section is limited
by systematic uncertainties, with statistical uncertainties
only contributing significantly near the low-$p_T$ thresholds where
yields are limited by trigger efficiency, and in the highest transverse momentum bin.

The total cross-section for inclusive $J/\psi\to\mu^+\mu^-$ production, multiplied by the branching fraction
into muons and under the FLAT production scenario for the central value, has been
measured for $J/\psi$ 
produced within $|y|<2.4$ and $p_T>7$~GeV 
to be:
\begin{align*}
Br(J/\psi\to\mu^+\mu^-) &\sigma(pp\to J/\psi X; |y_{J/\psi}|<2.4, p^{J/\psi}_T>7~\textrm{GeV}) \\ 
&\qquad = 81 \pm 1 \textrm{ (stat.)} \pm 10 \textrm{(syst.)} \pm ^{25}_{20} \textrm{ (spin)} \pm 3 \textrm{ (lumi.) nb}
\end{align*}
and for $J/\psi$ within $1.5<|y|<2$ and $p_T>1$~GeV to be:
\begin{align*}
Br(J/\psi\to\mu^+\mu^-) &\sigma(pp\to J/\psi X; 1.5<|y_{J/\psi}|<2, p^{J/\psi}_T>1~\textrm{GeV}) \\
&\qquad =510 \pm 70 \textrm{ (stat.)} \pm ^{80}_{120} \textrm{(syst.)} \pm ^{920}_{130} \textrm{ (spin)} \pm 20 \textrm{ (lumi.) nb.}
\end{align*}

\providecommand{\dt}{\ensuremath{\Delta t}\xspace}
\providecommand{\Jpsi}{\ensuremath{J/\psi}\xspace}
\providecommand{\Bbar}{\ensuremath{\bar{B}}\xspace}
\providecommand{\Bz}{\ensuremath{B^0}\xspace}
\providecommand{\Bzb}{\ensuremath{\bar{B}^0}\xspace}
\providecommand{\Bp}{\ensuremath{B^+}\xspace}
\providecommand{\Bm}{\ensuremath{B^-}\xspace}

\providecommand{\fsig}{\ensuremath{f_{\rm sig}}\xspace}
\providecommand{\mmumu}{\ensuremath{m_{\mu\mu}}\xspace}
\providecommand{\sigmmumu}{\ensuremath{\sigma_{m_{\mu\mu}}}\xspace}

\providecommand{\tildet}{\ensuremath{\tau}\xspace}

\providecommand{\sigtildet}{\ensuremath{\sigma_\tau}\xspace}

\providecommand{\fcore}{\ensuremath{f_{\rm core}}\xspace}
\providecommand{\Score}{\ensuremath{S_{\rm core}}\xspace}
\providecommand{\Stail}{\ensuremath{S_{\rm tail}}\xspace}
\providecommand{\fcorebgd}{\ensuremath{f_{\rm core, bgd}}\xspace}
\providecommand{\Scorebgd}{\ensuremath{S_{\rm core, bgd}}\xspace}
\providecommand{\Stailbgd}{\ensuremath{S_{\rm tail, bgd}}\xspace}

\providecommand{\BR}{\mbox{\ensuremath{\mathcal B}}}

\section{Measurement of the Non-Prompt {$J/\psi$} Fraction}

\label{section:ratio}

Experimentally, it is possible to distinguish $J/\psi$ from
prompt production and decays of heavier charmonium states from the 
$J/\psi$ produced in $B$-hadron decays (non-prompt production). The prompt decays occur very close to the primary vertex of the
parent proton-proton collision, while many of the $J/\psi$ mesons produced in $B$-hadron decays will have a measurably displaced
decay point due to the long lifetime of their $B$-hadron parent. 

From the measured distances between the primary vertices and corresponding $J/\psi$ decay vertices the fraction  $f_B$  of $J/\psi$ that originate 
from non-prompt sources, as defined in Equation~\ref{eqn:fraction}, can be inferred. An unbinned maximum likelihood fit is used to extract this fraction from the data.

\subsection{Pseudo-proper time}

The signed projection of the $J/\psi$ flight distance, $\vec{L}$, onto its transverse momentum, $\vec{p}_T^{J/\psi}$, is constructed according to the following formula
\begin{equation}
L_{xy} \equiv {\vec{L}}\cdot {\vec{p}_T^{J/\psi}} / p_T^{J/\psi},
\end{equation}
where $\vec{L}$ is the vector from the primary vertex to the
$J/\psi$ decay vertex and $\vec{p}_T^{J/\psi}$ is the transverse
momentum vector of the $J/\psi$. Here $L_{xy}$ measures the displacement of the $J/\psi$
vertex in the transverse plane.

The probability for the decay of a $B$-hadron as a function of
proper decay time $t$ follows an exponential distribution
\begin{equation}
p(t) = \frac{1}{\tau_B} \exp(-t/\tau_B),
\end{equation}
where $\tau_B$ is the lifetime of the $B$-hadron. 
For each decay the proper decay time can be calculated as
\begin{equation}
t = \frac{L}{ \beta \gamma},
\end{equation}
where $L$ is the distance between the $B$-hadron production and
decay point and $\beta\gamma$ is the Lorentz factor.
Taking the projection of the decay length and momentum on the
transverse plane for $B$-hadrons, one obtains
\begin{equation}
t = \frac{L_{xy}\ m_B}{p_T^{B}}.
\end{equation}
In this case, $L_{xy}$ is measured between the position of the
reconstructed secondary vertex and the primary vertex in the
event. The primary vertex is refitted with the two muon tracks
excluded, to avoid a bias. The uncertainty on $L_{xy}$ is calculated
from the covariance matrices of the primary and the secondary
vertices. The majority of the events contain only a single primary
vertex.  In the few that contain multiple vertices, the $J/\psi$ is
assigned to a primary vertex based on the use of the tracks by the
ATLAS reconstruction software; if both $J/\psi$ tracks are included in
the reconstruction of the same primary vertex, this is the one which
is assigned. In a small number of cases (fewer than $0.2\%$) the two
tracks making the $J/\psi$ candidate are included in the
reconstruction of different primary vertices. These candidates are discarded.

Since the $B$-hadron is not reconstructed completely, one does not
know its transverse momentum. Instead  the $J/\psi$
momentum is used to construct a variable called the ``pseudo-proper time''
\begin{equation}
\tau = \frac{L_{xy}\ m_{\textrm{PDG}}^{J/\psi}}{p_T^{J/\psi}}.
\end{equation}
\noindent
Here, the world average value of $m_{\textrm{PDG}}^{J/\psi}$ is used to reduce the correlation between the
fits that will be performed on the mass and the lifetime. Studies show that the
results are insensitive to this choice.

At large $p_T^{J/\psi}$, where most of the $B$-hadron transverse momentum is
carried by the $J/\psi$, the distribution of $\tau$ is
approximately exponential, with the $B$-hadron lifetime
as a parameter. At small $p_T^{J/\psi}$, the range of opening angles between the
$J/\psi$ and $B$-hadron momentum leads to a smearing of the underlying
exponential distribution.
\subsection{Fitting procedure}
The sample is divided into bins of $p_{T}$ and rapidity $y$ of the $J/\psi$ candidates. 
In each bin,
a maximum likelihood fit is performed in order to
determine the fraction of the non-prompt to inclusive $J/\psi$ production cross-sections in that particular bin. 
The mass and pseudo-proper time are simultaneously fitted in the entire mass region from $2.5$ to $3.5$ GeV, using the likelihood function:
\begin{equation} 
 L =  
   \prod_{i=1}^N  \left[ { f_{\textrm{sig}}  {P}_{\textrm{sig}}(\tau,\delta_{\tau}) {F}_{\textrm{sig}}(m_{\mu\mu},\delta_m) 
+ ( 1- f_{\textrm{sig}} ) } {P}_{\textrm{bkg}}(\tau,\delta_{\tau}){F}_{\textrm{bkg}}(m_{\mu\mu}) \right]
\end{equation}
 where $N$ is the total number of events in the $2.5-3.5$ GeV mass region and $f_{\textrm{sig}}$ is the fraction of signal $J/\psi$ candidates in this region determined from the fit.  ${P}_{\textrm{sig}}$ and  ${P}_{\textrm{bkg}}$ are
pseudo-proper time probability density distributions (PDFs) for the $J/\psi$ signal and background candidates respectively, and are described fully below. The $F_{\textrm{sig}}$, $F_{\textrm{bkg}}$ functions are the mass distribution models for signal and background. 
In summary, the input variables to the maximum likelihood fit to determine the production ratio
are the pseudo-proper time $\tau$, its uncertainty $\delta_{\tau}$, the di-muon mass $m_{\mu\mu}$ and its
uncertainty $\delta_m$.

\subsubsection{Invariant mass and pseudo-proper time probability density functions}
\label{sec:ratio_pdf}
For the signal, the mass is modelled with a Gaussian distribution: 
\begin{equation}
F_{\textrm{sig}}  (m_{\mu\mu}, \delta_m)  \equiv  \frac{1}{ \sqrt{2\pi}~S \delta_m } e^{\frac{-(m_{\mu\mu}-m_{J/\psi})^{2}}{2( S\delta_m)^{2}}}
\label{pdfsig}
\end{equation}
whose mean value  $m_{J/\psi}$ is the $J/\psi$  mass, determined in the fit, and whose width is the product  
$S\delta_m$, where $ \delta_m$  is the measured mass error calculated for each muon pair 
from the covariance matrix of the vertex reconstruction and $S$ is a global scale factor to account for a difference 
between $\delta_m$ and the mass resolution from the fit.  For the background, the mass distribution is assumed to follow a second-order 
polynomial function.
   
The pseudo-proper time PDF for $J/\psi$ signal candidates, ${P}_{\textrm{sig}}$, consists of two terms. 
One term describes the $J/\psi$ from $B$-hadron decays (${P}_B$), and the other describes the
$J/\psi$ from prompt decays (${P}_P$):
\begin{equation}
{P}_{\textrm{sig}}(\tildet,\delta_{\tau}) = f_B{P}_B(\tildet,\delta_{\tau}) + (1-f_B){P}_P(\tildet,\delta_{\tau}),
\end{equation}
where $f_B$ is the fraction of $J/\psi$ from $B$-hadron decays as defined in Equation~\ref{eqn:fraction}.

The pseudo-proper time distribution of the $J/\psi$ particles from $B$-hadron
decays ${P}_B(\tildet,\delta_{\tau})$ is an exponential function $E(\tau) = \exp(-\tildet/\tau_{\textrm{eff}})$ with a pseudo-proper time slope $\tau_{\textrm{eff}}$,
convolved with the pseudo-proper time resolution function $R(\tildet'-\tildet,\delta_{\tau})$:
\begin{equation}
{P}_B(\tildet,\delta_{\tau}) = R(\tildet'-\tildet,\delta_{\tau}) \otimes E(\tildet').
\end{equation}
Promptly produced $J/\psi$ particles decay at the primary vertex, and their pseudo-proper time distribution is thus given by
the resolution function:
\begin{equation}
{P}_P(\tildet,\delta_{\tildet}) = R(\tildet'-\tildet,\delta_{\tau})
\otimes \delta(\tildet') = R(\tau,\delta_{\tau}).
\end{equation}
The resolution function $R$ is a Gaussian distribution centred at $\tildet=0$
with a width $S_t\delta_{\tau}$, where $S_t$ is a scale factor (a parameter of the fit) and $\delta_{\tau}$ is the per-candidate uncertainty on \tildet, 
the measured pseudo-proper lifetime determined from the covariant error matrix of the tracks.

The pseudo-proper time PDF for background candidates ${P}_{\textrm{bkg}}$ 
consists of the sum of a long lived component modeled with an exponential function and a prompt component modeled
by a delta function and two symmetric exponential tails.  Each component is convolved with the Gaussian resolution function:
\begin{equation}
{P}_{\textrm{bkg}}(\tau,\delta_{\tau}) = \left(
(1- b_1- b_2) \delta(\tildet') + 
b_1 \exp\left(\frac{-\tildet'}{\tau_{\textrm{eff1}}}\right)
+  b_2 \exp\left(\frac{-|\tildet'|}{\tau_{\textrm{eff2}}}\right) 
\right) \otimes R_{\textrm{bkg}}(\tildet'-\tau,\delta_{\tau}),
\end{equation}
\noindent where $R_{\textrm{bkg}}(\tildet)$
is a Gaussian   
distribution centered at $\tildet=0$ with a width 
$S_{\textrm{bkg}} \delta_{\tau}$, where $S_{\textrm{bkg}}$  is a  scale factor (a parameter of the fit) and $\delta_{\tau}$ is the per-candidate uncertainty on the measured $\tau$. 
Parameters $\tau_{\textrm{eff1}}$ and $\tau_{\textrm{eff2}}$ are pseudo-proper time slopes of the two components of background, and  $b_1$ and $b_2$  are  the corresponding fractions of the background.  All four parameters ($\tau_{\textrm{eff1}}$, $\tau_{\textrm{eff2}}$, $b_1$ and $b_2$) are determined from the fit.

\subsubsection{Summary of free parameters}
The full list of the parameters of the fit are as follows:
\begin{itemize}
\item  $f_{\textrm{sig}}$  the fraction of signal $J/\psi$ candidates in the $2.5-3.5$ GeV mass region of the fit;  $m_{J/\psi}$ the mean value  of the $J/\psi$  mass;  the scale factor $S$ to account for a difference between $\delta_m$ and the mass resolution from the fit; 
\item  $f_B$  the fraction of $J/\psi$ from $B$-hadron decays; a pseudo-proper time slope $\tau_{\textrm{eff}}$ describing the  $B$-hadron decays; $S_t$ a scale factor to account for a difference between $\delta_{\tau}$ and the  $B$-hadron pseudo-proper time resolution from the fit;
\item  the slope parameters $\tau_{\textrm{eff1}}$, $\tau_{\textrm{eff2}}$ and $S_{\textrm{bkg}}$  describing the time evolution of the $J/\psi$ background, in analogy to the parameters of $B$-hadron decays, defined above;  $b_1$ and $b_2$, fractions of the two background components.

\end{itemize}

\subsection{Results of the likelihood fits}
\label{sec:timeresults}

\begin{figure}[htb]
  \begin{center}
\subfigure[$|y_{J/\psi}| < 0.75$]{\includegraphics[width=0.45\textwidth]{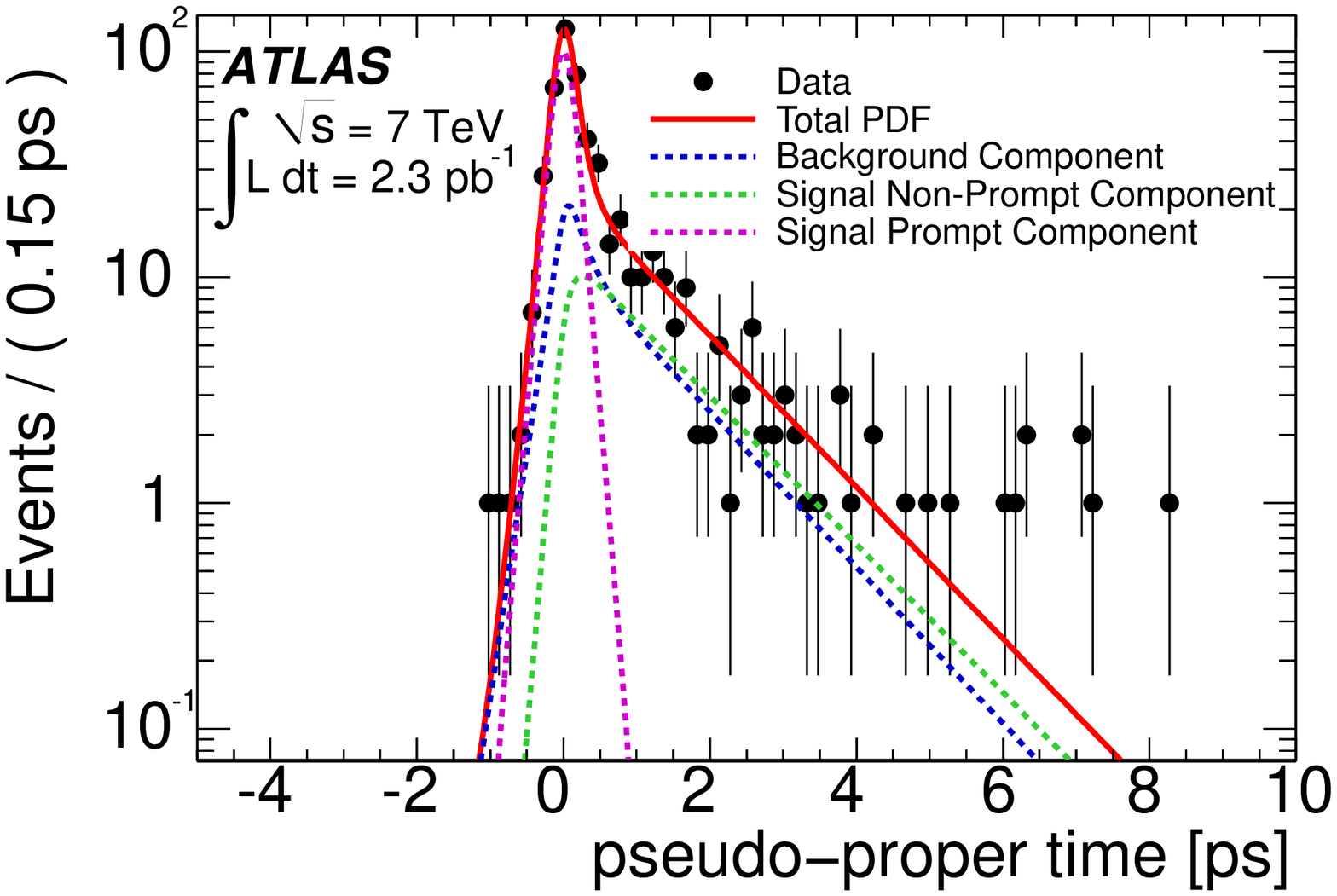}}
\subfigure[$|y_{J/\psi}| < 0.75$]{\includegraphics[width=0.45\textwidth]{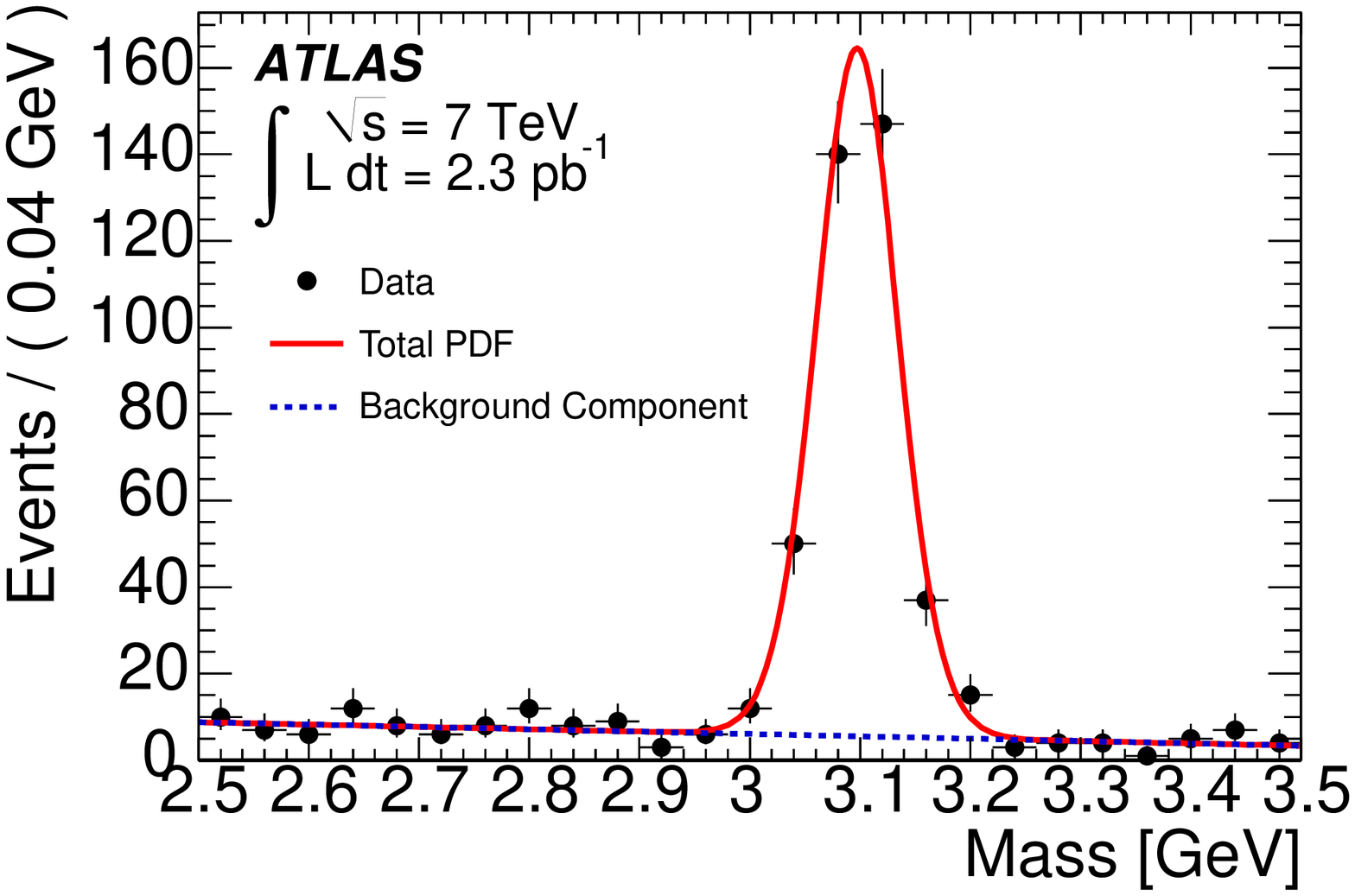}}
\subfigure[$2.0<|y_{J/\psi}| < 2.4$]{\includegraphics[width=0.45\textwidth]{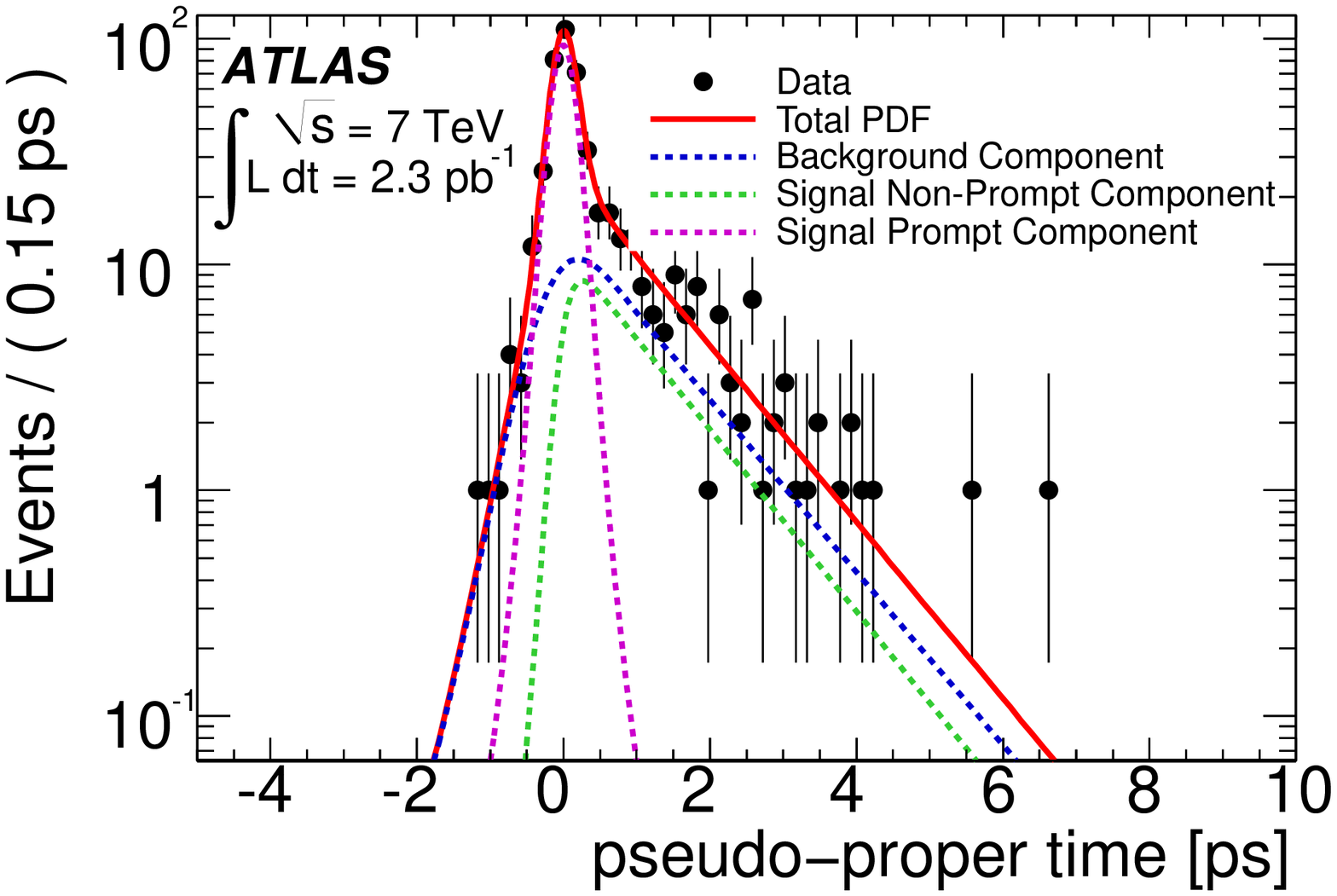}}
\subfigure[$2.0<|y_{J/\psi}| < 2.4$]{\includegraphics[width=0.45\textwidth]{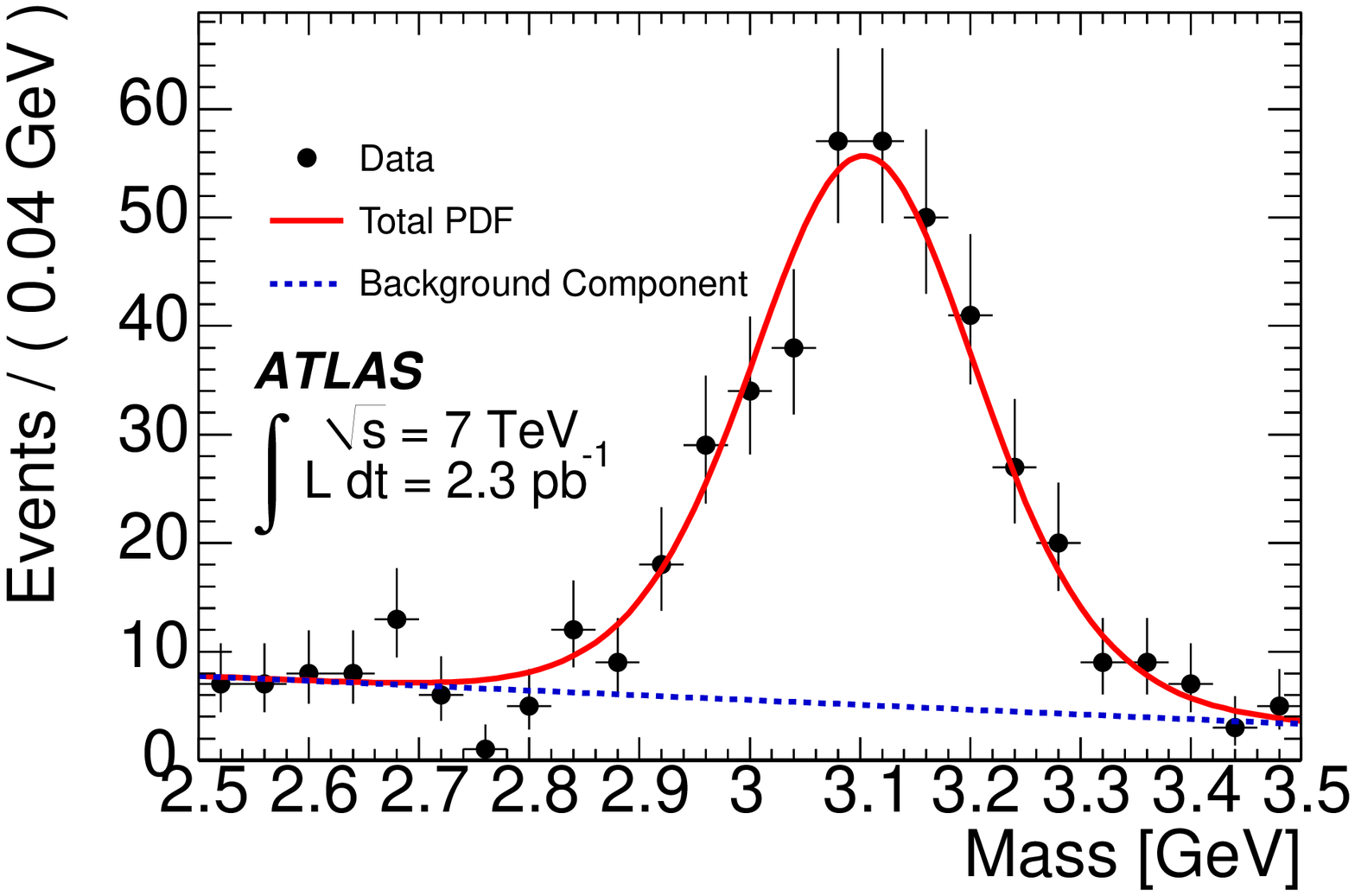}}
  \end{center}
  \caption{Pseudo-proper time distributions (left) of $J/\psi\to\mu^+\mu^-$
    candidates in the signal region, for a selected $p_T$ bin
    $9.5<p_T<10.0$~GeV in the most central and most forward rapidity regions. The points with error bars are data. The solid
    line is the result of the maximum likelihood unbinned fit to all
    di-muon pairs in the $2.5-3.5$ GeV mass region projected on the
    narrow mass window $2.9-3.3$ GeV. The invariant mass distributions
    which are simultaneously fitted with the pseudo-proper time are
    shown on the right for the same bins.
  }
 \label{fig:pstime_fits}
\end{figure}

The results of the likelihood fit to the pseudo-proper time distributions in a representative $p_T^{J/\psi}$ bin are shown in Figure \ref{fig:pstime_fits}. 
The figure shows the result of the unbinned maximum likelihood fits for the signal and background components projected onto the lifetime and invariant mass distributions.
From the results of the fit, it is possible to derive the non-prompt to inclusive production fraction as a function
of $p_T^{J/\psi}$. The $\chi^2$ probabilities and Kolmogorov-Smirnov test results for the fits across all analysis bins are found to be consistent with statistical expectations, with the lowest fit probability out of over 70 fits being 1\%.

\subsection{Systematic uncertainties}

Several studies performed to assess all relevant sources of systematic uncertainties on the measured fraction of non-prompt to inclusive $J/\psi$ decays are outlined below,
in order of importance.

\begin{enumerate}

\item
{\bf Spin-alignment of prompt ${\boldmath J/\psi}$:}
In general, spin-alignment may be different for prompt and non-prompt $J/\psi$, which may result in different acceptances in the two cases. The central value assumes they are the same (isotropic distribution in both angles, as for the inclusive cross-section central result), but four additional scenarios for the prompt component are also considered, as discussed in Section~\ref{sec:acceptance}.
The largest variations within the four models from FLAT is calculated for each bin in turn and assigned as an uncertainty envelope on prompt production. 

\item
{\bf Spin-alignment of non-prompt ${\boldmath J/\psi}$:}
The possible variation of spin-alignment in $B\to J/\psi X$ decays is expected to be much smaller than for prompt $J/\psi$ due to the averaging effect
caused by the admixture of various exclusive $B\to J/\psi X$ decays.
We assign an additional uncertainty on the non-prompt fraction (and non-prompt cross-section) for the difference in final result when using either an isotropic spin-alignment
assumption for non-prompt decays or maps reweighted to the CDF result\,\cite{CDF_bjpsi_pol} for $B\to J/\psi$ spin-alignment.
This contributes up to an additional 0.4\% uncertainty on the overall
(prompt and non-prompt) systematic due to spin-alignment on the fraction.

\item
{\bf Fit:}
A number of changes are applied to the fitting procedure,
and the fit is repeated in order to gauge the sensitivity of the fraction $f_B$ to the details of the fits:

\begin{itemize}
\item{
The central value for the fraction assumes a background model for the proper time distribution of the background that includes one exponential function with a negative slope and a symmetric double exponential term with the same absolute value, $\tau_{\textrm{eff2}}$, for the negative and positive slopes. To test the robustness of the result, this model is changed in two ways. First, the symmetric term is no longer required to be symmetric, so different values of the negative and positive slopes are allowed. Second, the sum of two asymmetric double exponentials is used, having the same negative decay constant but differing positive decay constants. The maximum deviation from the central value is taken as a systematic uncertainty.
}
\item {
The per-candidate Gaussian convolution function is changed to a per-candidate double Gaussian convolution, 
allowing different scale factors (to account for differences between the resolution returned by the tracking algorithm and measured resolution) for each Gaussian to be determined from the fit. 
Differences from the main fit are assigned as a systematic uncertainty.
}
\item {
The main result uses a second-order polynomial in the mass fit to describe the background. To test the sensitivity to this choice, the fits are repeated using instead polynomials
of degree one and three. Differences from the main fit are assigned as a systematic. 
}
\item {
The central result takes $J/\psi$ candidates in a mass range from $2.5$ to $3.5$~GeV, to avoid the mass region of the $\psi$(2S). In order to test the stability of the result and to increase the statistics in the side bands, the analysis is repeated with a mass range from 2 to 4~GeV, but excluding the region from $3.5$ to $3.8$~GeV. The result is stable compared to the statistical uncertainties, and so no systematic uncertainty is assigned for this source.
}
\item {
The analysis relies on a simultaneous fit to the proper time and mass distributions. The likelihood used assumes no correlation between the two quantities. To test the reliability of this assumption, the mean measured invariant mass is plotted as a function of the proper time. The resulting distribution is flat, except in the negative lifetime region and at very long proper lifetimes, where residual background dominates the sample and invalidates the test. Accordingly, no explicit systematic for this correlation is assigned.
}
\end{itemize}

\item 
{\bf Kinematic dependence:}
Differences in the acceptance of prompt and non-prompt $J/\psi$ due to their different momentum spectra, averaged across an analysis bin, can bias the fraction measurement.
A correction factor is calculated based on the acceptance maps with and without momentum reweighting to account for the differences between prompt and non-prompt $J/\psi$
and this correction assigned as a systematic uncertainty.

\item
{\bf Reconstruction efficiencies:}
The central result for the fraction assumes that
the reconstruction efficiencies are the same for non-prompt and prompt $J/\psi$ mesons and hence cancel in extracting the fraction.
This assumption is tested on Monte Carlo samples described in Section \ref{section:samples},
and no statistically significant shift is observed. Thus, no systematic uncertainty is assigned.

\item
{\bf Pile-up/multiple interactions:}
Some collisions result in the reconstruction of multiple primary vertices. The primary
vertex chosen determines the transverse decay displacement $L_{xy}$ used in
the proper time determination. The central value is obtained by taking
the primary vertex that is formed using both of the $J/\psi$ candidate
muons and rejecting cases where those candidates are associated
with different primary vertices. To assess the effect of this
procedure, two alternate methods where used. The first chooses the
primary vertex with the highest summed squared transverse momenta of the tracks
that form it. The second takes the same vertex, but rejects cases
where either of the muon candidates are not used in determining that
primary vertex. As no significant variation is seen in the results
from the two methods, no additional uncertainties are assigned due to
this source.

\end{enumerate}

The stability of the method used is checked using simplified Monte
Carlo trial experiment samples to perform various tests of the closure
of the analysis. The simultaneous mass and pseudo-proper time fit model is used to generate 100 simplified Monte Carlo experiments for each $p_{T}$ and $y$ bin. The number of events generated is approximately the same as the number of data events for the corresponding bin. For each event the invariant mass and pseudo-proper time values are generated randomly from the total PDF, while the per-candidate error on invariant mass and pseudo-proper time are sampled from the corresponding experimental data distributions. 

For each experiment, a fit of the total PDF on the simple Monte Carlo sample is performed. The pull, $\Delta$, defined as 
\begin{eqnarray}                                                                                       
\Delta = \frac{(f_{\textrm{generated}} - {f_{\textrm{extracted}})}}{\sigma(f_{\textrm{extracted}})}, \nonumber
\end{eqnarray}
is computed for each Monte Carlo experiment. Here $f_{\textrm{generated}}$ is the non-prompt fraction for the signal component  according to which the Monte Carlo samples are generated (i.e. the result of the fit of the global model to the experimental data), while $f_{\textrm{extracted}}$ and $\sigma(f_{\textrm{extracted}})$ are the value and uncertainty obtained from the fit.  The Gaussian mean and sigma are statistically compatible with zero and unity, respectively, in all bins, indicating that no bias or improper uncertainty estimate is introduced by the fit.

\subsection{Fraction of non-prompt { $J/\psi$}  as a function of { $J/\psi$} transverse momentum and rapidity} 
\begin{sidewaysfigure}[htbp]
  \begin{center}
      \includegraphics[width=0.49\textwidth]{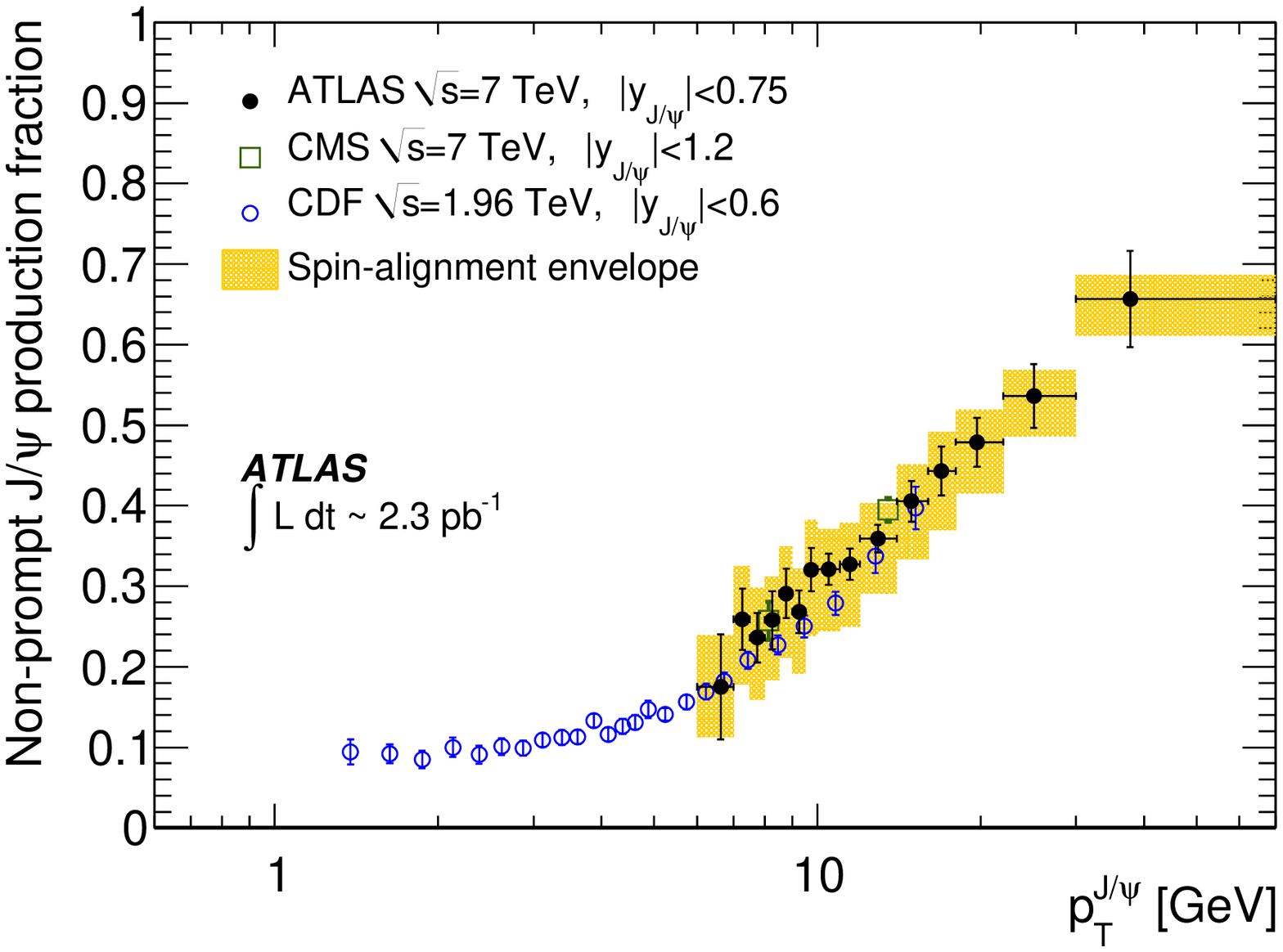}
      \includegraphics[width=0.49\textwidth]{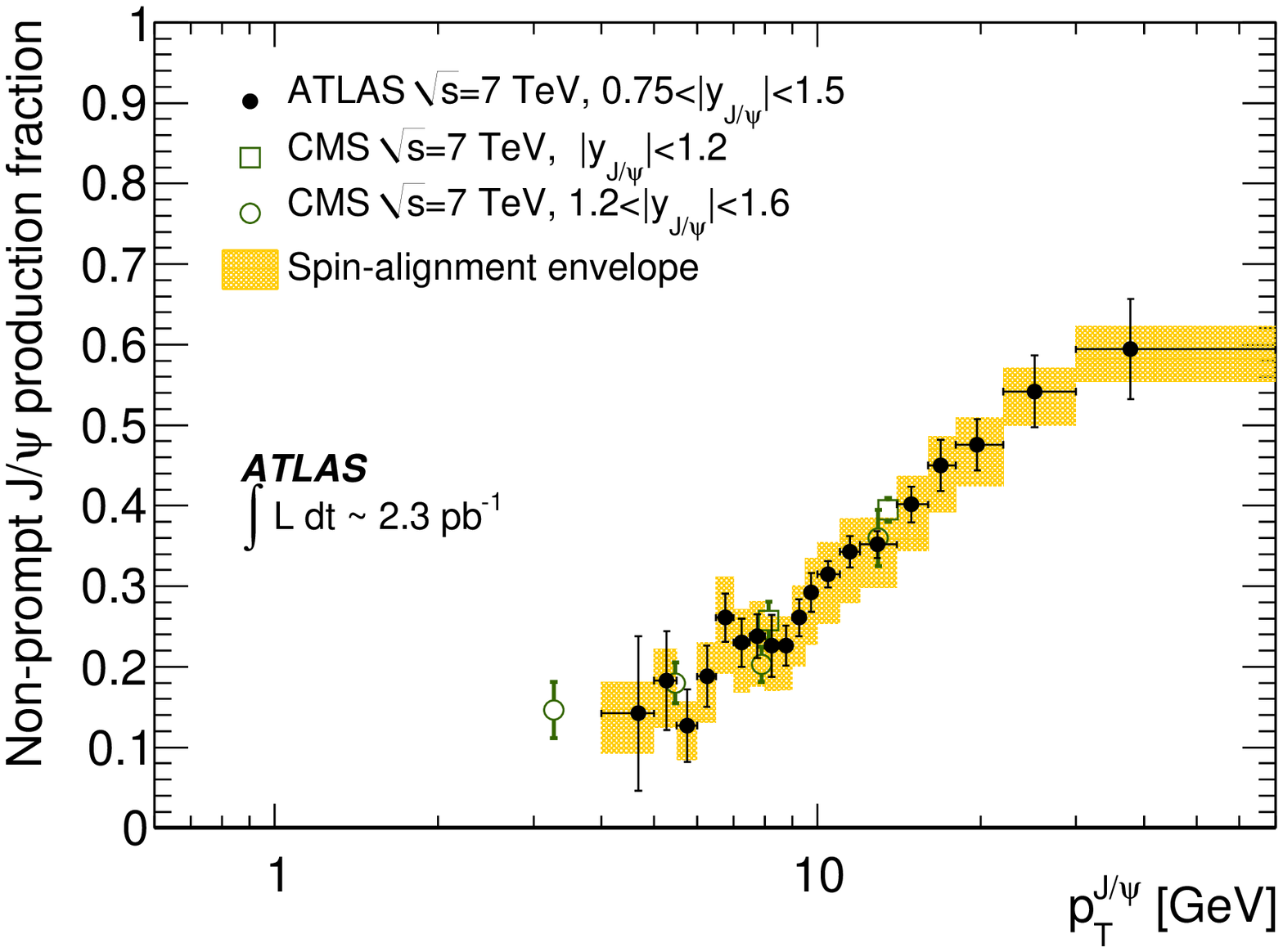}
      \includegraphics[width=0.49\textwidth]{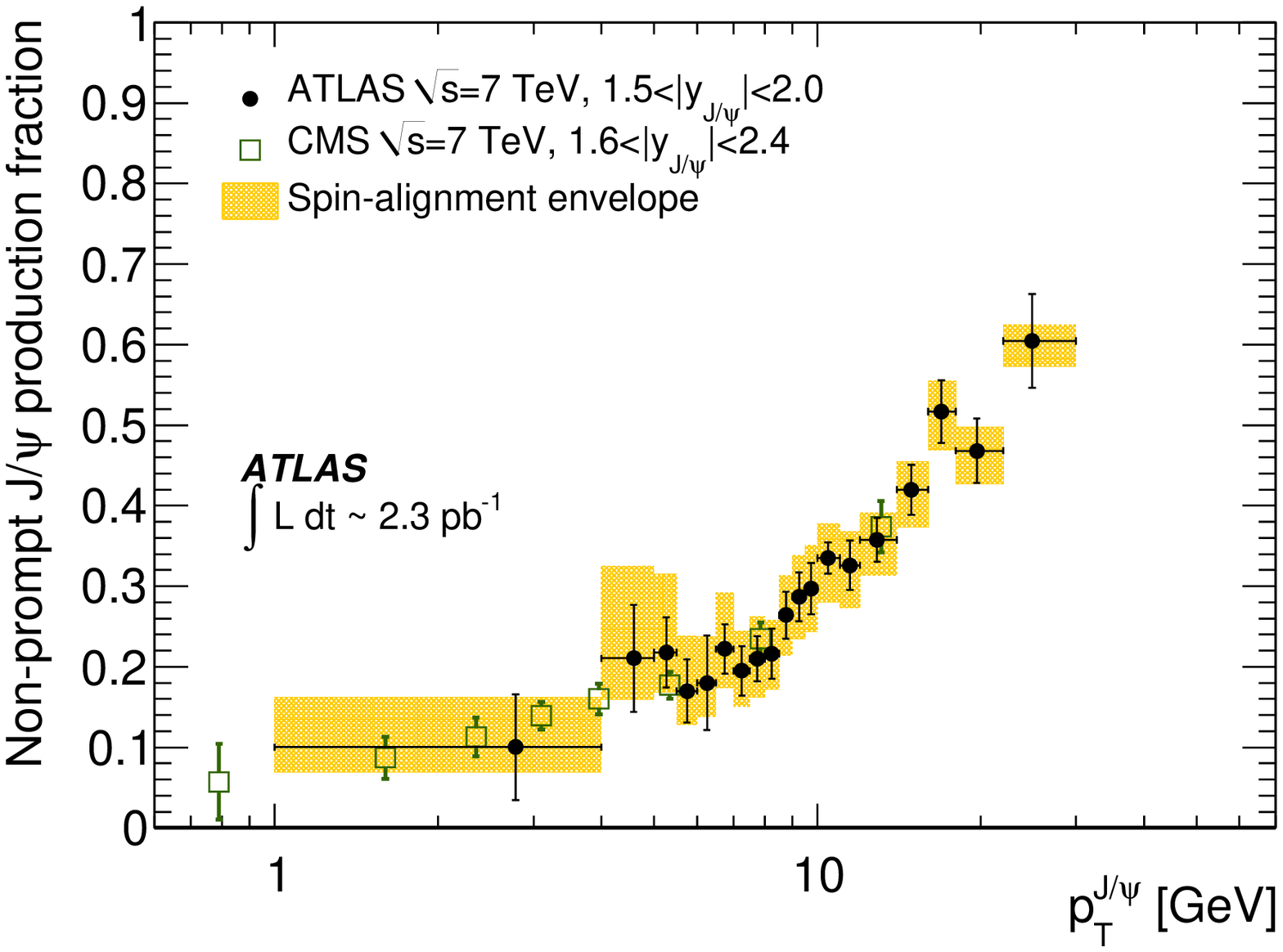}
      \includegraphics[width=0.49\textwidth]{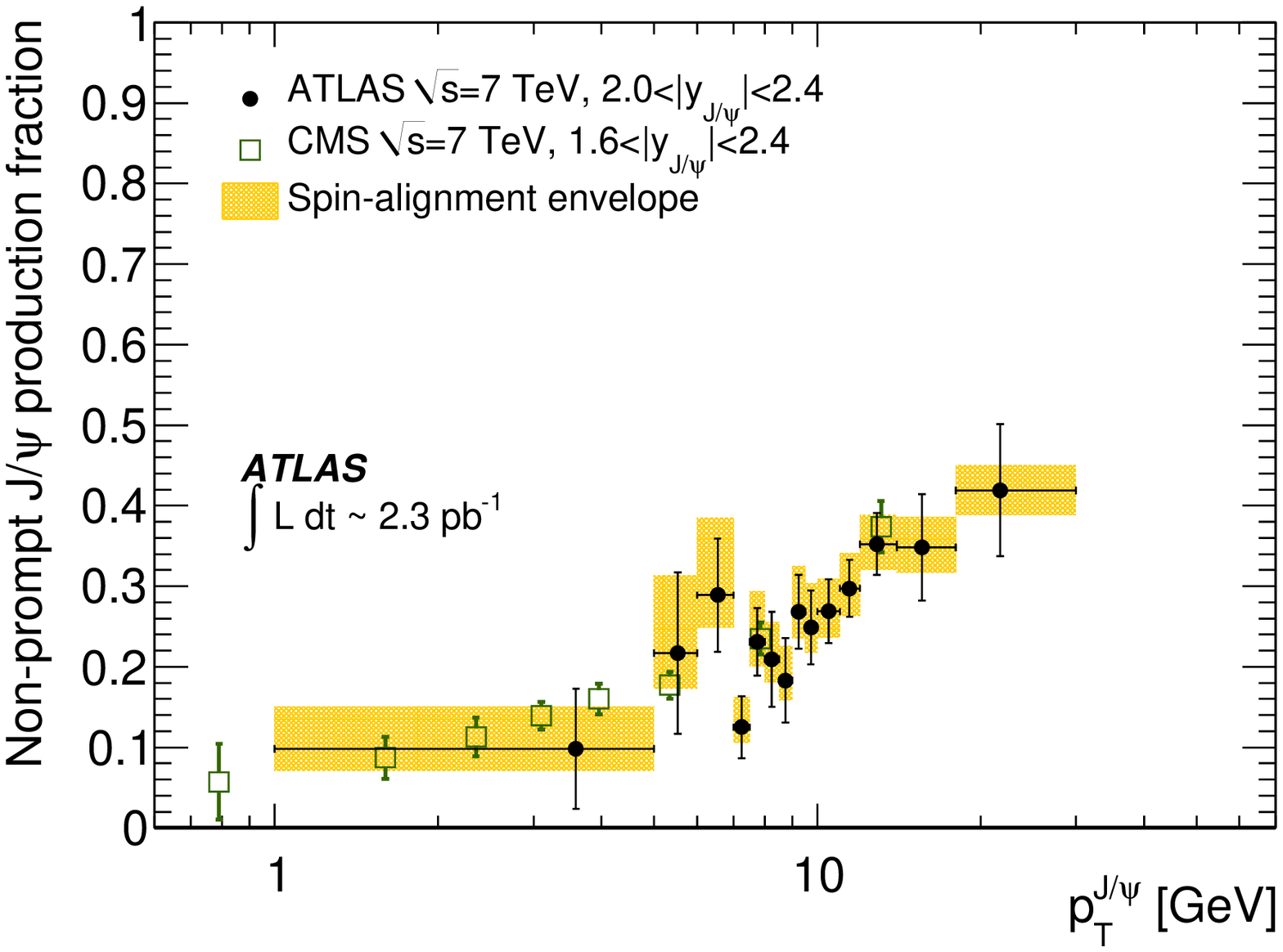}
    \caption{$J/\psi$ non-prompt to inclusive fractions as
      a function of $J/\psi$ transverse momentum.
      Overlaid is a band representing the variation of the result
      under various spin-alignment scenarios (see text) representing a
      theoretical uncertainty on the prompt and non-prompt $J/\psi$ components. The equivalent results from CMS
      \cite{CMS} and CDF \cite{CDF} are included. 
      \label{fig:fraction_result}
    }
  \end{center}
\end{sidewaysfigure}

Figure \ref{fig:fraction_result} and 
Tables~\ref{tab:AfractionMain0} to~\ref{tab:AfractionMain3} 
show the results of the differential non-prompt fraction
measurement
as a function of average $p_T^{J/\psi}$, in each of the four rapidity bins.
The uncertainty envelopes due to the unknown spin-alignment are
overlaid as solid bands. 

The measurements are compared with those of CMS~\cite{CMS} and CDF~\cite{CDF}
and build upon those results with finer rapidity binning, a much
extended rapidity coverage relative
to CDF and significantly increased $p_T$ reach relative to both experiments. Strong $p_T$
dependence of the fraction is observed: $\sim90\%$ of $J/\psi$ are
produced promptly at low $p_T$, but the fraction
of non-prompt $J/\psi$ rapidly increases at mid-$p_T$ from $\sim15\%$
at $7$~GeV to $\sim70\%$ at the highest accessible $p_T$ values.
No significant rapidity dependence is seen.
The ATLAS results exhibit good agreement with CMS results where they overlap, and also with the CDF measurements,
indicating that there is no strong dependence of the fraction on
collision energies.

\section{The Prompt and Non-Prompt Differential Production Cross-Sections}

The prompt and non-prompt $J/\psi$ production cross-sections can be derived from
the inclusive production cross-section and the non-prompt fraction. Where
necessary, $p_T$ bins in the inclusive
cross-section are merged to align bins in the prompt/non-prompt cross-section result
with those in the non-prompt fraction measurement.
The relative systematic uncertainties in each of the fraction and
inclusive cross-section measurement bins (merged where appropriate)
are taken to be uncorrelated, while the statistical uncertainties are combined taking correlations
into account. The spin alignment uncertainties are quoted independently of the experimental uncertainties.

\subsection{Non-prompt differential production cross-sections}

We assume the spin-alignment of a $J/\psi$ meson from a $B\to J/\psi X$ decay has
no net polar or azimuthal anisotropy for the central result, as the possible variation of spin-alignment 
in $B\to J/\psi X$ decays is expected to be much smaller than for prompt $J/\psi$ due to the averaging effect
caused by the admixture of various exclusive $B\to J/\psi X$ decays.
We assign a spin-alignment uncertainty on the non-prompt cross-section for the difference in the final result when using either an isotropic spin-alignment
assumption for non-prompt decays or maps reweighted to the CDF result\,\cite{CDF_bjpsi_pol} for $B\to J/\psi$ spin-alignment.

The total integrated cross-section for non-prompt $J/\psi$, multiplied by the branching
fraction into muons and under the ``FLAT" production scenario, has
been measured
for $J/\psi$ mesons produced within $|y|<2.4$ and $p_T>7$~GeV 
to be:
\begin{align*}
Br(J/\psi\to\mu^+\mu^-) &\sigma(pp\to B+X\to J/\psi X; |y_{J/\psi}|<2.4, p^{J/\psi}_T>7~\textrm{GeV}) \\
&\qquad = 23.0 \pm 0.6 \textrm{ (stat.)} \pm 2.8 \textrm{(syst.)} \pm 0.2 \textrm{ (spin)} \pm 0.8 \textrm{ (lumi.) nb}
\end{align*}
and for $J/\psi$ mesons produced with $1.5<|y|<2$ and $p_T>1$~GeV to be:
\begin{align*}
Br(J/\psi\to\mu^+\mu^-) &\sigma(pp\to B+X\to J/\psi X; 1.5<|y_{J/\psi}|<2, p^{J/\psi}_T>1~\textrm{GeV}) \\
&\qquad = 61 \pm 24 \textrm{ (stat.)} \pm 19 \textrm{ (syst.)} \pm 1 \textrm{ (spin)} \pm 2 \textrm{ (lumi.) nb.}
\end{align*}

\subsubsection{Comparisons with theoretical predictions}

\begin{sidewaysfigure}[htbp]
  \begin{center}
      \includegraphics[width=0.49\textwidth]{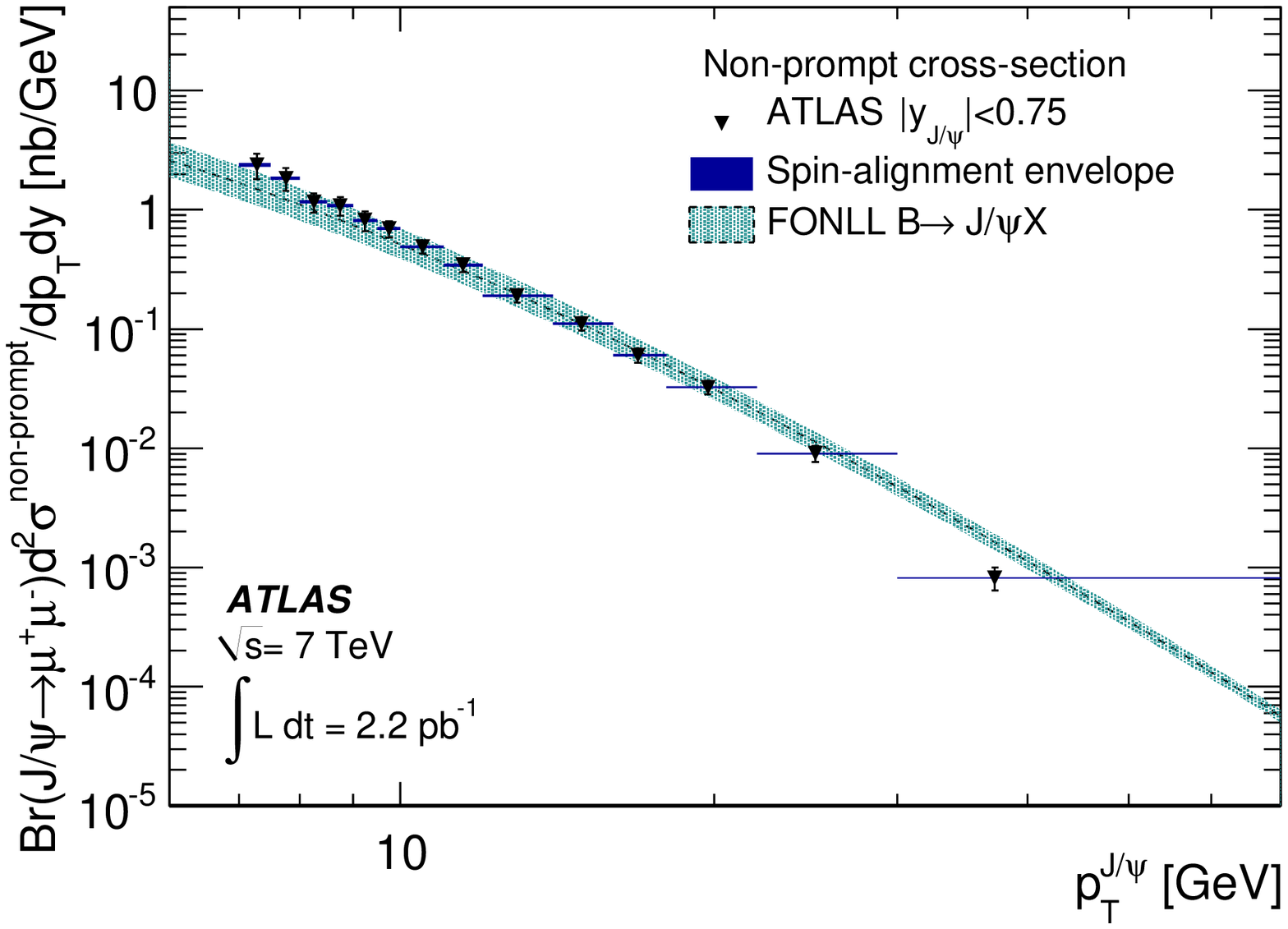}
      \includegraphics[width=0.49\textwidth]{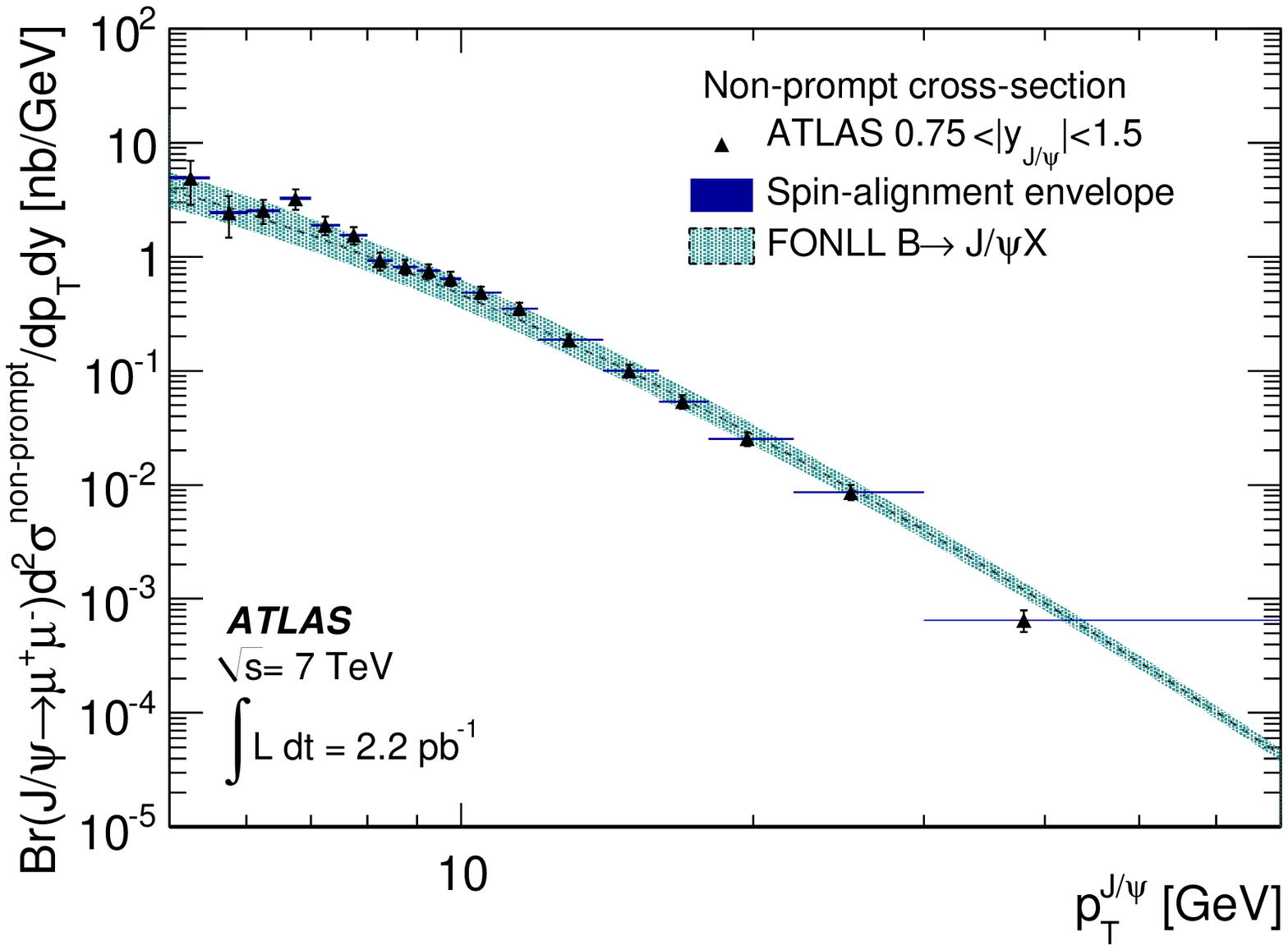}
      \includegraphics[width=0.49\textwidth]{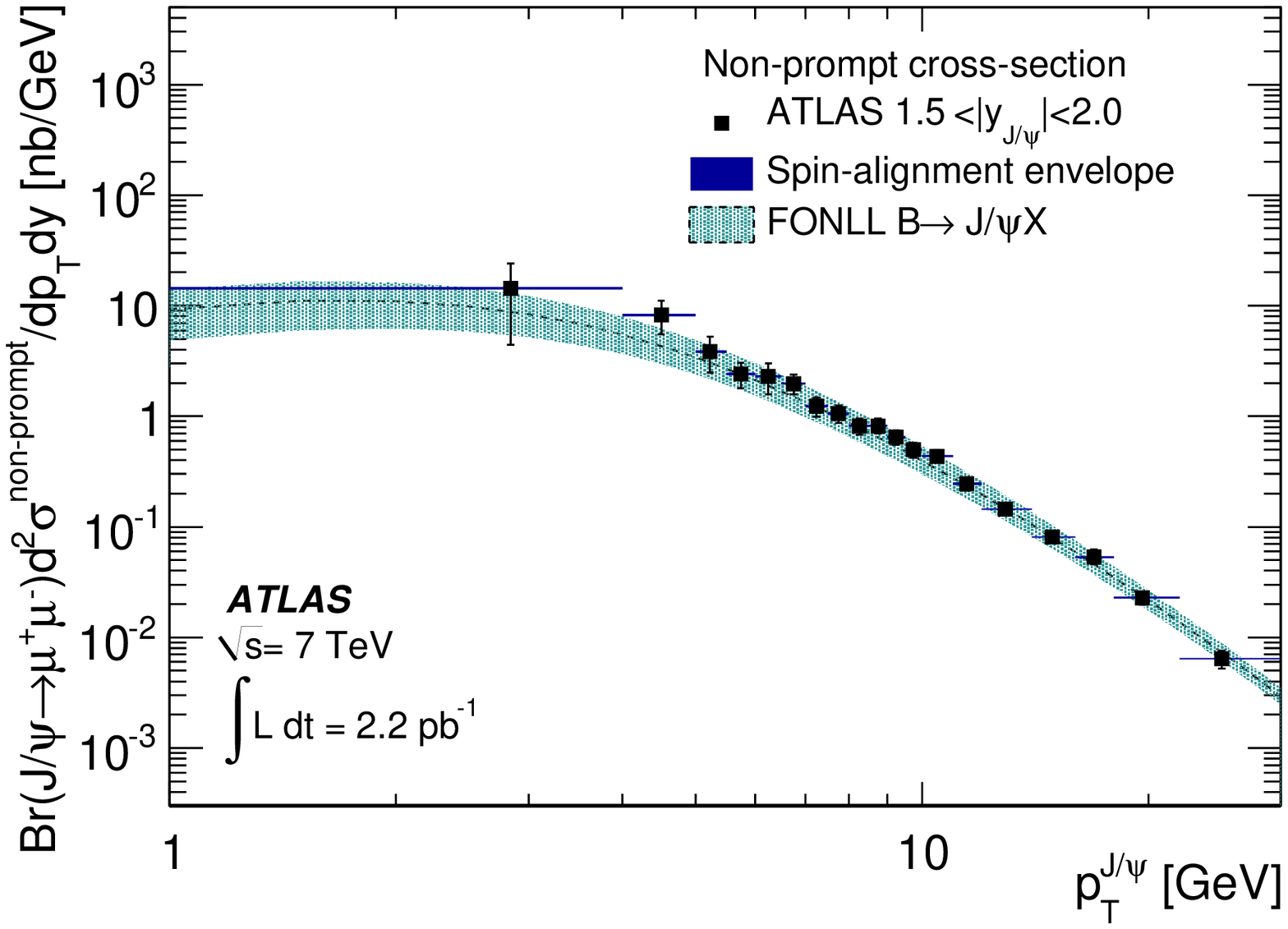}
      \includegraphics[width=0.49\textwidth]{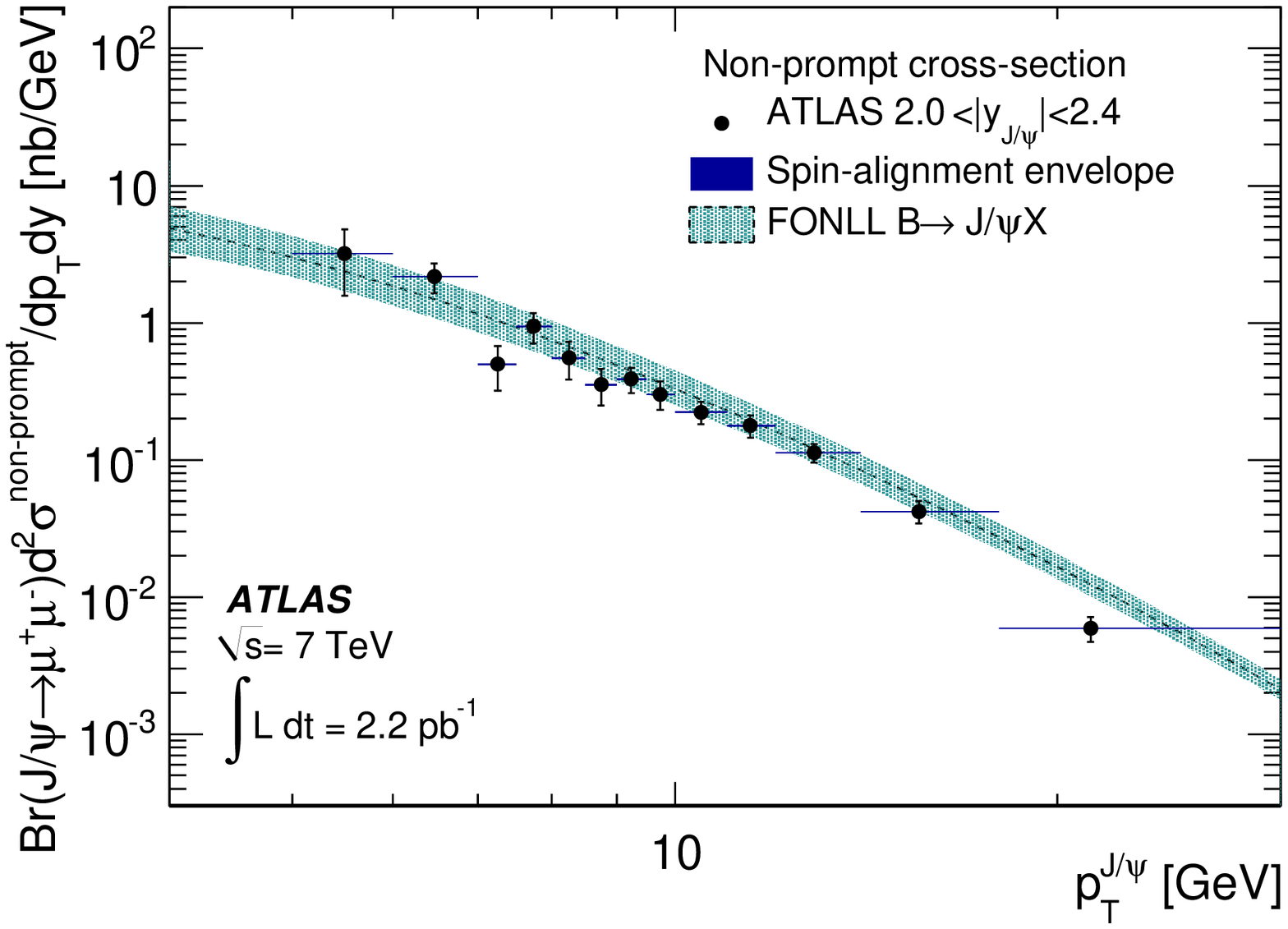}
    \caption{Non-prompt $J/\psi$ production cross-section as
      a function of $J/\psi$ transverse momentum, compared to predictions
from FONLL theory. Overlaid is a band representing the variation of the result
      under spin-alignment variation on the non-prompt $J/\psi$ component as described in the text. The central value
assumes an isotropic polarisation for both prompt and non-prompt components.
 The luminosity uncertainty (3.4\%) is not shown.
      \label{fig:nonprompt_xsec}
    }
  \end{center}
\end{sidewaysfigure}

ATLAS non-prompt $J/\psi$ production cross-section measurements are
compared to Fixed Order Next-to-Leading Logarithm (FONLL) calculations~\cite{Cacciari}
in Tables~\ref{tab:Anonpromptxsec_1} and ~\ref{tab:Anonpromptxsec_4} and in Figure~\ref{fig:nonprompt_xsec}. 
FONLL~v1.3.2 is used for these predictions,
using the CTEQ6.6\,\cite{CTEQ} parton density function set. 
FONLL predictions use a $B\to J/\psi X$ branching fraction of $Br(B\to J/\psi) = 0.0116$.
Uncertainty bands associated with the predictions come from
the input b-quark mass, varied within $4.75\pm 0.25$~GeV, renormalisation ($\mu_R$) and factorisation ($\mu_F$) 
scales (independently) varied within $0.5<\mu_{R,F}/m<2$ (with the additional constraint that $0.5<\mu_R/\mu_F<2$)
and parton density function uncertainties.
Good agreement is seen between the experimental data and the
theoretical prediction across the full range of rapidity and
transverse momentum considered.

\subsection{Prompt differential production cross-sections}
The prompt production cross-section is of direct interest for the study
of quarkonium production in QCD.
The spin-alignment state and $p_T$ dependence of the spin-alignment of
promptly produced $J/\psi$ particles are thought to be non-trivial, so the
spin-alignment uncertainty envelope on the inclusive
cross-section measurement is propagated into the prompt
cross-section measurement.
The prompt production cross-sections are presented in
Tables~\ref{tab:Apromptxsec_1} to~\ref{tab:Apromptxsec_4}. 

The total cross-section for prompt $J/\psi$ (times branching fraction
into muons) under the flat production scenario has been
measured
for $J/\psi$ produced within $|y|<2.4$ and $p_T>7$~GeV to be:
\begin{align*}
Br(J/\psi\to\mu^+\mu^-) &\sigma(pp\to\textrm{prompt } J/\psi X; |y|<2.4, p_T>7~\textrm{GeV}) \\
&\qquad = 59 \pm 1 \textrm{ (stat.)} \pm 8 \textrm{(syst.)} \pm {}^{9}_{6} \textrm{ (spin)} \pm 2 \textrm{ (lumi.) nb}
\end{align*}
and for $J/\psi$ within $1.5<|y|<2$ and $p_T>1$~GeV to be:
\begin{align*}
Br(J/\psi\to\mu^+\mu^-) &\sigma(pp\to\textrm{prompt } J/\psi X; 1.5<|y|<2, p_T>1~\textrm{GeV}) \\
&\qquad = 450 \pm 70 \textrm{ (stat.)} \pm ^{90}_{110} \textrm{(syst.)} \pm {}^{740}_{110} \textrm{ (spin)} \pm 20 \textrm{ (lumi.) nb.}
\end{align*}

\begin{sidewaysfigure}[htbp]
  \begin{center}
      \includegraphics[width=0.49\textwidth]{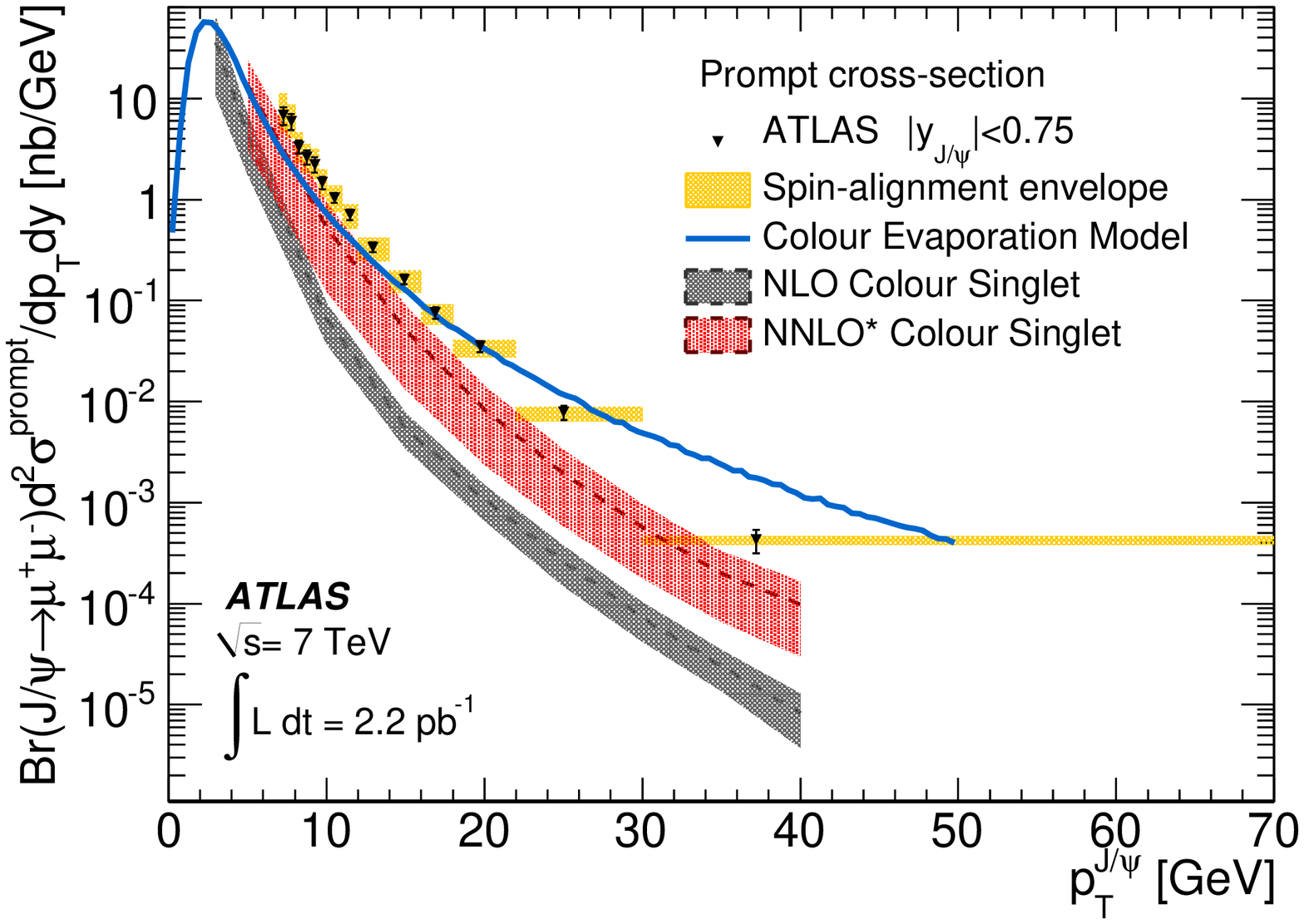}
      \includegraphics[width=0.49\textwidth]{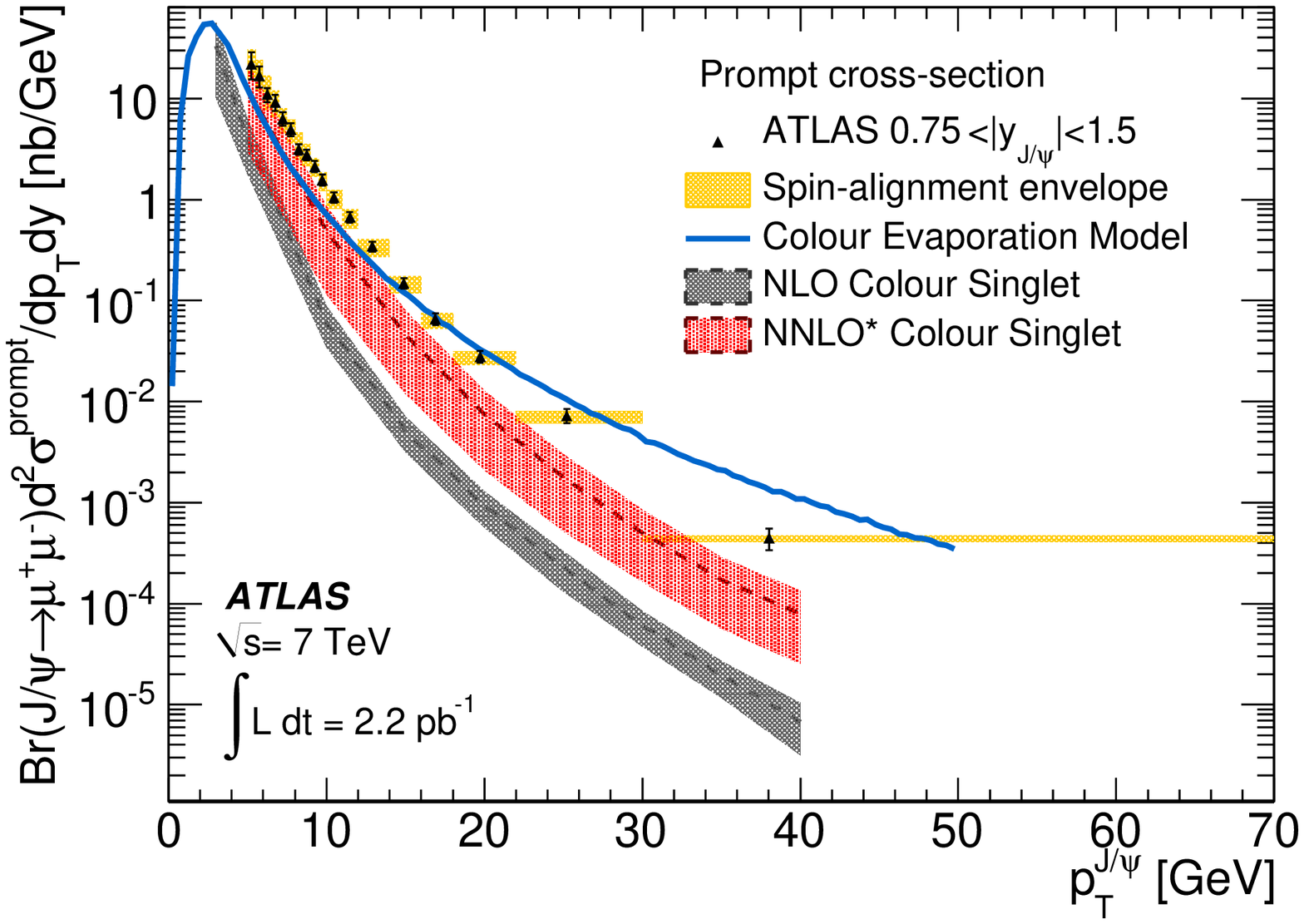}
      \includegraphics[width=0.49\textwidth]{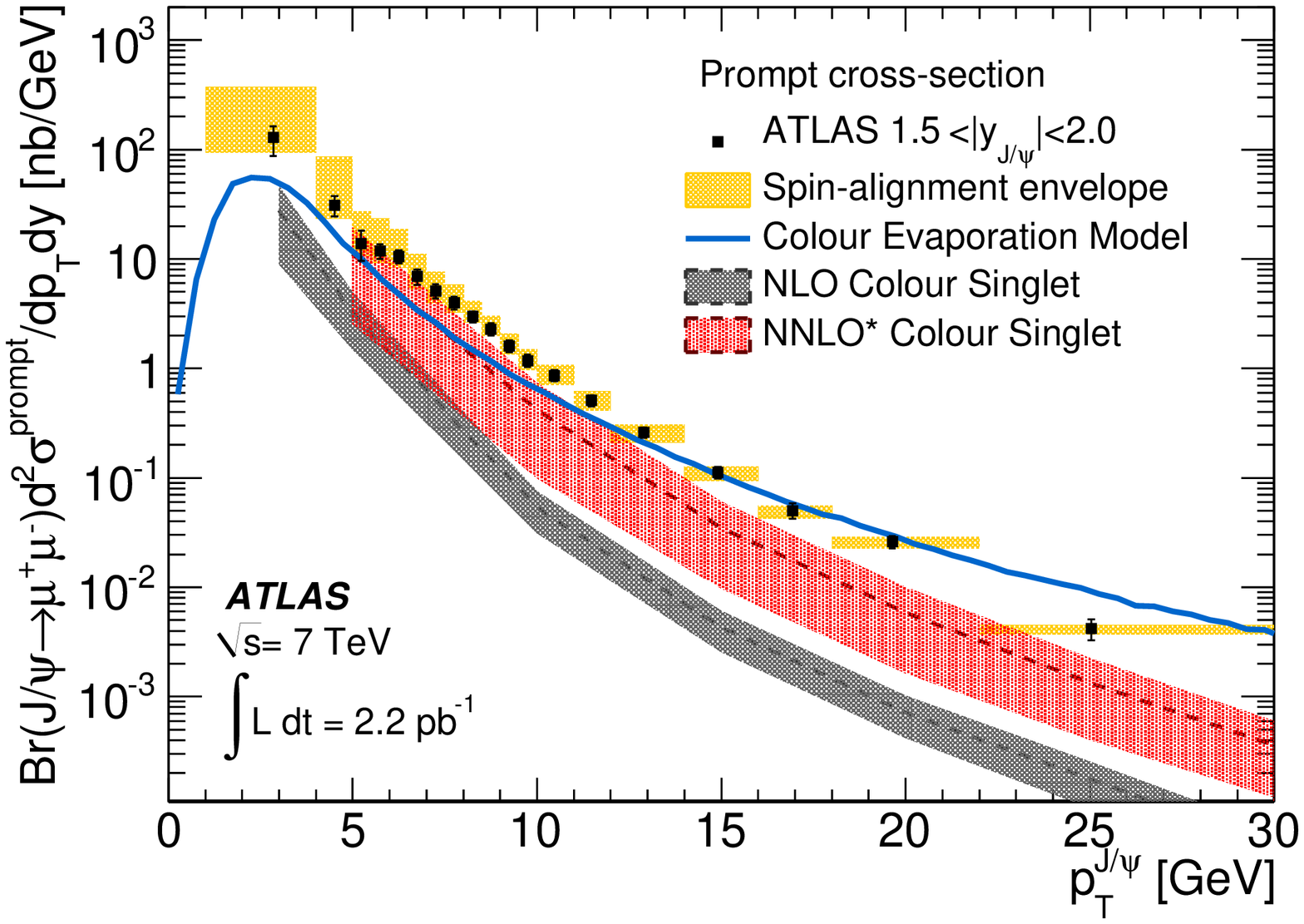}
      \includegraphics[width=0.49\textwidth]{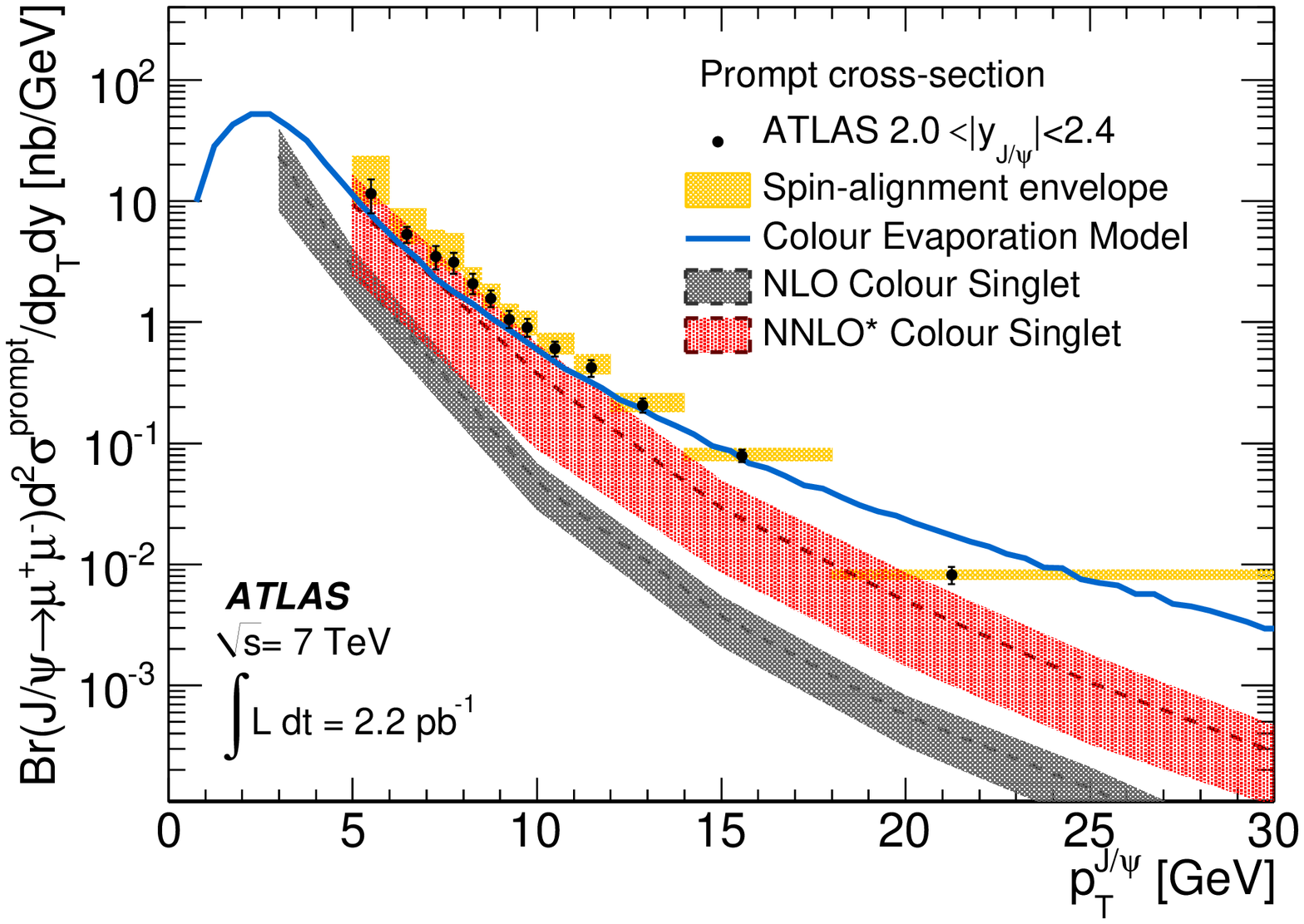}
    \caption{Prompt $J/\psi$ production cross-section as
      a function of $J/\psi$ transverse momentum in the four rapidity bins.
      Overlaid is a band representing the variation of the result
      under various spin-alignment scenarios (see text) representing a
      theoretical uncertainty on the prompt component. Predictions from NLO and NNLO$^\star$ calculations,
      and the Colour Evaporation Model are overlaid. The luminosity uncertainty (3.4\%) is not shown.
      \label{fig:prompt_xsec}
    }
  \end{center}
\end{sidewaysfigure}

\subsubsection{Comparisons with theoretical predictions}
\label{sec:models}
In Figure~\ref{fig:prompt_xsec} the prompt production data are 
compared to the predictions of the Colour Evaporation Model (CEM)~\cite{CEM_RHIC,CEM} for prompt $J/\psi$ production (with no uncertainties defined) and a calculation of the direct $J/\psi$ production cross-section in the Colour Singlet Model (CSM)~\cite{Lansberg,Lansberg2} at next-to-leading order (NLO) and a partial next-to-next-leading order calculation (NNLO$^\star$).

The Colour Evaporation Model predictions are produced
using the CTEQ6M parton density functions, a charm quark mass of 1.2~GeV and the
renormalisation and factorisation
scales set to $\mu_0= 2\sqrt{p^2_T + m^2_{Q} + k^2_T}$ (where $p_T$ is the
transverse momentum of the $J/\psi$ and $m_Q$ is the quark mass and $k_T$ is a phenomenological fit parameter set to $1.5$~GeV$^2$).
The CEM predictions include contributions from $\chi_c$ and $\psi$(2S)
feed-down and can be directly compared with the prompt $J/\psi$ data.
The normalisation of the CEM prediction is generally lower than in data and strongly diverges in shape from measured data, showing 
significant disagreement in the extended $p_T$ range probed by the measurement described in this paper.

The Colour Singlet NLO and NNLO$^\star$ predictions\footnote{The NNLO$^\star$ calculation is not a {\em full} 
next-to-next-to-leading order prediction, as it does not include all loop corrections to $pp\to Q+jjj$ (where $j$ is a light parton) up
to order $\alpha^5_s$. This limits the applicability of the calculation to values above a
particular ${J/\psi}$ $p_T$ threshold (due to soft and collinear divergences).}
for direct $J/\psi$ production use a charm quark mass of 1.5~GeV, the CTEQ6M parton density function set, and 
factorisation and renormalisation scales set to $\mu_0 = \sqrt{p^2_T + m^2_{Q}}$ (varied up and down by a factor of two to determine scale uncertainties). 
As the calculation is for direct production, corrections must be
applied for $\chi_c$ and $\psi$(2S) feed-down to bring the calculations
in direct comparison with data. To correct for feed-down, a flat $10\%$ correction is applied to account for the contribution of $\psi\textrm{(2S)}\to J/\psi\pi\pi$ and a 40\% correction is added to account for radiative $\chi_c$  decays. This yields a total correction of 50\%. The correction
factor is not well-determined from theory or experiment so is assigned 
a 100\% uncertainty.
This uncertainty is not included in the CSM theoretical uncertainty.

The NLO and NNLO$^\star$ predictions
are overlaid with the ATLAS measurements in Figure~\ref{fig:prompt_xsec} for each
rapidity region.
The dashed lines represent the central NLO and NNLO$^\star$ predictions
while the shaded areas show the range of the prediction due to
factorisation and renormalisation scale variation (although the upper
band of this uncertainty may not encapsulate the full range of
infrared uncertainties~\cite{NNLO_upsilon}).

The Colour Singlet Model predictions at NNLO$^\star$ show significant improvement in describing the $p_T$ dependence and normalisation 
of prompt $J/\psi$ production over NLO, and vast improvement over earlier LO predictions that are compared to Tevatron data, although it is clear that these predictions still fall short of fully describing the production mechanisms of prompt $J/\psi$, particularly
at the highest transverse momenta explored in this analysis.
The overall scale of the central prediction is somewhat low, but these discrepancies are similar in nature to those seen between NNLO$^\star$
calculations and $\psi$(2S) production as measured by CDF~\cite{Lansberg2, psi2s_cdf} at lower $p_T$ and centre-of-mass energy
and may be attributed to higher order corrections beyond NNLO$^\star$ that are still expected to be relatively significant for hidden charm
production.

\section{Summary}
\label{section:conclusion}

Results are reported on the measurement of the inclusive cross-section of $J/\psi\to\mu^+\mu^-$ production 
in proton-proton collisions at a collision energy of 7 TeV using the ATLAS detector with up to 2.3~pb$^{-1}$ of integrated luminosity. 
The inclusive cross-section is measured in bins of
rapidity $y$ and transverse momentum $p_T$ of $J/\psi$, covering the range
$|y| < 2.4$ and $1<p_T<70$\;GeV. The fraction of non-prompt $J/\psi$ mesons is also measured
as a function of $J/\psi$ transverse momentum and rapidity and using the above two measurements,
double-differential cross-sections are extracted separately for promptly-produced
$J/\psi$ mesons and those coming from $B$-hadron decays.

It is found that the measurements made by ATLAS and CMS are in good agreement with each other
in the overlapping range of moderate $p_T$ values and complement each other at high (ATLAS) and low (CMS)
  values of transverse momenta. 
The non-prompt production fraction results are also compared to those from CDF at lower energy and reasonable agreement is found, 
suggesting there is no strong dependence of the fraction on the collision energy.

The results are also compared to various theoretical calculations of
prompt as well as non-prompt $J/\psi$ production. In general, the theoretical
curves describe the non-prompt data well, but significant deviations are
observed in the prompt production spectra both in shape and normalisation, particularly at high transverse momenta. These measurements can thus provide  
input towards an improved understanding and theoretical description of $J/\psi$ hadronic production.

\section{Acknowledgements}
The authors would like to thank Jean-Philippe Lansberg and Ramona Vogt for
providing theoretical predictions for prompt production and for useful
discussions. They would also like to thank Matteo Cacciari for providing
predictions for the $B\to J/\psi X$ production cross-sections in the
FONLL scheme.

We thank CERN for the very successful operation of the LHC, as well as the
support staff from our institutions without whom ATLAS could not be
operated efficiently.

We acknowledge the support of ANPCyT, Argentina; YerPhI, Armenia; ARC,
Australia; BMWF, Austria; ANAS, Azerbaijan; SSTC, Belarus; CNPq and FAPESP,
Brazil; NSERC, NRC and CFI, Canada; CERN; CONICYT, Chile; CAS, MOST and
NSFC, China; COLCIENCIAS, Colombia; MSMT CR, MPO CR and VSC CR, Czech
Republic; DNRF, DNSRC and Lundbeck Foundation, Denmark; ARTEMIS, European
Union; IN2P3-CNRS, CEA-DSM/IRFU, France; GNAS, Georgia; BMBF, DFG, HGF, MPG
and AvH Foundation, Germany; GSRT, Greece; ISF, MINERVA, GIF, DIP and
Benoziyo Center, Israel; INFN, Italy; MEXT and JSPS, Japan; CNRST, Morocco;
FOM and NWO, Netherlands; RCN, Norway; MNiSW, Poland; GRICES and FCT,
Portugal; MERYS (MECTS), Romania; MES of Russia and ROSATOM, Russian
Federation; JINR; MSTD, Serbia; MSSR, Slovakia; ARRS and MVZT, Slovenia;
DST/NRF, South Africa; MICINN, Spain; SRC and Wallenberg Foundation,
Sweden; SER, SNSF and Cantons of Bern and Geneva, Switzerland; NSC, Taiwan;
TAEK, Turkey; STFC, the Royal Society and Leverhulme Trust, United Kingdom;
DOE and NSF, United States of America.

The crucial computing support from all WLCG partners is acknowledged
gratefully, in particular from CERN and the ATLAS Tier-1 facilities at
TRIUMF (Canada), NDGF (Denmark, Norway, Sweden), CC-IN2P3 (France),
KIT/GridKA (Germany), INFN-CNAF (Italy), NL-T1 (Netherlands), PIC (Spain),
ASGC (Taiwan), RAL (UK) and BNL (USA) and in the Tier-2 facilities
worldwide.

\bibliographystyle{atlasnote}

\providecommand{\href}[2]{#2}\begingroup\raggedright\endgroup
\renewcommand\floatpagefraction{.9}
\renewcommand\topfraction{.9}
\renewcommand\bottomfraction{.9}
\renewcommand\textfraction{.1}   
\setcounter{totalnumber}{50}
\setcounter{topnumber}{50}
\setcounter{bottomnumber}{50}
\clearpage

\begin{table}[htb]
\begin{center}
\caption{Inclusive $J/\psi$ production cross-sections as a function of $J/\psi$ $p_T$ in four rapidity ($|y|$) bins. 
The first uncertainty is statistical, the second is systematic and the third encapsulates any possible variation due to spin-alignment from the unpolarised ($\lambda_\theta = \lambda_\phi = \lambda_{\theta\phi} = 0$) central value.}
\label{tab:Ainclxsec1}
\footnotesize
\begin{tabular}{rr | rlll||r | rlll }
\hline
\hline
 &  & \multicolumn{9}{c}{$\frac{d^{2}\sigma}{dp_{T}dy}\cdot$Br$(J/\psi\to\mu^+\mu^-)$ [pb/GeV]}\\
$p_T$ & $\langle p_T \rangle$ & \multicolumn{4}{c||}{$2<|y|<2.4$}& $\langle p_T \rangle$ & \multicolumn{4}{c}{$1.5<|y|<2$} \\
(GeV) & (GeV) & Value & $\pm$ (stat.) & $\pm$ (syst.) & $\pm$ (spin)& (GeV) & Value & $\pm$ (stat.) & $\pm$ (syst.) & $\pm$ (spin) \\
\hline
1.0-4.0 &&&&&                                                                                 									& 2.8 & $ 143000 $ & $\pm 23000 $ & $\pm{}^{25000}_{39000}$ & $\pm{}^{274000}_{39000}$\\
4.0-5.0 &&&&&                                                                                 									& 4.5 & $ 39400 $ & $\pm 5500 $ & $\pm{}^{5700}_{5700}$ & $\pm{}^{69300}_{9700}$\\
5.0-5.5 & 5.3 & 	$ 15900 $ & $\pm 4300 $ & $\pm{}^{2800}_{2600}$ & $\pm{}^{28800}_{4300}$     	& 5.2 & $ 17600 $ & $\pm 3300 $ & $\pm{}^{3000}_{2600}$ & $\pm{}^{17300}_{4100}$\\
5.5-6.0 & 5.8 & 	$ 13500 $ & $\pm 3600 $ & $\pm{}^{1900}_{2200}$ & $\pm{}^{11400}_{2700}$     	& 5.7 & $ 14300 $ & $\pm 1200 $ & $\pm{}^{1700}_{1700}$ & $\pm{}^{14000}_{3100}$ \\
6.0-6.5 & 6.3 & 	$ 8800 $ & $\pm 1100 $ & $\pm{}^{1300}_{1200}$ & $\pm{}^{7900}_{2200}$        	& 6.3 & $ 12760 $ & $\pm 920 $ & $\pm{}^{1840}_{1690}$ & $\pm{}^{9970}_{2620}$\\
6.5-7.0 & 6.8 & 	$ 6290 $ & $\pm 700 $ & $\pm{}^{830}_{980}$ & $\pm{}^{5140}_{1360}$           	& 6.8 & $ 8910 $ & $\pm 610 $ & $\pm{}^{1270}_{1270}$ & $\pm{}^{5420}_{1990}$ \\
7.0-7.5 & 7.3 & 	$ 3990 $ & $\pm 500 $ & $\pm{}^{560}_{550}$ & $\pm{}^{2630}_{690}$            	& 7.2 & $ 6350 $ & $\pm 430 $ & $\pm{}^{860}_{860}$ & $\pm{}^{3130}_{1430}$\\
7.5-8.0 & 7.7 & 	$ 4070 $ & $\pm 450 $ & $\pm{}^{570}_{580}$ & $\pm{}^{2920}_{650}$            	& 7.7 & $ 5040 $ & $\pm 350 $ & $\pm{}^{590}_{520}$ & $\pm{}^{2260}_{900}$\\
8.0-8.5 & 8.3 & 	$ 2650 $ & $\pm 290 $ & $\pm{}^{460}_{390}$ & $\pm{}^{910}_{570}$             	& 8.3 & $ 3790 $ & $\pm 210 $ & $\pm{}^{440}_{430}$ & $\pm{}^{1490}_{450}$\\
8.5-9.0 & 8.7 & 	$ 1930 $ & $\pm 160 $ & $\pm{}^{260}_{260}$ & $\pm{}^{620}_{350}$             	& 8.7 & $ 3110 $ & $\pm 160 $ & $\pm{}^{420}_{360}$ & $\pm{}^{980}_{450}$\\
9.0-9.5 & 9.2 & 	$ 1450 $ & $\pm 130 $ & $\pm{}^{210}_{180}$ & $\pm{}^{480}_{210}$             	& 9.2 & $ 2260 $ & $\pm 110 $ & $\pm{}^{260}_{250}$ & $\pm{}^{640}_{370}$ \\
9.5-10.0 & 9.7 & 	$ 1208 $ & $\pm 94 $ & $\pm{}^{155}_{138}$ & $\pm{}^{440}_{166}$             		& 9.7 & $ 1674 $ & $\pm 85 $ & $\pm{}^{198}_{183}$ & $\pm{}^{450}_{296}$ \\
10.0-11.0 & 10.5 & $ 829 $ & $\pm 51 $ & $\pm{}^{96}_{92}$ & $\pm{}^{286}_{87}$               			& 10.5 & $ 1297 $ & $\pm 46 $ & $\pm{}^{146}_{139}$ & $\pm{}^{316}_{241}$ \\
11.0-12.0 & 11.5 & $ 598 $ & $\pm 43 $ & $\pm{}^{69}_{73}$ & $\pm{}^{174}_{71}$               			& 11.5 & $ 754 $ & $\pm 31 $ & $\pm{}^{90}_{83}$ & $\pm{}^{168}_{147}$\\
12.0-14.0 & 12.9 & $ 320 $ & $\pm 19 $ & $\pm{}^{38}_{36}$ & $\pm{}^{79}_{40}$                			& 12.9 & $ 404 $ & $\pm 15 $ & $\pm{}^{45}_{43}$ & $\pm{}^{74}_{75}$\\
14.0-16.0 & 14.9 & $ 164 $ & $\pm 12 $ & $\pm{}^{26}_{16}$ & $\pm{}^{33}_{21}$                			& 14.9 & $ 193 $ & $\pm 10 $ & $\pm{}^{21}_{19}$ & $\pm{}^{28}_{32}$ \\
16.0-18.0 & 16.9 & $ 77.8 $ & $\pm 8.2 $ & $\pm{}^{9.4}_{8.0}$ & $\pm{}^{14.1}_{9.9}$                   	& 16.9 & $ 103.0 $ & $\pm 6.9 $ & $\pm{}^{13.0}_{9.4}$ & $\pm{}^{12.0}_{15.5}$\\
18.0-22.0 & 19.7 & $ 29.9 $ & $\pm 3.3 $ & $\pm{}^{3.1}_{3.4}$ & $\pm{}^{3.7}_{3.8}$                    		& 19.6 & $ 48.9 $ & $\pm 3.2 $ & $\pm{}^{4.1}_{4.2}$ & $\pm{}^{4.9}_{6.5}$\\
22.0-30.0 & 24.9 & $ 6.2 $ & $\pm 1.1 $ & $\pm{}^{0.6}_{0.6}$ & $\pm{}^{0.6}_{0.7}$                      		& 25.0 & $ 10.6 $ & $\pm 1.1 $ & $\pm{}^{1.0}_{0.9}$ & $\pm{}^{0.8}_{1.2}$ \\
30.0-40.0 & 33.6 & $ 1.12 $ & $\pm 0.43 $ & $\pm{}^{0.10}_{0.28}$ & $\pm{}^{0.06}_{0.10}$       		& 34.1 & $ 2.22 $ & $\pm 0.40 $ & $\pm{}^{0.19}_{0.21}$ & $\pm{}^{0.13}_{0.22}$ \\
\hline
\hline
 &  & \multicolumn{9}{c}{$\frac{d^{2}\sigma}{dp_{T}dy}\cdot$Br$(J/\psi\to\mu^+\mu^-)$ [pb/GeV]}\\
$p_T$ & $\langle p_T \rangle$ & \multicolumn{4}{c||}{$0.75<|y|<1.5$}& $\langle p_T \rangle$ & \multicolumn{4}{c}{$|y|<0.75$} \\
(GeV) & (GeV) & Value & $\pm$ (stat.) & $\pm$ (syst.) & $\pm$ (spin)& (GeV) & Value & $\pm$ (stat.) & $\pm$ (syst.) & $\pm$ (spin) \\
\hline
5.0-5.5 & 5.3 & 	$ 26800 $ & $\pm 5600 $ & $\pm{}^{4100}_{3800}$ & $\pm{}^{10600}_{7900}$&&\\
5.5-6.0 & 5.8 & 	$ 19200 $ & $\pm 2800 $ & $\pm{}^{2700}_{2500}$ & $\pm{}^{8600}_{5700}$&&\\
6.0-6.5 & 6.2 & 	$ 13500 $ & $\pm 1100 $ & $\pm{}^{1700}_{1700}$ & $\pm{}^{7100}_{4000}$&&\\
6.5-7.0 & 6.7 & 	$ 12400 $ & $\pm 1100 $ & $\pm{}^{1700}_{1700}$ & $\pm{}^{3900}_{3600}$&&\\
7.0-7.5 & 7.2 & 	$ 8190 $ & $\pm 610 $ & $\pm{}^{1090}_{1040}$ & $\pm{}^{2220}_{2300}$       		& 7.3 & $ 9220 $ & $\pm 980 $ & $\pm{}^{1140}_{1150}$ & $\pm{}^{5770}_{2960}$\\
7.5-8.0 & 7.7 & 	$ 6500 $ & $\pm 400 $ & $\pm{}^{860}_{810}$ & $\pm{}^{1620}_{1770}$         		& 7.8 & $ 7780 $ & $\pm 720 $ & $\pm{}^{1000}_{990}$ & $\pm{}^{3540}_{2470}$ \\  
8.0-8.5 & 8.2 & 	$ 4080 $ & $\pm 280 $ & $\pm{}^{420}_{440}$ & $\pm{}^{1870}_{900}$          			& 8.3 & $ 4500 $ & $\pm 320 $ & $\pm{}^{510}_{530}$ & $\pm{}^{1730}_{1410}$ \\
8.5-9.0 & 8.7 & 	$ 3600 $ & $\pm 200 $ & $\pm{}^{390}_{390}$ & $\pm{}^{1040}_{800}$          			& 8.8 & $ 3720 $ & $\pm 270 $ & $\pm{}^{450}_{440}$ & $\pm{}^{1310}_{1150}$\\
9.0-9.5 & 9.3 & 	$ 2880 $ & $\pm 140 $ & $\pm{}^{330}_{320}$ & $\pm{}^{610}_{640}$           			& 9.2 & $ 3040 $ & $\pm 280 $ & $\pm{}^{360}_{360}$ & $\pm{}^{1240}_{840}$\\
9.5-10.0 & 9.7 & 	$ 2210 $ & $\pm 100 $ & $\pm{}^{250}_{240}$ & $\pm{}^{420}_{490}$          			& 9.8 & $ 2170 $ & $\pm 140 $ & $\pm{}^{230}_{230}$ & $\pm{}^{740}_{600}$ \\
10.0-11.0 & 10.5 & $ 1542 $ & $\pm 51 $ & $\pm{}^{176}_{174}$ & $\pm{}^{283}_{348}$         			& 10.5 & $ 1528 $ & $\pm 59 $ & $\pm{}^{160}_{160}$ & $\pm{}^{471}_{430}$\\
11.0-12.0 & 11.5 & $ 1022 $ & $\pm 35 $ & $\pm{}^{121}_{120}$ & $\pm{}^{187}_{234}$         			& 11.5 & $ 1051 $ & $\pm 39 $ & $\pm{}^{116}_{116}$ & $\pm{}^{288}_{293}$\\
12.0-14.0 & 12.9 & $ 531 $ & $\pm 16 $ & $\pm{}^{60}_{58}$ & $\pm{}^{94}_{118}$             				& 12.9 & $ 528 $ & $\pm 17 $ & $\pm{}^{56}_{56}$ & $\pm{}^{127}_{141}$ \\
14.0-16.0 & 14.9 & $ 249 $ & $\pm 10 $ & $\pm{}^{26}_{26}$ & $\pm{}^{40}_{52}$              				& 14.9 & $ 274 $ & $\pm 12 $ & $\pm{}^{27}_{27}$ & $\pm{}^{60}_{70}$\\
16.0-18.0 & 16.9 & $ 119.2 $ & $\pm 6.7 $ & $\pm{}^{11.9}_{11.7}$ & $\pm{}^{17.0}_{23.1}$   			& 16.9 & $ 136.2 $ & $\pm 7.5 $ & $\pm{}^{13.1}_{13.1}$ & $\pm{}^{26.5}_{32.1}$\\
18.0-22.0 & 19.7 & $ 53.3 $ & $\pm 3.0 $ & $\pm{}^{5.2}_{5.0}$ & $\pm{}^{6.7}_{9.6}$        				& 19.7 & $ 67.7 $ & $\pm 3.6 $ & $\pm{}^{6.4}_{6.3}$ & $\pm{}^{10.9}_{14.5}$\\
22.0-30.0 & 25.2 & $ 15.9 $ & $\pm 1.1 $ & $\pm{}^{1.8}_{1.6}$ & $\pm{}^{1.7}_{2.4}$        				& 25.0 & $ 16.9 $ & $\pm 1.4 $ & $\pm{}^{1.7}_{1.7}$ & $\pm{}^{2.2}_{3.0}$\\
30.0-40.0 & 33.9 & $ 3.16 $ & $\pm 0.43 $ & $\pm{}^{0.34}_{0.34}$ & $\pm{}^{0.27}_{0.39}$  	 		& 33.6 & $ 3.60 $ & $\pm 0.48 $ & $\pm{}^{0.38}_{0.39}$ & $\pm{}^{0.43}_{0.52}$ \\
40.0-70.0 & 48.8 & $ 0.407 $ & $\pm 0.084 $ & $\pm{}^{0.041}_{0.043}$ & $\pm{}^{0.022}_{0.017}$  		& 46.6 & $ 0.462 $ & $\pm 0.093 $ & $\pm{}^{0.055}_{0.055}$ & $\pm{}^{0.046}_{0.049}$ \\
\hline
\hline
\end{tabular}
\end{center}
\end{table}

\begin{table}[tbfhp]
\begin{center}
\vspace{-.5cm}
\caption{Non-prompt to inclusive production cross-section fraction $f_B$ as a function of $J/\psi$ $p_T$ for $|y|_{J/\psi}<0.75$ under the assumption that prompt and non-prompt $J/\psi$ production is unpolarised ($\lambda_\theta = 0$). The spin-alignment envelope spans the range of possible prompt cross-sections under various polarisation hypotheses, plus the range of non-prompt cross-sections within $\lambda_\theta = \pm 0.1$. The first uncertainty is statistical, the second uncertainty is systematic, the third number is the uncertainty due to spin-alignment.}
\label{tab:AfractionMain0}
\small
\begin{tabular}{rr | rlll }
\hline
\hline
 &  & \multicolumn{4}{c}{Non-prompt to inclusive production fraction}  \\
$p_T$ & $\langle p_T \rangle$ & \multicolumn{4}{c}{$|y|<0.75$}  \\
(GeV) & (GeV) & $f_B$ & $\pm$ (stat.) & $\pm$ (syst.) & $\pm$ (spin) \\
\hline
6.0-7.0 & 6.6 & $ 0.175 $ & $\pm 0.057 $ & $\pm 0.032 $ & $\pm{}^{0.064}_{0.062}$ \\
7.0-7.5 & 7.3 & $ 0.259 $ & $\pm 0.038 $ & $\pm 0.002 $ & $\pm{}^{0.066}_{0.080}$ \\
7.5-8.0 & 7.8 & $ 0.236 $ & $\pm 0.030 $ & $\pm 0.007 $ & $\pm{}^{0.061}_{0.076}$ \\
8.0-8.5 & 8.3 & $ 0.258 $ & $\pm 0.032 $ & $\pm 0.017 $ & $\pm{}^{0.054}_{0.074}$ \\
8.5-9.0 & 8.8 & $ 0.291 $ & $\pm 0.030 $ & $\pm 0.005 $ & $\pm{}^{0.058}_{0.079}$ \\
9.0-9.5 & 9.2 & $ 0.268 $ & $\pm 0.025 $ & $\pm 0.008 $ & $\pm{}^{0.054}_{0.076}$ \\
9.5-10.0 & 9.8 & $ 0.320 $ & $\pm 0.026 $ & $\pm 0.006 $ & $\pm{}^{0.062}_{0.083}$ \\
10.0-11.0 & 10.5 & $ 0.321 $ & $\pm 0.018 $ & $\pm 0.007 $ & $\pm{}^{0.050}_{0.077}$ \\
11.0-12.0 & 11.5 & $ 0.327 $ & $\pm 0.019 $ & $\pm 0.003 $ & $\pm{}^{0.051}_{0.078}$ \\
12.0-14.0 & 12.9 & $ 0.359 $ & $\pm 0.017 $ & $\pm 0.003 $ & $\pm{}^{0.044}_{0.069}$ \\
14.0-16.0 & 14.9 & $ 0.405 $ & $\pm 0.024 $ & $\pm 0.008 $ & $\pm{}^{0.046}_{0.072}$ \\
16.0-18.0 & 16.9 & $ 0.443 $ & $\pm 0.030 $ & $\pm 0.005 $ & $\pm{}^{0.048}_{0.073}$ \\
18.0-22.0 & 19.7 & $ 0.479 $ & $\pm 0.030 $ & $\pm 0.004 $ & $\pm{}^{0.040}_{0.063}$ \\
22.0-30.0 & 25.0 & $ 0.536 $ & $\pm 0.039 $ & $\pm 0.008 $ & $\pm{}^{0.032}_{0.050}$ \\
30.0-70.0 & 37.7 & $ 0.656 $ & $\pm 0.059 $ & $\pm 0.008 $ & $\pm{}^{0.030}_{0.045}$ \\
\hline
\hline
\end{tabular}
\end{center}
\vspace{-.5cm}
\end{table}
\vspace{-.5cm}

\begin{table}[tbfhp]\begin{center}
\caption{Non-prompt to inclusive production cross-section fraction $f_B$ as a function of $J/\psi$ $p_T$ for $0.75<|y|_{J/\psi}<1.5$ under the assumption that prompt and non-prompt $J/\psi$ production is unpolarised ($\lambda_\theta = 0$). The spin-alignment envelope spans the range of possible prompt cross-sections under various polarisation hypotheses, plus the range of non-prompt cross-sections within $\lambda_\theta = \pm 0.1$. The first uncertainty is statistical, the second uncertainty is systematic, the third number is the uncertainty due to spin-alignment.}
\label{tab:AfractionMain1}
\small
\begin{tabular}{rr | rlll }
\hline
\hline
 &  & \multicolumn{4}{c}{Non-prompt to inclusive production fraction}  \\
$p_T$ & $\langle p_T \rangle$ & \multicolumn{4}{c}{$0.75<|y|<1.5$}  \\
(GeV) & (GeV) & $f_B$ & $\pm$ (stat.) & $\pm$ (syst.) & $\pm$ (spin) \\
\hline
4.0-5.0 & 4.7 & $ 0.142 $ & $\pm 0.094 $ & $\pm 0.018 $ & $\pm{}^{0.039}_{0.049}$ \\
5.0-5.5 & 5.3 & $ 0.183 $ & $\pm 0.049 $ & $\pm 0.036 $ & $\pm{}^{0.039}_{0.058}$ \\
5.5-6.0 & 5.8 & $ 0.127 $ & $\pm 0.038 $ & $\pm 0.024 $ & $\pm{}^{0.030}_{0.043}$ \\
6.0-6.5 & 6.3 & $ 0.188 $ & $\pm 0.033 $ & $\pm 0.019 $ & $\pm{}^{0.042}_{0.057}$ \\
6.5-7.0 & 6.8 & $ 0.261 $ & $\pm 0.029 $ & $\pm 0.007 $ & $\pm{}^{0.051}_{0.069}$ \\
7.0-7.5 & 7.2 & $ 0.230 $ & $\pm 0.025 $ & $\pm 0.017 $ & $\pm{}^{0.041}_{0.061}$ \\
7.5-8.0 & 7.8 & $ 0.238 $ & $\pm 0.023 $ & $\pm 0.015 $ & $\pm{}^{0.043}_{0.062}$ \\
8.0-8.5 & 8.2 & $ 0.226 $ & $\pm 0.022 $ & $\pm 0.032 $ & $\pm{}^{0.036}_{0.055}$ \\
8.5-9.0 & 8.8 & $ 0.226 $ & $\pm 0.021 $ & $\pm 0.013 $ & $\pm{}^{0.036}_{0.055}$ \\
9.0-9.5 & 9.2 & $ 0.261 $ & $\pm 0.021 $ & $\pm 0.009 $ & $\pm{}^{0.040}_{0.060}$ \\
9.5-10.0 & 9.8 & $ 0.292 $ & $\pm 0.023 $ & $\pm 0.008 $ & $\pm{}^{0.043}_{0.064}$ \\
10.0-11.0 & 10.5 & $ 0.315 $ & $\pm 0.016 $ & $\pm 0.004 $ & $\pm{}^{0.040}_{0.061}$ \\
11.0-12.0 & 11.5 & $ 0.343 $ & $\pm 0.018 $ & $\pm 0.007 $ & $\pm{}^{0.041}_{0.064}$ \\
12.0-14.0 & 12.9 & $ 0.352 $ & $\pm 0.016 $ & $\pm 0.005 $ & $\pm{}^{0.033}_{0.054}$ \\
14.0-16.0 & 14.9 & $ 0.401 $ & $\pm 0.022 $ & $\pm 0.003 $ & $\pm{}^{0.035}_{0.058}$ \\
16.0-18.0 & 16.9 & $ 0.450 $ & $\pm 0.031 $ & $\pm 0.006 $ & $\pm{}^{0.036}_{0.058}$ \\
18.0-22.0 & 19.7 & $ 0.476 $ & $\pm 0.031 $ & $\pm 0.006 $ & $\pm{}^{0.033}_{0.052}$ \\
22.0-30.0 & 25.1 & $ 0.542 $ & $\pm 0.042 $ & $\pm 0.015 $ & $\pm{}^{0.029}_{0.042}$ \\
30.0-70.0 & 37.8 & $ 0.594 $ & $\pm 0.060 $ & $\pm 0.016 $ & $\pm{}^{0.029}_{0.040}$ \\
\hline
\hline
\end{tabular}
\end{center}\vspace{-.3cm}
\end{table}

\begin{table}[tbfhp]\begin{center}
\vspace{-.5cm}
\caption{Non-prompt to inclusive production cross-section fraction $f_B$ as a function of $J/\psi$ $p_T$ for $1.5<|y|_{J/\psi}<2$ under the assumption that prompt and non-prompt $J/\psi$ production is unpolarised ($\lambda_\theta = 0$). The spin-alignment envelope spans the range of possible prompt cross-sections under various polarisation hypotheses, plus the range of non-prompt cross-sections within $\lambda_\theta = \pm 0.1$. The first uncertainty is statistical, the second uncertainty is systematic, the third number is the uncertainty due to spin-alignment.}
\label{tab:AfractionMain2}
\small
\begin{tabular}{rr | rlll }
\hline
\hline
 &  & \multicolumn{4}{c}{Non-prompt to inclusive production fraction}  \\
$p_T$ & $\langle p_T \rangle$ & \multicolumn{4}{c}{$1.5<|y|<2$}  \\
(GeV) & (GeV) & $f_B$ & $\pm$ (stat.) & $\pm$ (syst.) & $\pm$ (spin) \\
\hline
1.0-4.0 & 2.8 & $ 0.100 $ & $\pm 0.053 $ & $\pm 0.039 $ & $\pm{}^{0.061}_{0.031}$ \\
4.0-5.0 & 4.6 & $ 0.210 $ & $\pm 0.042 $ & $\pm 0.051 $ & $\pm{}^{0.115}_{0.051}$ \\
5.0-5.5 & 5.3 & $ 0.218 $ & $\pm 0.043 $ & $\pm 0.006 $ & $\pm{}^{0.097}_{0.050}$ \\
5.5-6.0 & 5.8 & $ 0.170 $ & $\pm 0.034 $ & $\pm 0.019 $ & $\pm{}^{0.068}_{0.041}$ \\
6.0-6.5 & 6.3 & $ 0.180 $ & $\pm 0.034 $ & $\pm 0.048 $ & $\pm{}^{0.057}_{0.042}$ \\
6.5-7.0 & 6.8 & $ 0.222 $ & $\pm 0.028 $ & $\pm 0.013 $ & $\pm{}^{0.069}_{0.048}$ \\
7.0-7.5 & 7.3 & $ 0.195 $ & $\pm 0.025 $ & $\pm 0.017 $ & $\pm{}^{0.049}_{0.044}$ \\
7.5-8.0 & 7.8 & $ 0.210 $ & $\pm 0.024 $ & $\pm 0.014 $ & $\pm{}^{0.052}_{0.047}$ \\
8.0-8.5 & 8.2 & $ 0.216 $ & $\pm 0.022 $ & $\pm 0.022 $ & $\pm{}^{0.042}_{0.044}$ \\
8.5-9.0 & 8.8 & $ 0.264 $ & $\pm 0.023 $ & $\pm 0.018 $ & $\pm{}^{0.049}_{0.050}$ \\
9.0-9.5 & 9.2 & $ 0.287 $ & $\pm 0.026 $ & $\pm 0.015 $ & $\pm{}^{0.051}_{0.052}$ \\
9.5-10.0 & 9.7 & $ 0.297 $ & $\pm 0.028 $ & $\pm 0.015 $ & $\pm{}^{0.053}_{0.053}$ \\
10.0-11.0 & 10.5 & $ 0.335 $ & $\pm 0.019 $ & $\pm 0.004 $ & $\pm{}^{0.043}_{0.055}$ \\
11.0-12.0 & 11.5 & $ 0.326 $ & $\pm 0.026 $ & $\pm 0.017 $ & $\pm{}^{0.042}_{0.054}$ \\
12.0-14.0 & 12.9 & $ 0.357 $ & $\pm 0.022 $ & $\pm 0.015 $ & $\pm{}^{0.034}_{0.045}$ \\
14.0-16.0 & 14.9 & $ 0.420 $ & $\pm 0.029 $ & $\pm 0.011 $ & $\pm{}^{0.035}_{0.047}$ \\
16.0-18.0 & 16.9 & $ 0.517 $ & $\pm 0.038 $ & $\pm 0.007 $ & $\pm{}^{0.039}_{0.048}$ \\
18.0-22.0 & 19.7 & $ 0.468 $ & $\pm 0.038 $ & $\pm 0.012 $ & $\pm{}^{0.029}_{0.041}$ \\
22.0-30.0 & 24.9 & $ 0.605 $ & $\pm 0.058 $ & $\pm 0.005 $ & $\pm{}^{0.021}_{0.032}$ \\
\hline
\hline
\end{tabular}
\end{center}\vspace{-.3cm}
\end{table}

\begin{table}[tbfhp]\begin{center}
\vspace{-.5cm}
\caption{Non-prompt to inclusive production cross-section fraction $f_B$ as a function of $J/\psi$ $p_T$ for $2<|y|_{J/\psi}<2.4$ under the assumption that prompt and non-prompt $J/\psi$ production is unpolarised ($\lambda_\theta = 0$). The spin-alignment envelope spans the range of possible prompt cross-sections under various polarisation hypotheses, plus the range of non-prompt cross-sections within $\lambda_\theta = \pm 0.1$. The first uncertainty is statistical, the second uncertainty is systematic, the third number is the uncertainty due to spin-alignment.}
\label{tab:AfractionMain3}
\small
\begin{tabular}{rr | rlll }
\hline
\hline
 &  & \multicolumn{4}{c}{Non-prompt to inclusive production fraction}  \\
$p_T$ & $\langle p_T \rangle$ & \multicolumn{4}{c}{$2<|y|<2.4$}  \\
(GeV) & (GeV) & $f_B$ & $\pm$ (stat.) & $\pm$ (syst.) & $\pm$ (spin) \\
\hline
1.0-5.0 & 3.6 & $ 0.098 $ & $\pm 0.065 $ & $\pm 0.036 $ & $\pm{}^{0.053}_{0.027}$ \\
5.0-6.0 & 5.5 & $ 0.217 $ & $\pm 0.077 $ & $\pm 0.065 $ & $\pm{}^{0.096}_{0.044}$ \\
6.0-7.0 & 6.6 & $ 0.289 $ & $\pm 0.047 $ & $\pm 0.052 $ & $\pm{}^{0.096}_{0.041}$ \\
7.0-7.5 & 7.2 & $ 0.125 $ & $\pm 0.035 $ & $\pm 0.016 $ & $\pm{}^{0.037}_{0.019}$ \\
7.5-8.0 & 7.8 & $ 0.231 $ & $\pm 0.037 $ & $\pm 0.020 $ & $\pm{}^{0.063}_{0.031}$ \\
8.0-8.5 & 8.2 & $ 0.209 $ & $\pm 0.042 $ & $\pm 0.042 $ & $\pm{}^{0.046}_{0.028}$ \\
8.5-9.0 & 8.7 & $ 0.183 $ & $\pm 0.041 $ & $\pm 0.032 $ & $\pm{}^{0.042}_{0.024}$ \\
9.0-9.5 & 9.2 & $ 0.268 $ & $\pm 0.037 $ & $\pm 0.027 $ & $\pm{}^{0.057}_{0.033}$ \\
9.5-10.0 & 9.7 & $ 0.249 $ & $\pm 0.045 $ & $\pm 0.008 $ & $\pm{}^{0.055}_{0.032}$ \\
10.0-11.0 & 10.5 & $ 0.269 $ & $\pm 0.037 $ & $\pm 0.014 $ & $\pm{}^{0.040}_{0.033}$ \\
11.0-12.0 & 11.5 & $ 0.297 $ & $\pm 0.034 $ & $\pm 0.010 $ & $\pm{}^{0.045}_{0.034}$ \\
12.0-14.0 & 12.9 & $ 0.352 $ & $\pm 0.034 $ & $\pm 0.018 $ & $\pm{}^{0.037}_{0.033}$ \\
14.0-18.0 & 15.6 & $ 0.348 $ & $\pm 0.044 $ & $\pm 0.049 $ & $\pm{}^{0.038}_{0.032}$ \\
18.0-30.0 & 21.7 & $ 0.419 $ & $\pm 0.058 $ & $\pm 0.058 $ & $\pm{}^{0.032}_{0.031}$ \\
\hline
\hline
\end{tabular}
\end{center}\vspace{-.3cm}
\end{table}

\begin{table}[tbfhp]\begin{center}
\caption{Non-prompt $J/\psi$ production cross-sections as a function of $J/\psi$ $p_T$ for $|y|_{J/\psi}<0.75$ under the assumption that prompt and non-prompt $J/\psi$ production is unpolarised ($\lambda_\theta = 0$), and the spin-alignment envelope spans the range of non-prompt cross-sections within $\lambda_\theta = \pm 0.1$. The first uncertainty is statistical, the second uncertainty is systematic. Comparison is made to FONLL predictions.}
\label{tab:Anonpromptxsec_1}
\small
\begin{tabular}{rc | llll | l}
\hline
\hline
 &  & \multicolumn{5}{c}{$\frac{d^{2}\sigma^{non-prompt}}{dp_{T}dy}\cdot$Br$(J/\psi\to\mu^+\mu^-)$ [nb/GeV]}  \\
$p_T$ & $\langle p_T \rangle$ & \multicolumn{4}{c}{$|y|<0.75$}  \\
(GeV) & (GeV) & Value & $\pm$ (stat.) & $\pm$ (syst.) & $\pm $ (spin) & FONLL prediction \\
\hline
7.0-7.5 & 7.3 & $ 2.4 $ & $\pm 0.4 $ & $\pm {0.4} $ & $\pm0.08$ & $1.7\pm{}^{0.7}_{0.4} $ \\
7.5-8.0 & 7.8 & $ 1.8 $ & $\pm 0.3 $ & $\pm {0.3} $ & $\pm0.06$ & $1.4\pm{}^{0.5}_{0.3} $ \\
8.0-8.5 & 8.3 & $ 1.2 $ & $\pm 0.2 $ & $\pm {0.1} $ & $\pm0.03$ & $1.1\pm{}^{0.4}_{0.3} $ \\
8.5-9.0 & 8.8 & $ 1.1 $ & $\pm 0.1 $ & $\pm {0.1} $ & $\pm0.03$ & $0.9\pm{}^{0.3}_{0.2} $ \\
9.0-9.5 & 9.3 & $ 0.8 $ & $\pm 0.1 $ & $\pm {0.1} $ & $\pm0.02$ & $0.7\pm{}^{0.3}_{0.2} $ \\
9.5-10.0 & 9.8 & $ 0.69 $ & $\pm 0.07 $ & $\pm {0.08}$ & $\pm0.02$ & $0.62\pm{}^{0.22}_{0.15} $ \\
10.0-11.0 & 10.5 & $ 0.49 $ & $\pm 0.03 $ & $\pm {0.05} $ & $\pm0.01$ & $0.47\pm{}^{0.16}_{0.11} $ \\
11.0-12.0 & 11.5 & $ 0.34 $ & $\pm 0.02 $ & $\pm {0.04} $ & $\pm0.01$ & $0.34\pm{}^{0.11}_{0.08} $ \\
12.0-14.0 & 12.9 & $ 0.19 $ & $\pm 0.01 $ & $\pm {0.02} $ & $\pm0.004$ & $0.21\pm{}^{0.06}_{0.05} $ \\
14.0-16.0 & 14.9 & $ 0.111 $ & $\pm 0.008 $ & $\pm {0.011} $ & $\pm0.003$ & $0.117\pm{}^{0.033}_{0.024} $ \\
16.0-18.0 & 16.9 & $ 0.060 $ & $\pm 0.005 $ & $\pm {0.006} $ & $\pm0.002$ & $0.069\pm{}^{0.018}_{0.013} $ \\
18.0-22.0 & 19.7 & $ 0.032 $ & $\pm 0.003 $ & $\pm {0.003} $ & $\pm0.001$ & $0.035\pm{}^{0.008}_{0.006} $ \\
22.0-30.0 & 25.0 & $ 0.0091 $ & $\pm 0.0010 $ & $\pm {0.0011} $ & $\pm0.0002$ & $0.0109\pm{}^{0.0022}_{0.0018} $ \\
30.0-70.0 & 37.2 & $ 0.0008 $ & $\pm 0.0001 $ & $\pm {0.0001} $ & $\pm0.0000$ & $0.0009\pm{}^{0.0001}_{0.0001} $ \\
\hline
\hline
\end{tabular}
\end{center}\end{table}

\begin{table}[tbfhp]\begin{center}
\caption{Non-prompt $J/\psi$ production cross-sections as a function of $J/\psi$ $p_T$ for $0.75<|y|_{J/\psi}<1.5$ under the assumption that prompt and non-prompt $J/\psi$ production is unpolarised ($\lambda_\theta = 0$), and the spin-alignment envelope spans the range of non-prompt cross-sections within $\lambda_\theta = \pm 0.1$. The first uncertainty is statistical, the second uncertainty is systematic. Comparison is made to FONLL predictions.}
\label{tab:Anonpromptxsec_2}
\small
\begin{tabular}{rc | llll | l}
\hline
\hline
 &  & \multicolumn{5}{c}{$\frac{d^{2}\sigma^{non-prompt}}{dp_{T}dy}\cdot$Br$(J/\psi\to\mu^+\mu^-)$ [nb/GeV]}  \\
$p_T$ & $\langle p_T \rangle$ & \multicolumn{4}{c}{$0.75<|y|<1.5$}  \\
(GeV) & (GeV) & Value & $\pm$ (stat.) & $\pm$ (syst.) & $\pm $ (spin) & FONLL prediction \\
\hline
5.0-5.5 & 5.3 & $ 4.9 $ & $\pm 1.7 $ & $\pm {1.2} $ & $\pm0.15$ & $3.8\pm{}^{1.6}_{1.1} $ \\
5.5-6.0 & 5.8 & $ 2.4 $ & $\pm 0.8 $ & $\pm {0.5} $ & $\pm0.07$ & $3.0\pm{}^{1.3}_{0.8} $ \\
6.0-6.5 & 6.3 & $ 2.5 $ & $\pm 0.5 $ & $\pm {0.4} $ & $\pm0.07$ & $2.4\pm{}^{1.0}_{0.7} $ \\
6.5-7.0 & 6.8 & $ 3.3 $ & $\pm 0.5 $ & $\pm {0.5} $ & $\pm0.09$ & $1.9\pm{}^{0.6}_{0.5} $ \\
7.0-7.5 & 7.2 & $ 1.9 $ & $\pm 0.3 $ & $\pm {0.3} $ & $\pm0.05$ & $1.5\pm{}^{0.6}_{0.4} $ \\
7.5-8.0 & 7.8 & $ 1.6 $ & $\pm 0.2 $ & $\pm {0.2} $ & $\pm0.04$ & $1.2\pm{}^{0.5}_{0.3} $ \\
8.0-8.5 & 8.3 & $ 0.9 $ & $\pm 0.1 $ & $\pm {0.1} $ & $\pm0.02$ & $1.0\pm{}^{0.4}_{0.3} $ \\

8.5-9.0 & 8.8 & $ 0.81 $ & $\pm 0.09 $ & $\pm {0.09} $ & $\pm0.02$ & $0.83\pm{}^{0.30}_{0.20} $ \\
9.0-9.5 & 9.3 & $ 0.75 $ & $\pm 0.07 $ & $\pm {0.08} $ & $\pm0.02$ & $0.68\pm{}^{0.24}_{0.17} $ \\
9.5-10.0 & 9.8 & $ 0.65 $ & $\pm 0.06 $ & $\pm {0.07} $ & $\pm0.02$ & $0.56\pm{}^{0.20}_{0.13} $ \\
10.0-11.0 & 10.5 & $ 0.49 $ & $\pm 0.03 $ & $\pm {0.05} $ & $\pm0.01$ & $0.43\pm{}^{0.15}_{0.10} $ \\
11.0-12.0 & 11.5 & $ 0.35 $ & $\pm 0.02 $ & $\pm {0.04} $ & $\pm0.01$ & $0.30\pm{}^{0.10}_{0.07} $ \\
12.0-14.0 & 12.9 & $ 0.19 $ & $\pm 0.01 $ & $\pm {0.02} $ & $\pm0.00$ & $0.19\pm{}^{0.06}_{0.04} $ \\

14.0-16.0 & 14.9 & $ 0.100 $ & $\pm 0.007 $ & $\pm {0.011} $ & $\pm0.002$ & $0.104\pm{}^{0.029}_{0.021} $ \\
16.0-18.0 & 16.9 & $ 0.054 $ & $\pm 0.005 $ & $\pm {0.006} $ & $\pm0.001$ & $0.061\pm{}^{0.016}_{0.012} $ \\
18.0-22.0 & 19.7 & $ 0.025 $ & $\pm 0.002 $ & $\pm {0.003} $ & $\pm0.001$ & $0.030\pm{}^{0.007}_{0.006} $ \\

22.0-30.0 & 25.2 & $ 0.0086 $ & $\pm 0.0009 $ & $\pm {0.0010} $ & $\pm0.0001$ & $0.0093\pm{}^{0.0019}_{0.0015} $ \\
30.0-70.0 & 38.0 & $ 0.0007 $ & $\pm 0.0001 $ & $\pm {0.0001} $ & $\pm0.0000$ & $0.0007\pm{}^{0.0001}_{0.0001} $ \\
\hline
\hline
\end{tabular}
\end{center}\end{table}

\begin{table}[tbfhp]\begin{center}
\caption{Non-prompt $J/\psi$ production cross-sections as a function of $J/\psi$ $p_T$ for $1.5<|y|_{J/\psi}<2$ under the assumption that prompt and non-prompt $J/\psi$ production is unpolarised ($\lambda_\theta = 0$), and the spin-alignment envelope spans the range of non-prompt cross-sections within $\lambda_\theta = \pm 0.1$. The first uncertainty is statistical, the second uncertainty is systematic. Comparison is made to FONLL predictions.}
\label{tab:Anonpromptxsec_3}
\small
\begin{tabular}{rc | llll | l}
\hline
\hline
 &  & \multicolumn{5}{c}{$\frac{d^{2}\sigma^{non-prompt}}{dp_{T}dy}\cdot$Br$(J/\psi\to\mu^+\mu^-)$ [nb/GeV]}  \\
$p_T$ & $\langle p_T \rangle$ & \multicolumn{4}{c}{$1.5<|y|<2$}  \\
(GeV) & (GeV) & Value & $\pm$ (stat.) & $\pm$ (syst.) & $\pm $ (spin) & FONLL prediction \\
\hline
1.0-4.0 & 2.8 & $ 14.3 $ & $\pm 7.9 $ & $\pm {5.5} $ & $\pm0.37$ & $9.4\pm{}^{4.5}_{3.9} $ \\
4.0-5.0 & 4.5 & $ 8.3 $ & $\pm 2.0 $ & $\pm {1.9} $ & $\pm0.21$ & $4.9\pm{}^{2.2}_{1.6} $ \\
5.0-5.5 & 5.2 & $ 3.8 $ & $\pm 1.0 $ & $\pm {0.9} $ & $\pm0.09$ & $3.4\pm{}^{1.5}_{1.0} $ \\
5.5-6.0 & 5.7 & $ 2.4 $ & $\pm 0.5 $ & $\pm {0.4} $ & $\pm0.06$ & $2.7\pm{}^{1.1}_{0.7} $ \\
6.0-6.5 & 6.3 & $ 2.3 $ & $\pm 0.5 $ & $\pm {0.6} $ & $\pm0.05$ & $2.1\pm{}^{0.9}_{0.6} $ \\
6.5-7.0 & 6.8 & $ 2.0 $ & $\pm 0.3 $ & $\pm {0.3} $ & $\pm0.05$ & $1.7\pm{}^{0.7}_{0.4} $ \\
7.0-7.5 & 7.3 & $ 1.2 $ & $\pm 0.2 $ & $\pm {0.2} $ & $\pm0.03$ & $1.3\pm{}^{0.5}_{0.4} $ \\
7.5-8.0 & 7.8 & $ 1.1 $ & $\pm 0.1 $ & $\pm {0.1} $ & $\pm0.02$ & $1.1\pm{}^{0.4}_{0.3} $ \\

8.0-8.5 & 8.3 & $ 0.82 $ & $\pm 0.10 $ & $\pm {0.10} $ & $\pm0.02$ & $0.87\pm{}^{0.33}_{0.22} $ \\
8.5-9.0 & 8.8 & $ 0.82 $ & $\pm 0.08 $ & $\pm {0.10} $ & $\pm0.02$ & $0.71\pm{}^{0.26}_{0.18} $ \\
9.0-9.5 & 9.3 & $ 0.65 $ & $\pm 0.07 $ & $\pm {0.07} $ & $\pm0.01$ & $0.58\pm{}^{0.21}_{0.14} $ \\
9.5-10.0 & 9.8 & $ 0.50 $ & $\pm 0.05 $ & $\pm {0.05} $ & $\pm0.01$ & $0.48\pm{}^{0.17}_{0.11} $ \\
10.0-11.0 & 10.5 & $ 0.43 $ & $\pm 0.03 $ & $\pm {0.05} $ & $\pm0.01$ & $0.36\pm{}^{0.12}_{0.08} $ \\
11.0-12.0 & 11.5 & $ 0.25 $ & $\pm 0.02 $ & $\pm {0.03} $ & $\pm0.01$ & $0.25\pm{}^{0.08}_{0.06} $ \\
12.0-14.0 & 12.9 & $ 0.14 $ & $\pm 0.01 $ & $\pm {0.01} $ & $\pm0.002$ & $0.16\pm{}^{0.05}_{0.03} $ \\

14.0-16.0 & 14.9 & $ 0.081 $ & $\pm 0.007 $ & $\pm {0.009} $ & $\pm0.001$ & $0.085\pm{}^{0.024}_{0.017} $ \\
16.0-18.0 & 16.9 & $ 0.053 $ & $\pm 0.005 $ & $\pm {0.007} $ & $\pm0.001$ & $0.049\pm{}^{0.013}_{0.009} $ \\
18.0-22.0 & 19.6 & $ 0.023 $ & $\pm 0.002 $ & $\pm {0.002} $ & $\pm0.0000$ & $0.024\pm{}^{0.006}_{0.004} $ \\

22.0-30.0 & 25.0 & $ 0.0064 $ & $\pm 0.0009 $ & $\pm {0.00081} $ & $\pm0.0001$ & $0.0071\pm{}^{0.0014}_{0.0012} $ \\
\hline
\hline
\end{tabular}
\end{center}\end{table}

\begin{table}[tbfhp]\begin{center}
\caption{Non-prompt $J/\psi$ production cross-sections as a function of $J/\psi$ $p_T$ for $2<|y|_{J/\psi}<2.4$ under the assumption that prompt and non-prompt $J/\psi$ production is unpolarised ($\lambda_\theta = 0$), and the spin-alignment envelope spans the range of non-prompt cross-sections within $\lambda_\theta = \pm 0.1$. The first uncertainty is statistical, the second uncertainty is systematic. Comparison is made to FONLL predictions.}
\label{tab:Anonpromptxsec_4}
\small
\begin{tabular}{rc | llll | l}
\hline
\hline
 &  & \multicolumn{5}{c}{$\frac{d^{2}\sigma^{non-prompt}}{dp_{T}dy}\cdot$Br$(J/\psi\to\mu^+\mu^-)$ [nb/GeV]}  \\
$p_T$ & $\langle p_T \rangle$ & \multicolumn{4}{c}{$2<|y|<2.4$}  \\
(GeV) & (GeV) & Value & $\pm$ (stat.) & $\pm$ (syst.) & $\pm $ (spin) & FONLL prediction \\
\hline
5.0-6.0 & 5.5 & $ 3.2 $ & $\pm 1.3 $ & $\pm {1.0} $ & $\pm0.04$ & $2.7\pm{}^{1.1}_{0.7} $ \\
6.0-7.0 & 6.5 & $ 2.2 $ & $\pm 0.4 $ & $\pm {0.4} $ & $\pm0.02$ & $1.7\pm{}^{0.7}_{0.4} $ \\
7.0-7.5 & 7.3 & $ 0.5 $ & $\pm 0.2 $ & $\pm {0.1} $ & $\pm0.01$ & $1.2\pm{}^{0.5}_{0.3} $ \\
7.5-8.0 & 7.8 & $ 0.9 $ & $\pm 0.2 $ & $\pm {0.2} $ & $\pm0.01$ & $0.9\pm{}^{0.4}_{0.2} $ \\
8.0-8.5 & 8.3 & $ 0.6 $ & $\pm 0.1 $ & $\pm {0.1} $ & $\pm0.01$ & $0.7\pm{}^{0.3}_{0.2} $ \\

8.5-9.0 & 8.8 & $ 0.35 $ & $\pm 0.09 $ & $\pm {0.06} $ & $\pm0.01$ & $0.60\pm{}^{0.22}_{0.15} $ \\
9.0-9.5 & 9.2 & $ 0.39 $ & $\pm 0.06 $ & $\pm {0.06} $ & $\pm0.01$ & $0.49\pm{}^{0.18}_{0.12} $ \\
9.5-10.0 & 9.7 & $ 0.30 $ & $\pm 0.06 $ & $\pm {0.04} $ & $\pm0.005$ & $0.40\pm{}^{0.14}_{0.10} $ \\
10.0-11.0 & 10.5 & $ 0.22 $ & $\pm 0.03 $ & $\pm {0.03} $ & $\pm0.003$ & $0.30\pm{}^{0.10}_{0.07} $ \\
11.0-12.0 & 11.5 & $ 0.18 $ & $\pm 0.02 $ & $\pm {0.02} $ & $\pm0.003$ & $0.21\pm{}^{0.07}_{0.05} $ \\
12.0-14.0 & 12.9 & $ 0.11 $ & $\pm 0.01 $ & $\pm {0.01} $ & $\pm0.002$ & $0.13\pm{}^{0.04}_{0.03} $ \\

14.0-18.0 & 15.6 & $ 0.042 $ & $\pm 0.006 $ & $\pm {0.005} $ & $\pm0.0004$ & $0.053\pm{}^{0.015}_{0.011} $ \\

18.0-30.0 & 21.3 & $ 0.0059 $ & $\pm 0.0010 $ & $\pm {0.0007} $ & $\pm0.0001$ & $0.0097\pm{}^{0.0022}_{0.0017} $ \\
\hline
\hline
\end{tabular}
\end{center}\end{table}

\begin{table}[tbfhp]\begin{center}
\caption{Prompt $J/\psi$ production cross-sections as a function of $J/\psi$ $p_T$ for $|y|_{J/\psi}<0.75$. The central value assumes unpolarised ($\lambda_\theta = 0$) prompt and non-prompt production, and the spin-alignment envelope spans the range of possible prompt cross-sections under various polarisation hypotheses. The first quoted uncertainty is statistical, the second uncertainty is systematic. Comparison is made to the Colour Evaporation Model prediction.}
\label{tab:Apromptxsec_1}
\small
\begin{tabular}{rc | llll | l }
\hline
\hline
 &  & \multicolumn{5}{c}{$\frac{d^{2}\sigma^{prompt}}{dp_{T}dy}\cdot$Br$(J/\psi\to\mu^+\mu^-)$ [nb/GeV]}  \\
$p_T$ & $\langle p_T \rangle$ & \multicolumn{5}{c}{$|y|<0.75$}  \\
(GeV) & (GeV) & Value & $\pm$ (stat.) & $\pm$ (syst.) & $\pm $ (spin) & CEM prediction \\
\hline
7.0-7.5 & 7.3 & $ 6.8 $ & $\pm 0.8 $ & $\pm {}^{1.1}_{1.1} $ & $\pm {}^{4.3}_{2.2}$ & $2.8$ \\
7.5-8.0 & 7.8 & $ 5.9 $ & $\pm 0.6 $ & $\pm {}^{0.9}_{0.9} $ & $\pm {}^{2.7}_{1.9}$ & $2.2$ \\
8.0-8.5 & 8.3 & $ 3.3 $ & $\pm 0.3 $ & $\pm {}^{0.4}_{0.4} $ & $\pm {}^{1.3}_{1.0}$ & $1.7$ \\
8.5-9.0 & 8.8 & $ 2.6 $ & $\pm 0.2 $ & $\pm {}^{0.4}_{0.4} $ & $\pm {}^{0.9}_{0.8}$ & $1.3$ \\
9.0-9.5 & 9.2 & $ 2.2 $ & $\pm 0.2 $ & $\pm {}^{0.3}_{0.3} $ & $\pm {}^{0.9}_{0.6}$ & $1.0$ \\
9.5-10.0 & 9.8 & $ 1.5 $ & $\pm 0.1 $ & $\pm {}^{0.2}_{0.2} $ & $\pm {}^{0.5}_{0.4}$ & $0.8$ \\

10.0-11.0 & 10.5 & $ 1.04 $ & $\pm 0.05 $ & $\pm {}^{0.11}_{0.11} $ & $\pm {}^{0.32}_{0.29}$ & $0.60$ \\
11.0-12.0 & 11.5 & $ 0.71 $ & $\pm 0.03 $ & $\pm {}^{0.08}_{0.08} $ & $\pm {}^{0.19}_{0.20}$ & $0.41$ \\
12.0-14.0 & 12.9 & $ 0.34 $ & $\pm 0.01 $ & $\pm {}^{0.04}_{0.04} $ & $\pm {}^{0.08}_{0.09}$ & $0.24$ \\

14.0-16.0 & 14.9 & $ 0.163 $ & $\pm 0.010 $ & $\pm {}^{0.016}_{0.016} $ & $\pm {}^{0.036}_{0.042}$ & $0.128$ \\
16.0-18.0 & 16.9 & $ 0.076 $ & $\pm 0.006 $ & $\pm {}^{0.008}_{0.008} $ & $\pm {}^{0.015}_{0.018}$ & $0.071$ \\
18.0-22.0 & 19.7 & $ 0.035 $ & $\pm 0.003 $ & $\pm {}^{0.004}_{0.004} $ & $\pm {}^{0.006}_{0.008}$ & $0.035$ \\

22.0-30.0 & 25.0 & $ 0.0078 $ & $\pm 0.0009 $ & $\pm {}^{0.0009}_{0.0009} $ & $\pm {}^{0.0010}_{0.0014}$ & $0.0109$ \\
30.0-70.0 & 37.2 & $ 0.0004 $ & $\pm 0.0001 $ & $\pm {}^{0.0001}_{0.0001} $ & $\pm {}^{0.0000}_{0.0000}$ & $0.0008$ \\
\hline
\hline
\end{tabular}
\end{center}\end{table}

\begin{table}[tbfhp]\begin{center}
\caption{Prompt $J/\psi$ production cross-sections as a function of $J/\psi$ $p_T$ for $0.75<|y|_{J/\psi}<1.5$. The central value assumes unpolarised ($\lambda_\theta = 0$) prompt and non-prompt production, and the spin-alignment envelope spans the range of possible prompt cross-sections under various polarisation hypotheses. The first quoted uncertainty is statistical, the second uncertainty is systematic. Comparison is made to the Colour Evaporation Model prediction.}
\label{tab:Apromptxsec_2}
\small
\begin{tabular}{rc | llll | l }
\hline
\hline
 &  & \multicolumn{5}{c}{$\frac{d^{2}\sigma^{prompt}}{dp_{T}dy}\cdot$Br$(J/\psi\to\mu^+\mu^-)$ [nb/GeV]}  \\
$p_T$ & $\langle p_T \rangle$ & \multicolumn{5}{c}{$0.75<|y|<1.5$}  \\
(GeV) & (GeV) & Value & $\pm$ (stat.) & $\pm$ (syst.) & $\pm $ (spin) & CEM prediction \\
\hline
5.0-5.5 & 5.3 & $ 21.9 $ & $\pm 4.7 $ & $\pm {}^{4.8}_{4.6} $ & $\pm {}^{8.7}_{6.5}$ & $10.4$ \\
5.5-6.0 & 5.8 & $ 16.8 $ & $\pm 2.6 $ & $\pm {}^{3.0}_{2.9} $ & $\pm {}^{7.5}_{5.0}$ & $7.2$ \\
6.0-6.5 & 6.2 & $ 11.0 $ & $\pm 1.0 $ & $\pm {}^{1.4}_{1.4} $ & $\pm {}^{5.8}_{3.2}$ & $5.2$ \\
6.5-7.0 & 6.7 & $ 9.2 $ & $\pm 0.9 $ & $\pm {}^{1.4}_{1.4} $ & $\pm {}^{2.9}_{2.6}$ & $3.7$ \\
7.0-7.5 & 7.2 & $ 6.3 $ & $\pm 0.5 $ & $\pm {}^{0.8}_{0.8} $ & $\pm {}^{1.7}_{1.8}$ & $2.8$ \\
7.5-8.0 & 7.7 & $ 5.0 $ & $\pm 0.3 $ & $\pm {}^{0.6}_{0.6} $ & $\pm {}^{1.2}_{1.4}$ & $2.1$ \\
8.0-8.5 & 8.2 & $ 3.2 $ & $\pm 0.2 $ & $\pm {}^{0.3}_{0.3} $ & $\pm {}^{1.4}_{0.7}$ & $1.6$ \\
8.5-9.0 & 8.7 & $ 2.8 $ & $\pm 0.2 $ & $\pm {}^{0.3}_{0.3} $ & $\pm {}^{0.8}_{0.6}$ & $1.3$ \\
9.0-9.5 & 9.3 & $ 2.1 $ & $\pm 0.1 $ & $\pm {}^{0.2}_{0.2} $ & $\pm {}^{0.5}_{0.5}$ & $1.0$ \\

9.5-10.0 & 9.7 & $ 1.57 $ & $\pm 0.09 $ & $\pm {}^{0.17}_{0.17} $ & $\pm {}^{0.30}_{0.35}$ & $0.79$ \\
10.0-11.0 & 10.5 & $ 1.06 $ & $\pm 0.04 $ & $\pm {}^{0.12}_{0.12} $ & $\pm {}^{0.19}_{0.24}$ & $0.59$ \\
11.0-12.0 & 11.5 & $ 0.67 $ & $\pm 0.03 $ & $\pm {}^{0.08}_{0.08} $ & $\pm {}^{0.12}_{0.15}$ & $0.39$ \\
12.0-14.0 & 12.9 & $ 0.34 $ & $\pm 0.01 $ & $\pm {}^{0.04}_{0.04} $ & $\pm {}^{0.06}_{0.08}$ & $0.23$ \\

14.0-16.0 & 14.9 & $ 0.149 $ & $\pm 0.008 $ & $\pm {}^{0.016}_{0.016} $ & $\pm {}^{0.024}_{0.031}$ & $0.120$ \\
16.0-18.0 & 16.9 & $ 0.066 $ & $\pm 0.005 $ & $\pm {}^{0.007}_{0.007} $ & $\pm {}^{0.009}_{0.013}$ & $0.067$ \\
18.0-22.0 & 19.7 & $ 0.028 $ & $\pm 0.002 $ & $\pm {}^{0.003}_{0.003} $ & $\pm {}^{0.004}_{0.005}$ & $0.032$ \\

22.0-30.0 & 25.2 & $ 0.0073 $ & $\pm 0.0008 $ & $\pm {}^{0.0008}_{0.0008} $ & $\pm {}^{0.0008}_{0.0011}$ & $0.0100$ \\
30.0-70.0 & 38.0 & $ 0.0004 $ & $\pm 0.0001 $ & $\pm {}^{0.0001}_{0.0001} $ & $\pm {}^{0.0000}_{0.0000}$ & $0.0007$ \\
\hline
\hline
\end{tabular}
\end{center}\end{table}

\begin{table}[tbfhp]\begin{center}
\caption{Prompt $J/\psi$ production cross-sections as a function of $J/\psi$ $p_T$ for $1.5<|y|_{J/\psi}<2$. The central value assumes unpolarised ($\lambda_\theta = 0$) prompt and non-prompt production, and the spin-alignment envelope spans the range of possible prompt cross-sections under various polarisation hypotheses. The first quoted uncertainty is statistical, the second uncertainty is systematic. Comparison is made to the Colour Evaporation Model prediction.}
\label{tab:Apromptxsec_3}
\small
\begin{tabular}{rc | llll | l }
\hline
\hline
 &  & \multicolumn{5}{c}{$\frac{d^{2}\sigma^{prompt}}{dp_{T}dy}\cdot$Br$(J/\psi\to\mu^+\mu^-)$ [nb/GeV]}  \\
$p_T$ & $\langle p_T \rangle$ & \multicolumn{5}{c}{$1.5<|y|<2$}  \\
(GeV) & (GeV) & Value & $\pm$ (stat.) & $\pm$ (syst.) & $\pm $ (spin) & CEM prediction \\
\hline
1.0-4.0 & 2.8 & $ 129 $ & $\pm 22 $ & $\pm {}^{25}_{35} $ & $\pm {}^{246}_{35}$ & $43$ \\
4.0-5.0 & 4.5 & $ 31.1 $ & $\pm 4.6 $ & $\pm {}^{4.5}_{4.5} $ & $\pm {}^{54.7}_{7.7}$ & $17.7$ \\
5.0-5.5 & 5.2 & $ 13.8 $ & $\pm 2.7 $ & $\pm {}^{3.4}_{3.2} $ & $\pm {}^{13.5}_{3.2}$ & $10.0$ \\
5.5-6.0 & 5.7 & $ 11.8 $ & $\pm 1.1 $ & $\pm {}^{1.5}_{1.5} $ & $\pm {}^{11.6}_{2.6}$ & $6.7$ \\
6.0-6.5 & 6.3 & $ 10.5 $ & $\pm 0.9 $ & $\pm {}^{1.2}_{1.1} $ & $\pm {}^{8.2}_{2.2}$ & $4.8$ \\
6.5-7.0 & 6.8 & $ 6.9 $ & $\pm 0.5 $ & $\pm {}^{1.0}_{1.0} $ & $\pm {}^{4.2}_{1.6}$ & $3.4$ \\
7.0-7.5 & 7.2 & $ 5.1 $ & $\pm 0.4 $ & $\pm {}^{0.7}_{0.7} $ & $\pm {}^{2.5}_{1.2}$ & $2.6$ \\
7.5-8.0 & 7.7 & $ 4.0 $ & $\pm 0.3 $ & $\pm {}^{0.5}_{0.4} $ & $\pm {}^{1.8}_{0.7}$ & $1.9$ \\
8.0-8.5 & 8.3 & $ 3.0 $ & $\pm 0.2 $ & $\pm {}^{0.3}_{0.3} $ & $\pm {}^{1.2}_{0.4}$ & $1.5$ \\
8.5-9.0 & 8.7 & $ 2.3 $ & $\pm 0.1 $ & $\pm {}^{0.3}_{0.2} $ & $\pm {}^{0.7}_{0.3}$ & $1.2$ \\

9.0-9.5 & 9.2 & $ 1.61 $ & $\pm 0.09 $ & $\pm {}^{0.17}_{0.17} $ & $\pm {}^{0.46}_{0.26}$ & $0.89$ \\
9.5-10.0 & 9.7 & $ 1.18 $ & $\pm 0.08 $ & $\pm {}^{0.13}_{0.12} $ & $\pm {}^{0.32}_{0.21}$ & $0.72$ \\
10.0-11.0 & 10.5 & $ 0.86 $ & $\pm 0.04 $ & $\pm {}^{0.10}_{0.09} $ & $\pm {}^{0.21}_{0.16}$ & $0.53$ \\
11.0-12.0 & 11.5 & $ 0.51 $ & $\pm 0.03 $ & $\pm {}^{0.05}_{0.05} $ & $\pm {}^{0.11}_{0.10}$ & $0.35$ \\
12.0-14.0 & 12.9 & $ 0.26 $ & $\pm 0.01 $ & $\pm {}^{0.03}_{0.02} $ & $\pm {}^{0.05}_{0.04}$ & $0.21$ \\

14.0-16.0 & 14.9 & $ 0.112 $ & $\pm 0.008 $ & $\pm {}^{0.012}_{0.011} $ & $\pm {}^{0.016}_{0.019}$ & $0.106$ \\
16.0-18.0 & 16.9 & $ 0.050 $ & $\pm 0.005 $ & $\pm {}^{0.007}_{0.005} $ & $\pm {}^{0.006}_{0.008}$ & $0.057$ \\
18.0-22.0 & 19.6 & $ 0.026 $ & $\pm 0.003 $ & $\pm {}^{0.002}_{0.002} $ & $\pm {}^{0.003}_{0.004}$ & $0.028$ \\

22.0-30.0 & 25.0 & $ 0.0042 $ & $\pm 0.0007 $ & $\pm {}^{0.0005}_{0.0005} $ & $\pm {}^{0.0003}_{0.0005}$ & $0.0084$ \\
\hline
\hline
\end{tabular}
\end{center}\end{table}

\begin{table}[tbfhp]\begin{center}
\caption{Prompt $J/\psi$ production cross-sections as a function of $J/\psi$ $p_T$ for $2<|y|_{J/\psi}<2.4$. The central value assumes unpolarised ($\lambda_\theta = 0$) prompt and non-prompt production, and the spin-alignment envelope spans the range of possible prompt cross-sections under various polarisation hypotheses. The first quoted uncertainty is statistical, the second uncertainty is systematic. Comparison is made to the Colour Evaporation Model prediction.}
\label{tab:Apromptxsec_4}
\small
\begin{tabular}{rc | llll | l }
\hline
\hline
 &  & \multicolumn{5}{c}{$\frac{d^{2}\sigma^{prompt}}{dp_{T}dy}\cdot$Br$(J/\psi\to\mu^+\mu^-)$ [nb/GeV]}  \\
$p_T$ & $\langle p_T \rangle$ & \multicolumn{5}{c}{$2<|y|<2.4$}  \\
(GeV) & (GeV) & Value & $\pm$ (stat.) & $\pm$ (syst.) & $\pm $ (spin) & CEM prediction \\
\hline
5.0-6.0 & 5.5 & $ 11.5 $ & $\pm 2.5 $ & $\pm {}^{2.5}_{2.6} $ & $\pm {}^{12.1}_{2.0}$ & $7.8$ \\
6.0-7.0 & 6.5 & $ 5.6 $ & $\pm 0.6 $ & $\pm {}^{0.6}_{0.6} $ & $\pm {}^{3.3}_{0.9}$ & $3.9$ \\
7.0-7.5 & 7.3 & $ 3.5 $ & $\pm 0.5 $ & $\pm {}^{0.6}_{0.5} $ & $\pm {}^{2.3}_{0.6}$ & $2.3$ \\
7.5-8.0 & 7.7 & $ 3.1 $ & $\pm 0.4 $ & $\pm {}^{0.5}_{0.4} $ & $\pm {}^{2.2}_{0.5}$ & $1.8$ \\
8.0-8.5 & 8.3 & $ 2.1 $ & $\pm 0.3 $ & $\pm {}^{0.3}_{0.3} $ & $\pm {}^{0.7}_{0.5}$ & $1.4$ \\
8.5-9.0 & 8.7 & $ 1.6 $ & $\pm 0.2 $ & $\pm {}^{0.2}_{0.2} $ & $\pm {}^{0.5}_{0.3}$ & $1.1$ \\
9.0-9.5 & 9.2 & $ 1.1 $ & $\pm 0.1 $ & $\pm {}^{0.1}_{0.1} $ & $\pm {}^{0.3}_{0.2}$ & $0.9$ \\

9.5-10.0 & 9.7 & $ 0.91 $ & $\pm 0.09 $ & $\pm {}^{0.13}_{0.12} $ & $\pm {}^{0.33}_{0.12}$ & $0.68$ \\
10.0-11.0 & 10.5 & $ 0.61 $ & $\pm 0.05 $ & $\pm {}^{0.07}_{0.07} $ & $\pm {}^{0.21}_{0.06}$ & $0.47$ \\
11.0-12.0 & 11.5 & $ 0.42 $ & $\pm 0.04 $ & $\pm {}^{0.05}_{0.06} $ & $\pm {}^{0.12}_{0.05}$ & $0.32$ \\
12.0-14.0 & 12.9 & $ 0.21 $ & $\pm 0.02 $ & $\pm {}^{0.02}_{0.02} $ & $\pm {}^{0.05}_{0.03}$ & $0.18$ \\

14.0-18.0 & 15.6 & $ 0.079 $ & $\pm 0.007 $ & $\pm {}^{0.007}_{0.005} $ & $\pm {}^{0.012}_{0.007}$ & $0.071$ \\
18.0-30.0 & 21.2 & $ 0.008 $ & $\pm 0.001 $ & $\pm {}^{0.001}_{0.001} $ & $\pm {}^{0.001}_{0.001}$ & $0.012$ \\
\hline
\hline
\end{tabular}
\end{center}\end{table}

\newpage

\begin{flushleft}
{\Large The ATLAS Collaboration}

\bigskip

G.~Aad$^{\rm 48}$,
B.~Abbott$^{\rm 111}$,
J.~Abdallah$^{\rm 11}$,
A.A.~Abdelalim$^{\rm 49}$,
A.~Abdesselam$^{\rm 118}$,
O.~Abdinov$^{\rm 10}$,
B.~Abi$^{\rm 112}$,
M.~Abolins$^{\rm 88}$,
H.~Abramowicz$^{\rm 153}$,
H.~Abreu$^{\rm 115}$
% ...
\bigskip

% ...
$^{10}$ Institute of Physics, Azerbaijan Academy of Sciences, Baku, Azerbaijan\\
$^{11}$ Institut de F\'isica d'Altes Energies and Universitat Aut\`onoma  de Barcelona and ICREA, Barcelona, Spain\\
% ...
$^{48}$ Fakult\"{a}t f\"{u}r Mathematik und Physik, Albert-Ludwigs-Universit\"{a}t, Freiburg i.Br., Germany\\
$^{49}$ Section de Physique, Universit\'e de Gen\`eve, Geneva, Switzerland\\
% ...
$^{88}$ Department of Physics and Astronomy, Michigan State University, East Lansing MI, United States of America\\
% ...
$^{111}$ Homer L. Dodge Department of Physics and Astronomy, University of Oklahoma, Norman OK, United States of America\\
$^{112}$ Department of Physics, Oklahoma State University, Stillwater OK, United States of America\\
% ...
$^{115}$ LAL, Univ. Paris-Sud and CNRS/IN2P3, Orsay, France\\
% ...
$^{118}$ Department of Physics, Oxford University, Oxford, United Kingdom\\
% ...
$^{153}$ Raymond and Beverly Sackler School of Physics and Astronomy, Tel Aviv University, Tel Aviv, Israel\\
% ...
\end{flushleft}

\end{document}